\documentclass[11pt,hyperref,twoside]{cernyrep}

\usepackage{booktabs}
\usepackage{subfiles}
\usepackage{subfigure}
\usepackage[export]{adjustbox}
\usepackage[svgnames,table]{xcolor}
\usepackage[backend=biber,natbib=true,style=ieee]{biblatex}

\hypersetup{
  urlcolor=Blue,
  citecolor=Black,
  linkcolor=Black,
}

\usepackage{fancyhdr}
\fancyhfoffset{4 mm}
\fancypagestyle{ARTTITLE}{%
\fancyhf{} 
\lhead{Proceedings of the CERN--Accelerator--School course on
{\it Introduction to Accelerator Physics}}
\lfoot{Available online at \url{https://cas.web.cern.ch/previous-schools}}
\rfoot{\thepage\hspace*{3mm}}
 
}

\begin{document}

\title{Collective Effects - an introduction} \author{Kevin Li} \institute{CERN, Geneva, Switzerland}
\maketitle
\thispagestyle{ARTTITLE}

\begin{abstract}
  Collective effects in particle accelerators are one of the key constituents for determining the ultimate particle accelerator performance. Their role is becoming increasingly important as particle accelerators are being pushed ever closer towards the intensity and beam brightness frontiers. They are slightly peculiar in their nature as their impact and significance depend not only on external fields but also on the beam properties themselves. This results in a highly coupled and convoluted system. In these lectures we will give a brief overview over collective effects in particle accelerators in general. We will cover the topics in a highly conceptual and illustrative manner. The goal will be for the students to get an intuitive impression on the nature and the aftermath of collective effects. The lectures will cover different types of collective effects along with their manifestation in accelerators and briefly outline the limitations they impose along with a few means for potential mitigation techniques. 
\end{abstract}

\section{Introduction and concepts}
\label{sec:intro}

  Collective effects in particle accelerators and beam dynamics are another interesting and important topic; in fact, they are one of the key limiting phenomena in particle accelerators with respect to the reachable beam intensity and brightness. As such, collective effects play an important role in determining the ultimate performance of a particle accelerator.
  
  Collective effects are ideally taken into account at the design stage of a particle accelerator; they should be revisited for any upgrade that foresees an increase of the intensity or the beam brightness of the particle accelerator. As many of today's accelerators are being pushed towards their limits in terms of intensity and beam brightness, either by design, or through upgrades (i.e., the LHC Injectors Upgrade - LIU Project \cite{Damerau:1976692}), collective effects are becoming increasingly important. It is, therefore, essential to have a good understanding of the origin and the nature, as well as the impacts and the consequences of collective effects on particle accelerators and beam dynamics. This will help to ultimately make good design choices and foresee adequate mitigation methods against collective effects in order to ensure the ultimate performance reach of any particle accelerator. 
  
  Collective effects are challenging to model due to their intrinsically self-consistent nature. The charged particle beam dynamics can not be treated within a static environment of electromagnetic fields. Instead, the charged particle beam affects and alters its environment, changing the electromagnetic fields which, in turn, affects the charged particle beam. Thus, charged particle beams feed back onto themselves via the machine environment, forming a loop which needs to be solved self-consistently. This makes the theoretical modeling and the treatment and the understanding of collective effects in general rather difficult.
  
  The goal of these lectures will not be to give a full and in-depth theoretical understanding of the beam dynamics of collective effects. Rather, we will try and give an intuitive and illustrative introduction to collective effects and their signatures. Thus, a basic understanding of collective effects should be obtained. We will outline their origin and nature. We will show some of their signatures and discuss their impact on the machine and on the beam. In the course of these lectures, we will also, just very briefly, outline the theoretical description and treatment of collective effects before then moving to specific examples of collective effects in actual particle accelerators. We will illustrate the different types on real world examples mostly from the CERN accelerators complex.
  
  The lectures are organized as follows: Section~\ref{sec:multiparticle} introduces multi-particle systems and some of the concepts linked to multi-particle systems in absence of collective effects. The section will discuss some of the different forms for the representation of multi-particle systems, coherent and incoherent motion and a few effects particular to multi-particle systems such as fileamentation and decoherence. In Section~\ref{sec:spacecharge} we will move to a first actual form from of collective effects, namely spacecharge. We will discuss direct and indirect spacecharge effects and their implications for beam dynamics and machine operation. Next, in section~\ref{sec:wakesimpedances} we will look into the concept of wake fields and impedances and show how these can used to treat more complex machine environments and to evaluate the influence of the beam on the machine environment and vice-versa. Finally, in section~\ref{sec:instabilities} we treat the subject of coherent beam instabilities and investigate some of the different forms and signatures with which these can occur. Mostly transverse coherent instabilities will be discussed, but we will briefly address also instabilities in the longitudinal plane at the very end. There are a few excellent books that treat collective effects in particle accelerators and beam dynamics in much more depth; these are for instance \cite{Chao:246480, Ng:1012829, Wolski:1622200}, just to name a few.

\newpage

\section{Multi-particle systems}
\label{sec:multiparticle}

  This section will treat the description and the dynamics of multi-particle systems in absence of collective effects. Multi-particle systems are indeed to be distinguished from single particle systems; they have their own specific features and implications which one should be aware of. Sometimes, multi-particle effects are mistaken for being collective effects; we will try to make a clear distinction here. Simply put, multi-particle effects can be observed with zero-charge particles, whereas collective effects only come into play, once finite-charge particles are considered.
  
  We will briefly revise some of the basic equations to describe he most simple single particle dynamics. After this we will move towards the description of multi-particle systems. We will discuss coherent and incoherent motion and show some typical observations linked to multi-particle systems such as decoherence and filamentation.

\subsection{Accelerator and particle beam coordinates - a reminder}

\begin{figure}
      \centering
      \includegraphics[width=0.495\linewidth]{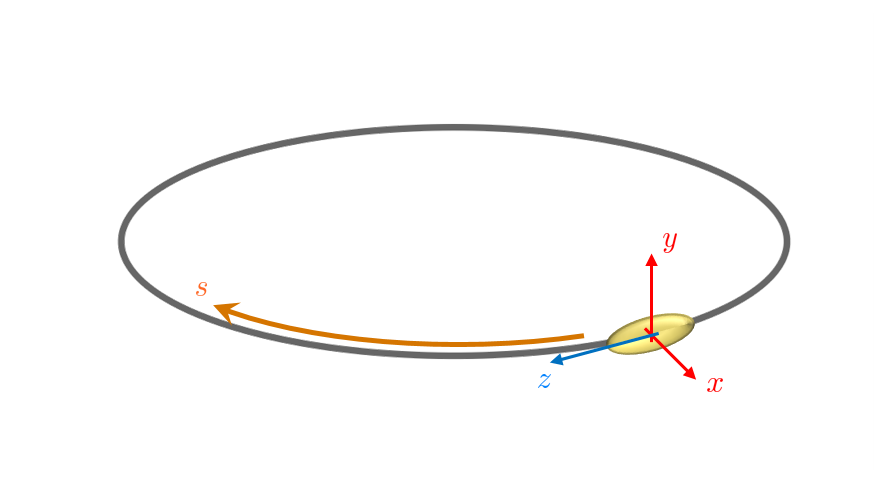}
      \includegraphics[width=0.495\linewidth]{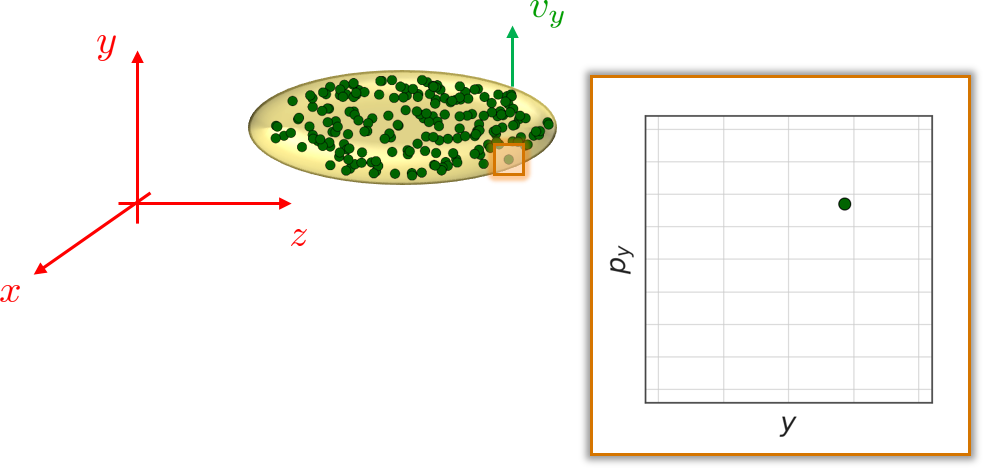}
      \caption{Accelerator and particle beam coordinates - notations and conventions used in these lectures. Positional coordinates in the accelerators are indicated on the left, phase space coordinates are shown on the right.}
      \label{fig:coordinates}
  \end{figure}

  For the sake of completeness and consistence we will briefly put together the notations and conventions used to describe accelerator and beam coordinates. All coordinates are illustrated in Fig~\ref{fig:coordinates}. The accelerator coordinates are:
  \begin{itemize}
      \item $s$: position along the accelerator
      \item $C$: accelerator circumference
  \end{itemize}
  The particle bunch coordinates are:
  \begin{itemize}
      \item $(x, y)$: transverse position with respect to orbit
      \item $z$: longitudinal position with respect to reference (synchronous) particle
  \end{itemize}
  The phase space coordinates are denoted as:
  \begin{itemize}
      \item $\left[ \begin{pmatrix} x\\p_x \end{pmatrix}, \begin{pmatrix} y\\p_y \end{pmatrix}, \begin{pmatrix} z\\p_z \end{pmatrix} \right] \in \bm{\Omega} \subset \mathbb{R}^N$: representation of particles as unique state in phase space
  \end{itemize}
  
  For those students who are not familiar with phase space coordinates and phase space dynamics, they are advised to familiarize themselves with these concepts, which are very well described in \cite{Goldstein:572817}, for example.

\subsection{Multi-particle systems and their representation}

  Single particle dynamics essentially treats individual charged particles within fixed external electromagnetic fields. In particular, single particle dynamics is entirely oblivious to any effects subject to, or involving any particle distributions. Of course, these external electromagnetic fields can be arbitrarily complex, non-linear, periodic or else time-dependent. Very rich dynamics can already evolve from these configurations as becomes clear when looking into the topic of non-linear beam dynamics and resonances, for instance. There are a variety of different sorts of fixed external electromagnetic fields. Examples include for instance the fields generated by the slowly ramping bending magnets, pulsed magnets such as injection or extraction kicker, or RF cavities used for bunching and for shaping of the longitudinal bunch profile. In the most simple case where the external fields are composed of simple alternating quadrupoles (plus the bending magnets to define the beam orbit), the resulting equation of motion in the transverse plane is given by Hill's equation describing the betatron motion
  
  \begin{align}
    x'' + K^2(s) x^2 &= 0\,, \quad K(s) = K(s + C)\,,\\
    x &= \sqrt{2 J \beta_x(s)}\,\cos\left(\psi(s)\right)\,,\\
    x' &= \sqrt{\frac{2 J}{\beta_x(s)}} \left(\sin\left(\psi(s)\right) - \frac{\beta'_x}{2} \cos\left(\psi(s)\right)\right) 
  \end{align}
  
  Equally simple, the longitudinal equation of motion for a single charged particle passing periodically through and RF cavity can be written as
  \begin{align}
    z' &= -\eta \delta\,,\\
    \delta' &= \frac{e V_\text{RF}}{m \gamma \beta^2 c^2 C}\,\sin\left(\frac{2\pi h}{C} z\right)
  \end{align}

\begin{figure}
      \centering
      \includegraphics[width=0.4\linewidth]{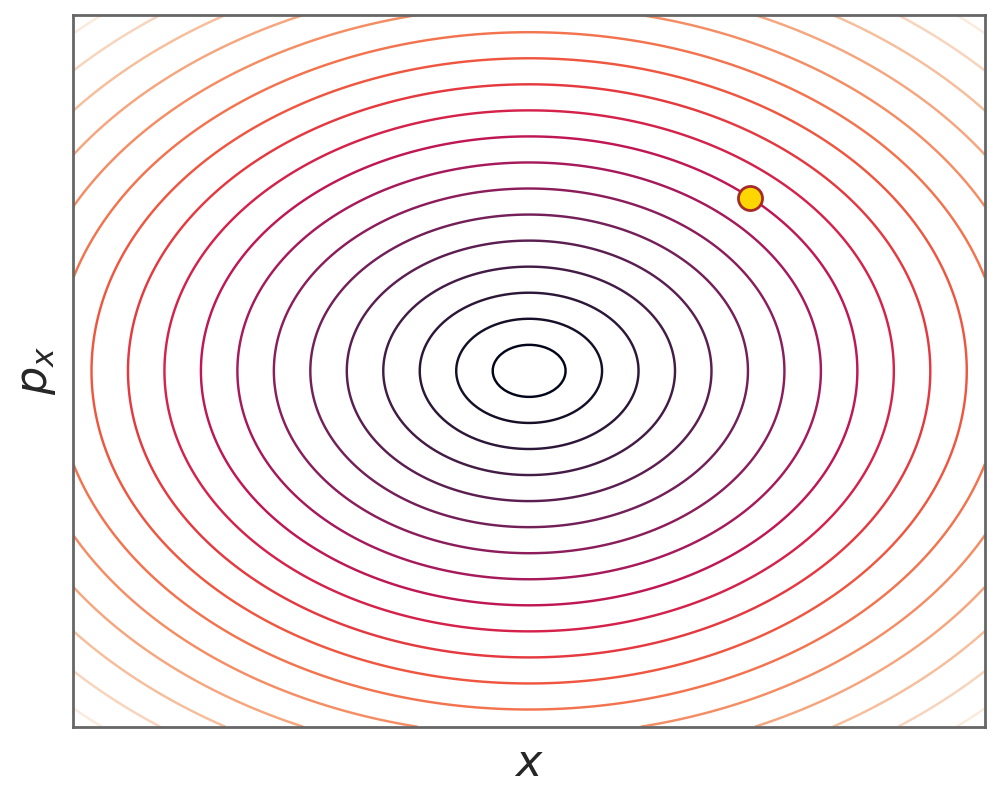}
      \includegraphics[width=0.4\linewidth]{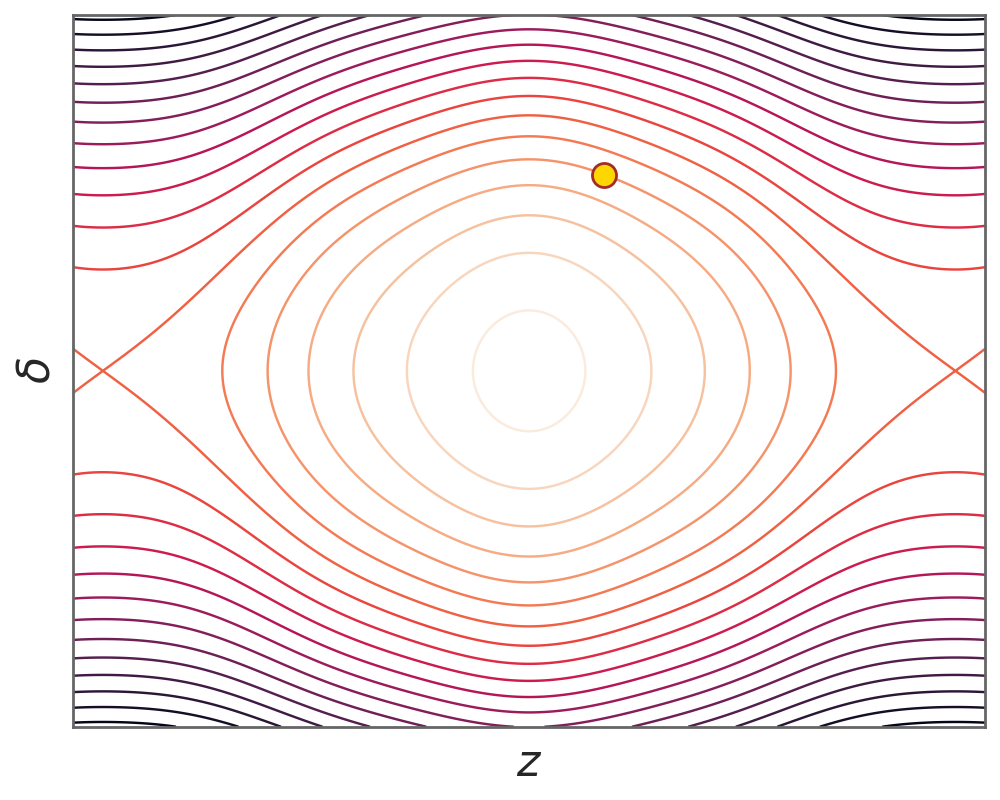}
      \caption{The phase space portrait for the single particle motion in the transverse (left) and the longitudinal (right) planes resulting in betatron and synchrotron motion, respectively.}
      \label{fig:single_particle_motion}
  \end{figure}

  Fig.~\ref{fig:single_particle_motion} shows the resulting phase space portraits for the two cases mentioned above.

  From single particle dynamics and the Hamilton equations of motion it becomes clear that a single particle, in terms of its dynamics, is fully and uniquely described by its generalized coordinates and momenta. Hence, it is perfectly well represented by its single particle state $\bm{\left(\vec{p}, \vec{q}\right)_1}$ as
  \begin{align}
      \left(\vec{p}, \vec{q}\right)_1 \equiv & \left(x_1, p_{x1}, y_1, p_{y1}, z_1, p_{z1}\right)
  \end{align}
  
  Consequently, a multi-particle state $\bm{\left(\vec{p}, \vec{q}\right)_N}$ is simply represented as the concatenation of multiple single particle states as
  \begin{align}\label{eq:mp_state_1}
      \left(\vec{p}, \vec{q}\right)_N \equiv & \left( x_1, p_{x1}, y_1, p_{y1}, z_1, p_{z1}, \right.\\\nonumber
        & \,\left. x_2, p_{x2}, y_2, p_{y2}, z_2, p_{z2}, \right.\\\nonumber
        & \,\left. \ldots \ldots \right.\\\nonumber
        & \left. x_N, p_{xN}, y_N, p_{yN}, z_N, p_{zN} \right)
  \end{align}
  
\begin{figure}
      \centering
      \includegraphics[width=0.8\linewidth]{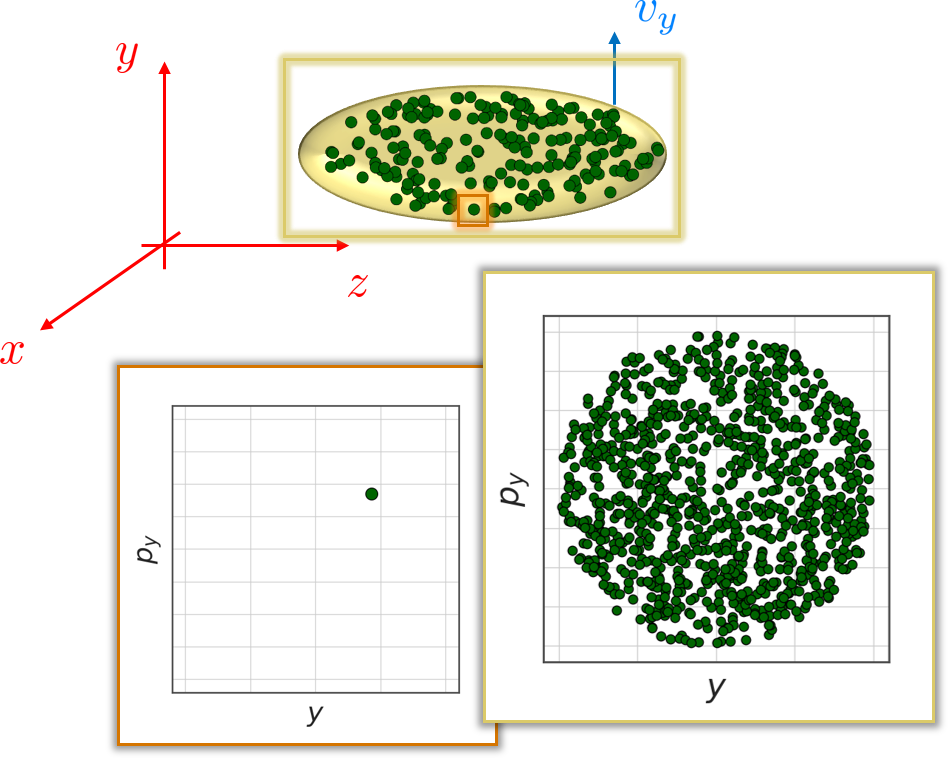}
      \caption{A simple graphical representation of a single- and a multi-particle state - $\bm{\left(\vec{p},\vec{q}\right)_1}$ and $\bm{\left(\vec{p},\vec{q}\right)_N}$ - defined by their generalized coordinates and momenta. The top image shows a particle bunch in real space, the bottom plots depict the (vertical) phase space of the single particle (orange) and the multi-particle system (yellow).}
      \label{fig:repr_graphical}
  \end{figure}

  Fig~\ref{fig:repr_graphical} shows a graphical representation of a single and a multi-particle state (it's projection onto one plane in phase space).
  
  Instead of describing a multi-particle state by Eq.~(\ref{eq:mp_state_1}), it is equivalent - but for theoretical studies often more convenient - to use as representation the probability density function $\bm{\Psi}$, which using the BBCKY-hierarchy \cite{Huang:295894} can be reduced to the single particle probability density functions $\bm{\psi}$, defined by the following properties
  \begin{align}
      P\bigr|_{(\vec{q}, \vec{p}); t} &= \frac{1}{N}\,\psi(\vec{q},\vec{p},t)\,,\\
      1 &= \frac{1}{N} \int \psi(\vec{q},\vec{p}, t)\,d\vec{q}d\vec{p}\,.
  \end{align}
  Here $P$ is the probability at any time $t$ to find a given particle in the state $\bm{(\vec{p}, \vec{q})}$. The probability to find the particle somewhere in the entire phase space $\bm{\Omega}$ is $1$; this defines the normalization of the single particle probability density function $\bm{\psi}$.
  
\begin{figure}
      \centering
      \includegraphics[width=0.85\linewidth]{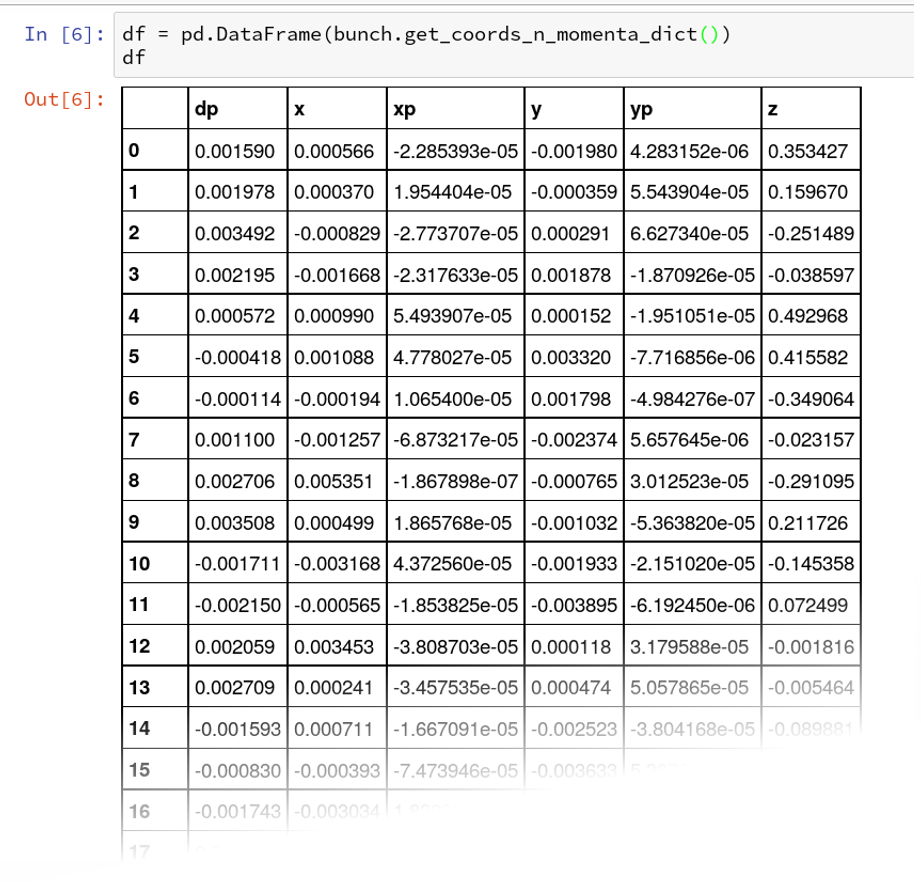}
      \caption{The numerical representation of a multi-particle state $\bm{\left(\vec{p},\vec{q}\right)_N}$. This is ultimately just a table of numbers; the table contains six columns for the six phase space variables and a number of rows which corresponds to the number of particles of the multi-particle system.}
      \label{fig:repr_numerical}
  \end{figure}

  For numerical studies, a multi-particle state $\bm{\left(\vec{p},\vec{q}\right)_N}$ is really simply described by a table of numbers containing the $6$ phase space coordinates of the $N$ particles of the multi-particle system. Fig.~\ref{fig:repr_numerical} shows a representation of such a multi-particle state, the way it is used for numerical modeling.
  
  We have thus seen three different possible representations of multi-particle states; graphical, theoretical and numerical. All these representations are used for different purposes. Before moving on, we will use the theoretical representation of multi-particle states (described as $\bm{\left(\vec{p},\vec{q}\right)_N}$) to elaborate further on the statistical properties of multi-particle systems (described by $\bm{\psi}$).
  
  We will now call the collection of all particles of a multi-particle system a multi-particle ensemble. The statistical properties of this multi-particle ensemble are then all defined by means of the single particle probability density function or, equivalently, the particle distribution function $\bm{\psi}$, e.g.,
  \begin{align}
      N &= \int \psi(\vec{q}, \vec{p})\,d\vec{q}d\vec{p}\\
      \langle x \rangle &= \int x\cdot \psi(\vec{q}, \vec{p})\,d\vec{q}d\vec{p}\\
      \sigma_x^2 &= \int \left( x - \langle x \rangle \right)^2\cdot \psi(\vec{q}, \vec{p})\,d\vec{q}d\vec{p}\,;
  \end{align}
  these are really just the statistical moments of the multi-particle ensemble (here shown for the horizontal plane; similar relations hold for the other planes). One particularly important quantity that is always used to describe the beam size in particle accelerators is the (normalized) statistical emittance which is defined by
  \begin{align}\label{eq:statistical_emittance}
      \varepsilon_x^{n,\textrm{rms}} = \sqrt{\sigma_x^2 \sigma_{p_x}^2 - (\sigma_x\sigma_{p_x})^2}\,.
  \end{align}
  The emittance essentially quantifies the area that a multi-particle ensemble occupies in phase space. It is a key quantity to describe the beam quality in an accelerator.
  
  An interesting theorem linked to this area, that a multi-particle ensemble occupies in phase space, is Liouville's theorem. This theorem states that in a typical accelerator environment (beams subject to external electromagnetic fields) a multi-particle ensemble evolves in phase space like an incompressible fluid. Liouville's theorem can be formally expressed again by means of the particle distribution function $\bm{\psi}$, as
  \begin{align}\label{eq:liouville}
       \frac{\partial \psi}{\partial t} + \sum_i^N \left( \frac{\partial \psi}{\partial q_i}\frac{\partial q_i}{\partial t} - \frac{\partial \psi}{\partial p_i}\frac{\partial p_i}{\partial t} \right) = 0
  \end{align}
  
\begin{figure}
      \centering
      \includegraphics[width=0.85\linewidth]{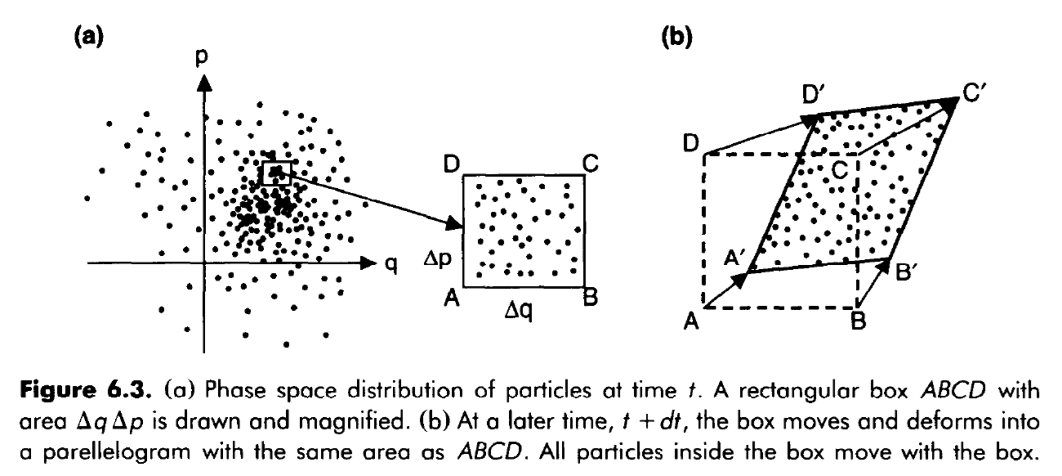}
      \caption{Evolution of the occupied area in phase space according to Liouville's theorem. The total area behaves like an incompressible fluid and it preserved. Image taken from \cite{Chao:246480}}
      \label{fig:liouville}
  \end{figure}

  In particular, this means that the total occupied are in phase space does not change\footnote{Note that the statistical emittance defined in Eq.~(\ref{eq:statistical_emittance}) is not strictly speaking the total occupied area but rather an elliptical envelope around this area; thus the normalized emittance is not necessarily always conserved, as we will see later}. This heavily limits the degrees of freedom for the motion of a multi-particle ensemble in phase space. Fig.~\ref{fig:liouville} shows and illustration of a possible evolution of a multi-particle ensemble in phase space.
  
  Liouville's theorem is a consequence of Hamilton's equations of motion which miraculously hold for charged particles in a particle accelerator (for more details on Hamiltonian systems see \cite{Huang:295894} or \cite{Goldstein:572817}). Hamilton's equations of motion are written as
  \begin{align}
      \frac{d q_i}{dt} &= \frac{\partial H}{\partial p_i}\,\\
      \frac{d p_i}{dt} &= -\frac{\partial H}{\partial q_i}\,,
  \end{align}
  for a given particle $i$. This can also be written more compactly as
  \begin{equation}
      \frac{d}{dt}\vec{x} = J\,\nabla_{\vec{x}} H\,,\quad J = \begin{pmatrix}0 & I\\-I & 0\end{pmatrix}
  \end{equation}
  Here $H$ is the Hamiltonian of the particle accelerator. This function $H$ fully describes the impact of the particle accelerator environment on the dynamics of the multi-particle ensemble. Thus, with knowledge of the accelerator Hamiltonian, the beam dynamics of a beam propagating within is in principle known. $J$ is the symplectic structure matrix and is, in a sense, responsible for the conservation of the occupied area of a multi-particle ensemble evolving in phase space. $\vec{x}$ here is the full multi-particle system state space vector - i.e., $\bm{\left(\vec{p},\vec{q}\right)}$:
  \[\vec{x} = (q_1, q_2, \ldots q_N, p_1, p_2, \ldots, p_N)\,.\]
  
  Looking now towards collective effects and their description for beam dynamics, one can, deriving from Liouville's theorem, immediately write down Vlasov's equation, which usually forms the basis of any analytically treatment of collective effects in beam dynamics
  \begin{align}\label{eq:vlasov}
      \frac{\partial \psi}{\partial t} = & \left\{H, \psi\right\} = \left\{H_\text{ext} + H_\text{coll}, \psi_0 + \psi_\text{pert}\right\}\,.
  \end{align}
  Here, $H$ is the accelerator Hamiltonian, $\psi$ is the particle distribution function and $\{\cdot,\cdot\}$ is the Poisson bracket. Again, Vlasov's equation in Eq.~(\ref{eq:vlasov}) along with all notations and the Poisson brackets are well introduced and explained in \cite{Huang:295894} or \cite{Goldstein:572817}. At this point, we just would like to retain the elegant formal expressiveness of Eq.~(\ref{eq:vlasov}); it states that the time-evolution of the single particle distribution function $\psi$ is given by the Hamiltonian acting on it via the Poisson brackets - to make this more explicit, sometimes one defines $\{H, \cdot\}$ as an operator $:H:$ such that Eq.~(\ref{eq:vlasov}) can be rewritten as
  \begin{align}\label{eq:vlasov2}
      \frac{\partial \psi}{\partial t} = & :H:\,\psi\,.
  \end{align}
  Moving back to Eq.~(\ref{eq:vlasov}), the accelerator Hamiltonian has been split into a Hamiltonian describing the fixed external fields $H_\text{ext}$ and the collective effects part $H_\text{coll}$. Similarly, the particle distribution function has been split into an equilibrium distribution $\psi_0$ and a perturbation $\psi\text{pert}$. $\psi_0$ is the solution to the equation without collective effects and corresponds to the stationary equilibrium distribution of a particle beam subject to the fixed external fields, only (in most cases). $\psi_\text{pert}$ is a perturbation on top of the stationary equilibrium distribution induced by collective effects. In this representation of Eq~.(\ref{eq:vlasov}) the separation of the external and the collective effects, as well as the manifestation of collective effects as a perturbation on the equilibrium state, becomes very obvious. The remarkable observation here is that if know these "magic" functions $H$ (and, in fact, we very often have a pretty good idea), most of the collective behavior of the beam can be (semi-)analytically calculated (see, e.g., N. Mounet \textit{''Direct Vlasov Solvers''} in \cite{Schmickler:2723239}).
  
  We will stop at this point with our theoretical excursion. The main takeaways should be that we understand the different representations of multi-particle systems; this includes their representation as the particle distribution function $\bm{\psi}$, which is usually what is chosen for the analytical treatment. We can use this representation to express important statistical properties of multi-particle ensembles. We know the (normalized) statistical emittance as one of the key properties of particle beams. We understand that the motion of charged particles in particle accelerators is described by the Hamilton equations of motion. We know the statement of Liouville's theorem. We have seen Vlasov's equation and how this can be used as starting point to embark into the analytical treatment of collective effects in beam dynamics.

\subsection{Incoherent and coherent motion}
  
  After having having worked with the representation of multi-particle systems and having now, hopefully, formed a good intuitive picture of multi-particle systems, we will now discuss some dynamics effects which are peculiar to multi-particle systems. They are not observed for single particles. We will discuss the differences of incoherent and coherent motion. We will see some implications of these differences and discuss phenomena such as decoherence and filamentation.
  
  When considering multi-particle systems one makes a difference between the incoherent ("microscopic") and the coherent ("macroscopic") motion. Loosely speaking, the incoherent motion is the motion of each individual particle within the multi-particle system. The coherent motion is the motion, that the full macro-particle system executes as an entity. Formally, one may regard the incoherent motion as the evolution of the individual particle coordinates, whereas the coherent motion is the evolution of the center-of-mass coordinates (or any other of it's statistical moments) of the full beam. Most generally, any motion involving the entire beam as an entity, i.e., all parts of the beam move in a fixed relation to one another, can be regarded as a coherent motion.
  
\begin{figure}
      \centering
      \includegraphics[width=\linewidth]{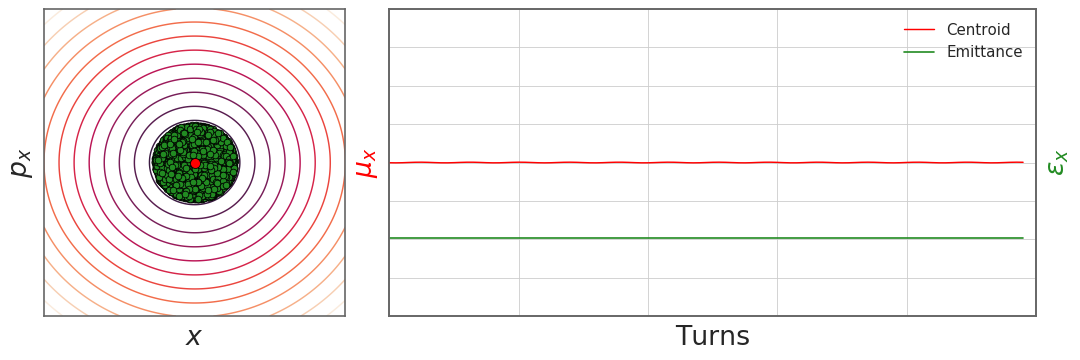}
      \caption{A beam injected on orbit. The beam's center of mass does not show any strong oscillatory motion, the beam size remains constant as shown on the right hand side. We say the beam is matched to the accelerator optics.}
      \label{fig:injection_center}
  \end{figure}

  As an example, we can imagine a particle beam injected on orbit into a perfectly linear machine. This is depicted in Fig.~\ref{fig:injection_center} where the horizontal phase space of the injected beam is shown. With the beam's center of mass (red) injected on-orbit, the full beam itself will remain circulating on-orbit around the machine hardly exhibiting any visible transverse motion. On the other hand, the individual particles (green) farther outside of the beam have actually been injected off-orbit and will consequentially perform transverse oscillations around the machine orbit. Thus, whereas there is hardly any coherent motion of the beam, there is lot's of oscillatory incoherent motion going on within the beam. The beam size, however, essentially does not change.
 
\begin{figure}
      \centering
      \includegraphics[width=0.5\linewidth]{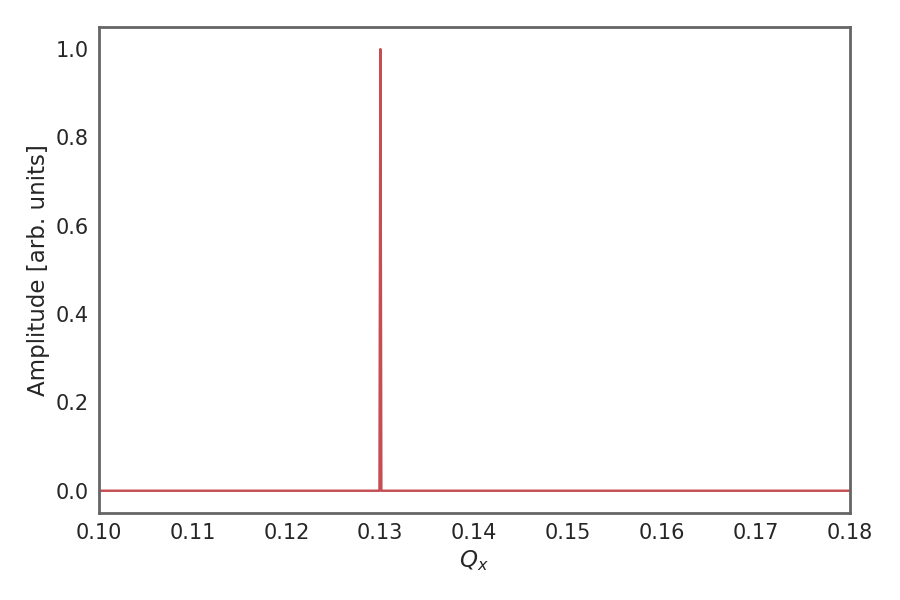}
      \includegraphics[width=0.475\linewidth]{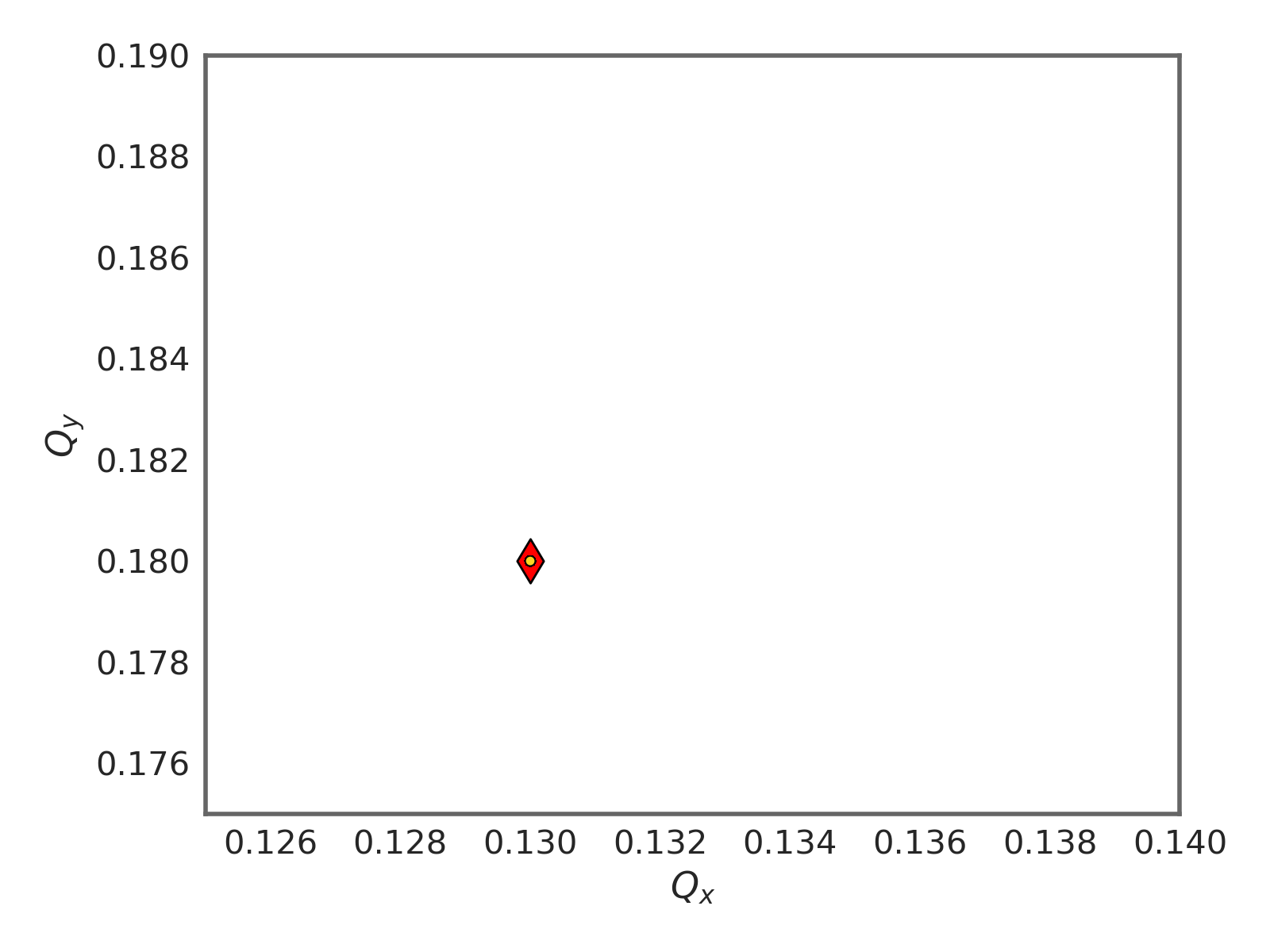}
      \caption{The tune evaluated by means of an FFT over the coherent centroid motion (left) and over each individual particle in the horizontal and in the vertical plane (right); it is evident from the plot on the right hand side, that all individual particle oscillate at the same frequency on the horizontal and the vertical planes, respectively, and together with the coherent frequency.}
      \label{fig:tunes_linear}
  \end{figure}

  One can now consider the same beam injected into the same perfectly linear machine but off-orbit. In this case, the beam's center of mass will now perform oscillations around the machine orbit; as a matter of fact, the beam's center of mass motion will be exactly the same as the motion of the individual particles within the beam. Thus, the coherent and the incoherent motion will be the same for this case. The beam size still remains constant. This situation is shown in Fig.~\ref{fig:injection_offcenter} below. One can take a Fast Fourier Transform (FFT) of the coherent motion in order to get the oscillation frequency components and to evaluate the coherent tune. One obtains a sharp line as depicted on the left hand side of Fig.~\ref{fig:tunes_linear}. If one does the same for each of the individual particles within the beam using the horizontal and the vertical motion to evaluate the corresponding horizontal and vertical tunes, and one represents each particle as a point in tune space, one obtains an image as shown on the right hand side in Fig.~\ref{fig:tunes_linear}; clearly, all individual particles within the beam oscillate at the same frequency. 
  
  \begin{figure}
      \centering
      \includegraphics[width=\linewidth]{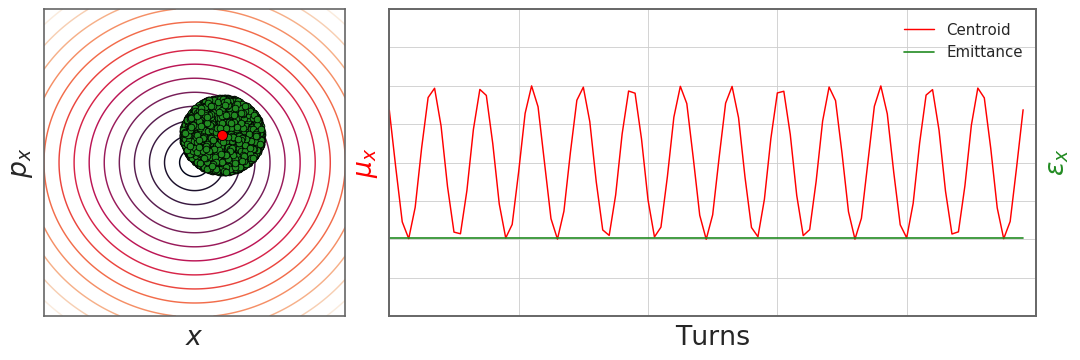}
      \caption{A beam injected off orbit into a linear machine. The beam's center of mass now features a clear oscillatory motion. This is the same motion that is performed also by the individual particles within the beam. The beam size still remains constant as the machine is linear. We say the beam has been injected with a mismatch into the accelerator.}
      \label{fig:injection_offcenter}
  \end{figure}
  
\begin{figure}
      \centering
      \includegraphics[width=\linewidth]{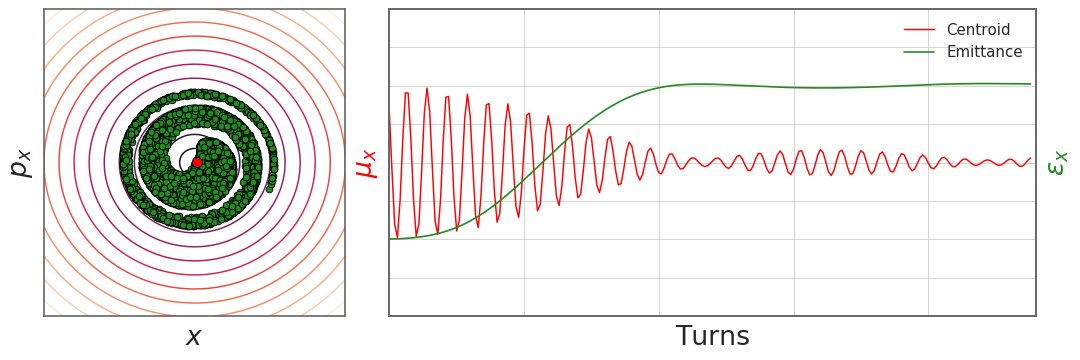}
      \caption{A beam injected off orbit into a non-linear machine. The beam's center of mass motion now decoheres as the beam filaments in phase space. At the same time, a blow-up of the beam's emittance is observed.}
      \label{fig:injection_filamentation}
  \end{figure}

\begin{figure}
      \centering
      \includegraphics[width=0.5\linewidth]{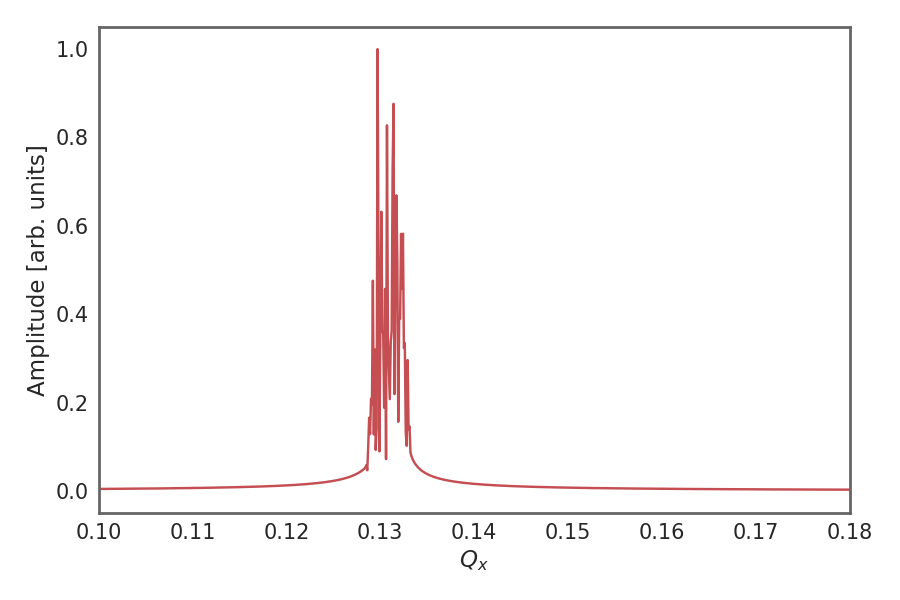}
      \includegraphics[width=0.475\linewidth]{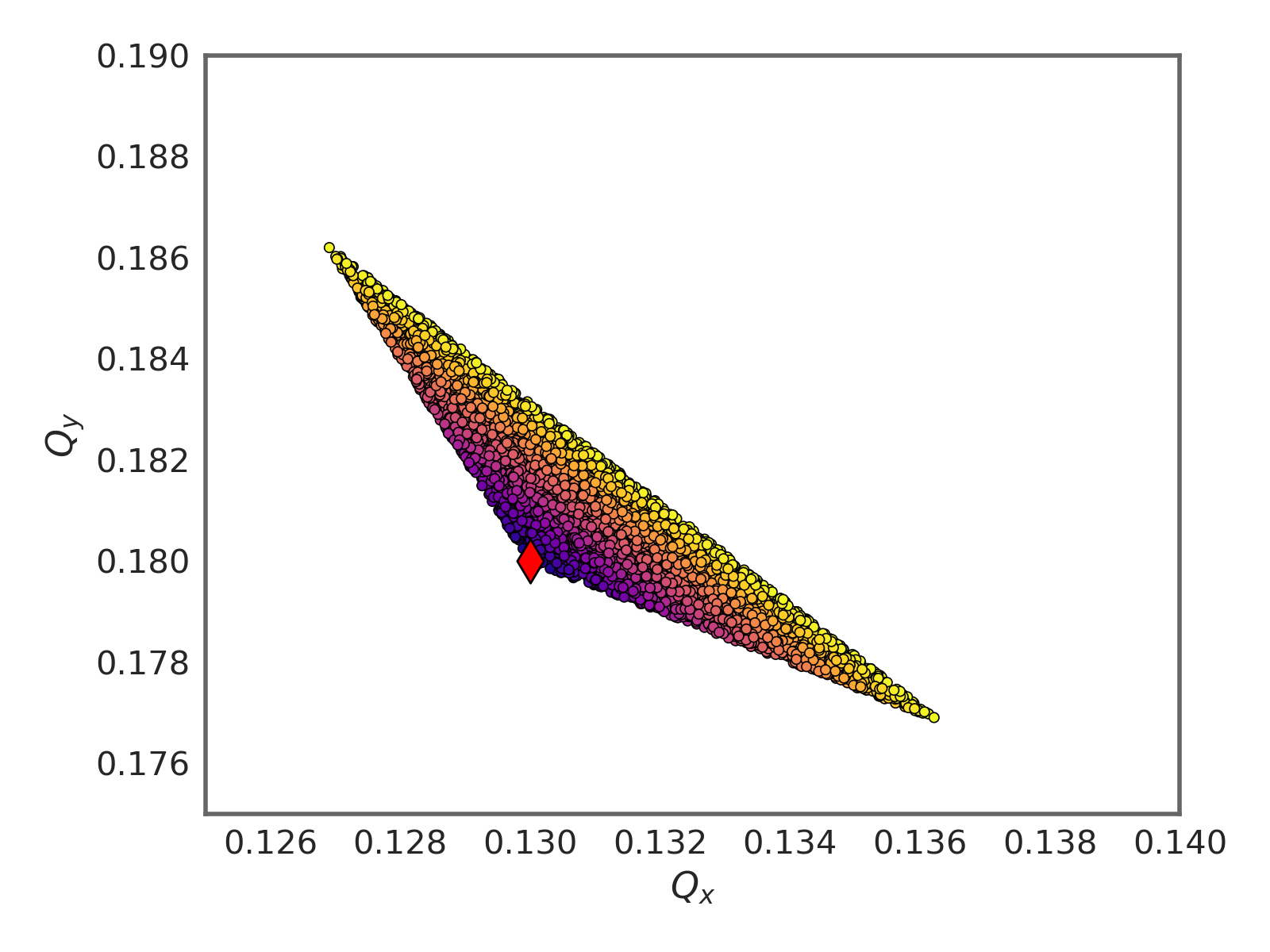}
      \caption{The tune evaluated by means of an FFT over the coherent centroid motion (left) now shows a spread over several frequencies. Performing adn FFT of the motion of each individual particle, and plotting the resulting horizontal and vertical tune in tune space gives the image shown on the right hand side. This is the typical tune footprint induced by octupoles in a machine.}
      \label{fig:tunes_filamentation}
  \end{figure}

  The situation becomes interesting if one now considers the same scenario as above, but within a machine that contains non-linearities. In particular, we assume the machine contains octupoles, which lead to detuning with amplitude; that is, particles farther away from the magnetic center will exhibit a different tune compared to particles in the magnetic center which will experience no detuning. If we now imagine the beam injected off-center into the machine and observe its evolution, we will see a behaviour similar to the one depicted in Fig.~\ref{fig:injection_filamentation}. We can make a few remarkable observations; the injected beam initially oscillates around the machine orbit. At the same time, the individual particles withing the beam oscillate around the same orbit but, due to the detuning with amplitude, they do this at different tunes. In the example below, particles farther away from the center have a larger tune. As a result, these particles rotate faster in phase space. This leads to a spiraling of the beam in phase space as particles farther outside overtake the particles in the center resulting in a phase space depicted on the left hand side of Fig.~\ref{fig:injection_filamentation}. We call this spiraling effect the filamentation of the beam. An important consequence linked to this effect is the emittance blow-up and the decoherence of the beam which can be clearly seen on the right hand side of Fig.~\ref{fig:injection_filamentation}. As the beam "smears out" in phase space particles distribute over a larger envelope area which leads to the emittance increase.\footnote{It is worth noting, however, that the actual occupied area in phase space does not change and thus Liouville's theorem is not violated; the blow-up simply originates from the rms-based definition of the emittance.} At the same time, the beam as a whole "homogenizes" as the center of mass moves closer towards the orbit and thus the center-of-mass oscillation amplitude reduces. We call this the decoherence of the beam. If one evaluates the tune from the coherent motion of the beam one no longer obtains a sharp tune line, but rather a tune distribution as shown in Fig.~\ref{fig:tunes_filamentation}. A tune analysis of each individual particle renders the plot shown on the right hand side in Fig.~\ref{fig:tunes_filamentation} where, again, each particle is plotted by means of its horizontal and vertical tune. The plot shows a typical tune spread that is obtained from octupoles. While this tune spread can lead to filamentation and emittance blow-up if a beam is not well injected, it also provides an important stabilization mechanism against coherent instabilities. This is a separate topic, however.  
  
\begin{figure}
      \centering
      \includegraphics[width=\linewidth]{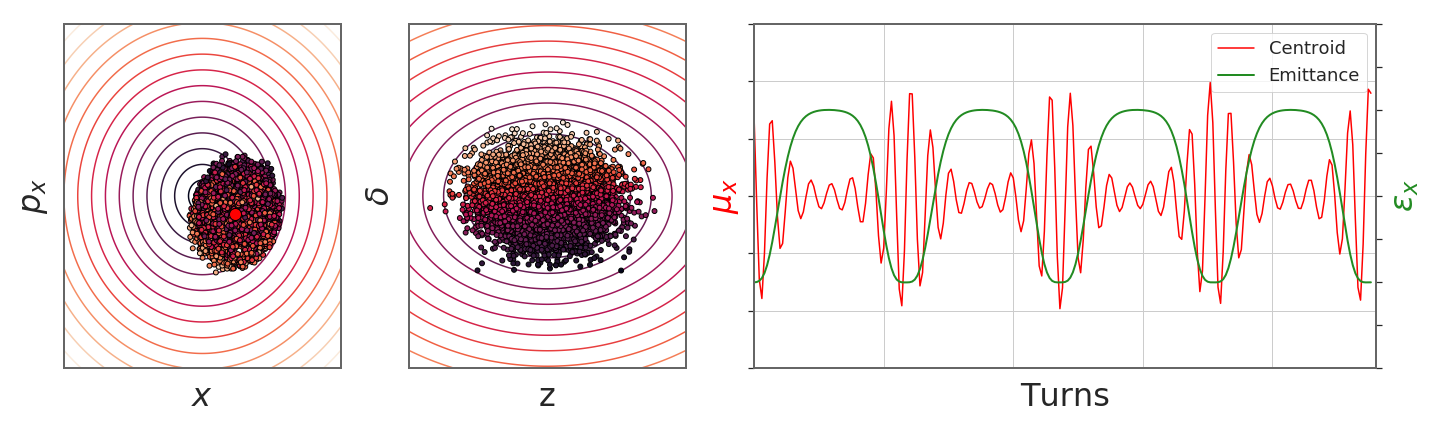}
      \caption{Typical decoherence -- recoherence behaviour for a beam injected off orbit into a machine with finite chromaticity. the beam fully recoheres as all particle are re-aligned in both the transverse and the longitudinal phase space after one full synchtrotron period.}
      \label{fig:injection_chroma1}
  \end{figure}

  In a last example we will take into account chromatic effects within the machine. This will lead to a detuning as a function of the off-momentumness of a particle. Precisely, the machine chromaticity will lead to a detuning of a particle depending on its momentum offset with respect to the reference or the design momentum of the machine, for which a particle is always on the design orbit of the machine. This actually provides a coupling between the transverse and the longitudinal planes. A beam injected off-orbit into a machine with a finite chromaticity will again at first exhibit betatron oscillations around the machine orbit. If the beam has an energy spread (or equivalently, a momentum spread), the higher and the lower momentum particles will rotate at different rates in transverse phase space. This will again cause the to beam smear out in phase space as particles re-distribute over a larger envelope area and, at the same time, cause the beam as a whole to homogenizes as the center of mass moves closer towards the orbit and the center-of-mass oscillation amplitude reduces. However, due to the synchrotron motion (which we will assume to be linear, for now) particle continuously keep changing their momentum offset; after one full synchrotron period, all particles have relined-up in the transverse and the longitudinal phase space. As a consequence, the beam has re-cohered, the beam emittance and oscillation amplitudes have recovered back to their initial values. This is well illustrated in Fig.~\ref{fig:injection_chroma1}. This decoherence -- recoherence effect is typical for beam with an energy spread in machines with finite chromaticity. 

  There are different sources for these non-linearities leading to particle detuning within a machine; some of them are actually used deliberately to help combat collective effects and coherent instabilities. Two of the most important quantities have already been mentioned above. These are

  \begin{itemize}
      \item Amplitude detuning
      \begin{itemize}
          \item Controlled with octupoles – provides (incoherent) tune spread
          \item Leads to absorption of coherent power into the incoherent spectrum suppressing the emenation of coherent instabilities
      \end{itemize}
      
      \item Chromaticity
      \begin{itemize}
          \item Controlled with sextupoles – provides chromatic shift of bunch spectrum wrt. impedance
          \item Changes interaction of beam with impedance
          \item Damping or excitation of headtail modes
      \end{itemize}
  \end{itemize}

This concludes the first chapter of these lectures. We have seen and also worked with multi-particle systems $\bm{\bigl[\vec{p},\vec{q}\bigr]}$, their representation, their analytical treatment and some of their fundamental properties and behaviour. We have discussed incoherent and coherent motions. We have looked at a few examples that nicely highlighted the difference between incoherent and coherent options and illustrate the effects of filamentation, decoherence and emittance blow-up. It is to be noted that none of these effects have anything to do with collective effects, as the particle charges have never entered the equation so far.
\clearpage\pagebreak
\section{Space charge}
\label{sec:spacecharge}

  We will now see a first ''real'' collective effect, namely the space charge effect. Here, the beam charge plays a central role. The larger the charge, the larger the effect. This is what makes it, essentially, an actual collective effect. 
  
  We will first discuss the meaning and the impact of direct space charge effects and then move over to describing in more detail the indirect space charge effects. We will also show means of how to mitigate space charge effects. For some of these impacts we will give a more formal description, for others we will stick to an intuitive picture. Lots of the material presented here has been taken directly from \cite{Schindl:941316}.
  
\begin{figure}[htbp]
      \centering
      \subfigure[Direct space charge in open space]{
        \includegraphics[width=0.435\linewidth]{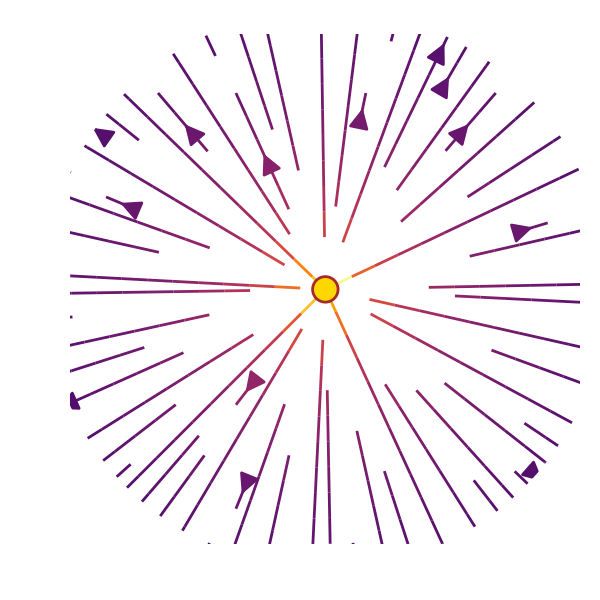}}
      \subfigure[Indirect space charge with material boundaries]{
        \includegraphics[width=0.450\linewidth]{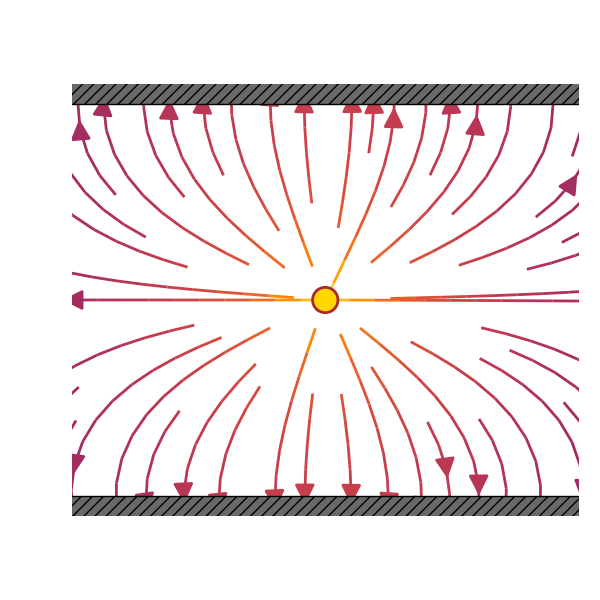}}
      \caption{Illustration of direct (left) and indirect (right) space charge effects by means of the electric field lines of a particle in open space and within a bounded region.}
      \label{fig:fields}
  \end{figure}

  Looking at Fig.~\ref{fig:fields}, we can already get an impression of the differences between direct and indirect space charge effects. Direct space charge describes the interaction of particles among each other in free space. Indirect space charge, in contrast, specifies the interaction of particles via the surrounding environment in a bounded space such as a metallic wall, for instance. 

\subsection{Direct space charge – impact on machine performance}

  To start understanding the direct space charge effects one often begins by considering two point-like charged particles carrying an equal charge of $q_1 = q_2 = q$ travelling parallel next to each other at equal velocity of some fraction of the speed of light $v_1 = v_2 = v$. We can denote one particle as the source particle and the other particle as the witness particle. We are interested in the total forces induced by the source particle and experienced by the witness particle  - actually, this approach is fundamental to any consideration dealing with collective effects and we will encounter it again later. If both particles are at rest, the witness particle feels the Coulomb force exerted by the source particle. If the two equally charged particles travel close to the speed of light and in the same direction, Ampere's law leads to an attraction of the two particles which can nearly compensate the Coulomb force. Thus, at rest, the direct space charge effect is the strongest and diminishes as particle approach the speed of light. 
  
\begin{figure}[htbp]
      \centering
      \includegraphics[width=0.8\linewidth]{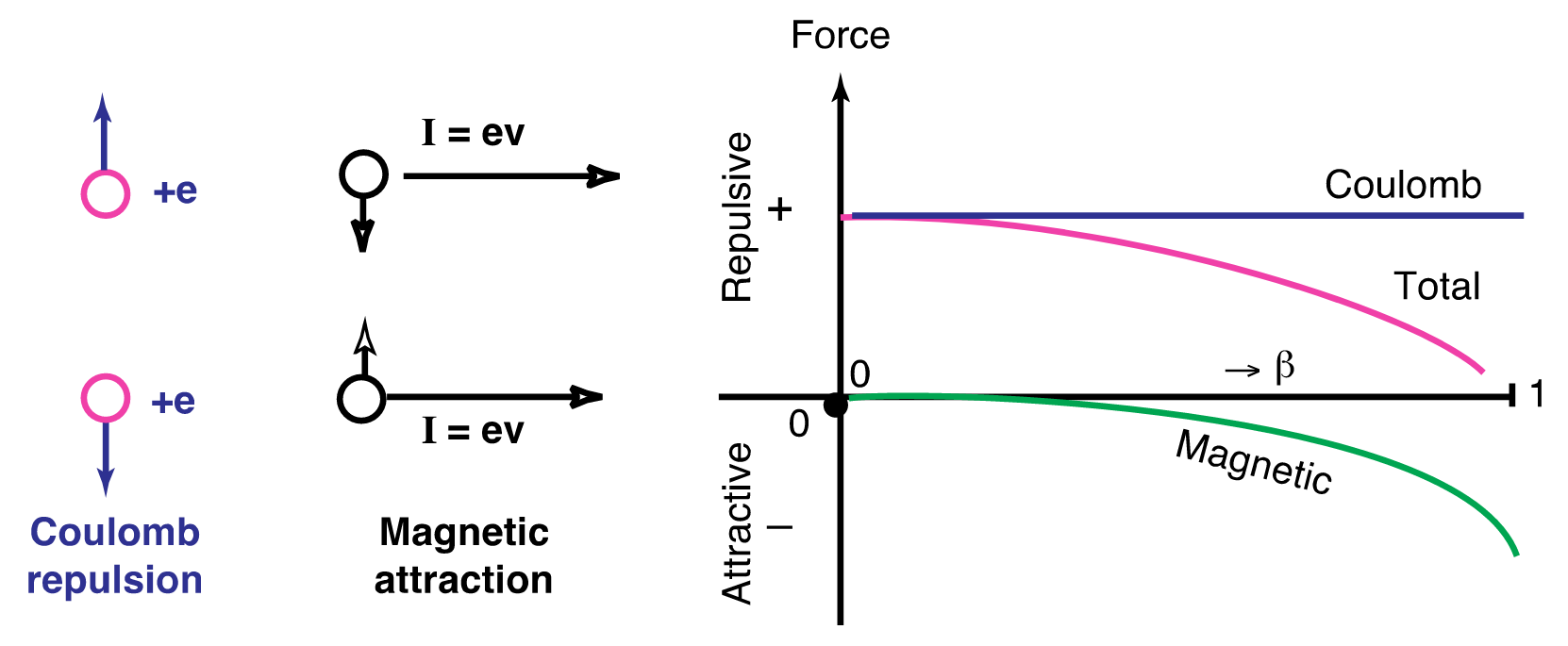}
      \caption{Two point-like charged particles propagating parallel next to each other at a given velocity. The net forces experienced by the particles is the sum of the repulsive Coulomb force and the attractive magnetic force. As the velocity approaches the speed of light, the two forces cancel each other.}
      \label{fig:lorentz}
  \end{figure}

  More formally, one can consider a situation as depicted in Fig.~\ref{fig:lorentz}; here, the electric and magnetic fields generated by the source particle travelling at velocity $v$ and separated by a distance $r$ can be written in the lab frame as
  \begin{align}
      E_r & = \frac{e}{4\pi \epsilon_0} \frac{\gamma}{r^2}\,, & B_\phi & = \frac{\beta E_r}{c}\,.
  \end{align}
  With this, the Lorenz force experienced by the witness particle can be written in the lab frame as
  \begin{align}\label{eq:lorentz}
      F_r &= e \left( E_r - v B_\phi \right) = e \left( E_r - \beta^2 E_r \right) = \frac{E_r}{\gamma^2} = \frac{e}{4\pi\epsilon_0\gamma r^2}\,.
  \end{align}
  On the dependence of the relativistic $\gamma$ factor in Eq.~\ref{eq:lorentz} it becomes immediately clear that the direct space charge effect among particles diminishes, as these particles approach the speed of light where $\gamma\rightarrow\infty$. 
  
  We can now move from point-like charged particles to a uniformly charged coasting beam (which can be seen as uniformly charged cylinders of infinite length). We can again compute the electric and magnetic fields experienced by particles within the coasting beam. These can be computed from Gauss' and Stokes' law, respectively, as
  \begin{align}\label{eq:e-fields_coasting}\nonumber
      \iiint \vec{\nabla} \cdot \vec{E}\,dV &= \iint \vec{E}\,d\vec{S}\,,\quad \vec{\nabla} \cdot \vec{E} = \frac{\rho}{\epsilon_0}\,\\\nonumber
      \Rightarrow \pi r^2 l \frac{\rho}{\epsilon_0} &= 2\pi r l E_r\,,\\
      \Rightarrow E_r &= \frac{\lambda}{2\pi\epsilon_0}\frac{r}{a^2}\,,\quad r<a
  \end{align}
  
  and as
  
  \begin{align}\label{eq:b-fields_coasting}\nonumber
      \iint \vec{\nabla} \times \vec{B}\,d\vec{S} &= \oint \vec{B}\,d\vec{s}\,,\quad \vec{\nabla} \times \vec{B} = \mu_0 \vec{J}\,,\\\nonumber
      \Rightarrow&\: \pi r^2 \mu_0 J = 2\pi r B_\phi\,,\\
      \Rightarrow&\: B_\phi = \frac{\lambda\beta}{2\pi\epsilon_0 c}\frac{r}{a^2}\,,\quad r<a\,.
  \end{align}
  The total force experienced by particles within the coasting beam is again simply the Lorentz force which, using the aforementioned electric and magnetic fields, is written as
  \begin{align}\label{eq:lorentz_coasting}
      F_r &= e \left( E_r - v_s B_\phi \right) = \frac{e\lambda}{2\pi\epsilon_0} \left( 1 - \beta^2  \right) \frac{r}{a^2} = \frac{\lambda e}{2\pi\epsilon_0 \gamma^2a^2}\,r\,.
  \end{align}
  Similarly to the chase of the two point-like particles it again becomes clear from Eq.~(\ref{eq:lorentz_coasting}) how the direct space charge effect among particles diminishes as these particles approach the speed of light as $\gamma\rightarrow\infty$.
  
  One can now make a further analysis in order to understand how the particle motion is affected for particles travelling within the coasting beam. There, the direct space charge forces are written in the horizontal and the vertical plane as
  \begin{align}\label{eq:lorentz_plane}
      F_x &= \frac{e\lambda}{2\pi\epsilon_0\gamma^2 a^2}\,x \,, & F_y &= \frac{\lambda e}{2\pi\epsilon_0\gamma^2 a^2}\,y \,,
  \end{align}
  Clearly, from Eqs.~(\ref{eq:lorentz_plane}) it can be seen that the forces in either plane are linear in the offsets, and always in the positive direction of the respective plane. This means these forces actually act as a defocusing quadrupole in both planes. Consequentially, we can look at the vertical plane, for instance, and write down Hill's equation governing the transverse motion, taking into account an additional term originating from the direct space charge forces. In this case, Hill's equation becomes
  \begin{align}
      y'' + \left( K_y(s) + K_y^\text{SC}(s) \right)\, y = 0\,,
  \end{align}
  where the extra focusing is determined by
  \begin{align}
      K_y^\text{SC} &= -\frac{1}{m\gamma\beta^2 c^2} \frac{F_y^\text{SC}}{y} = -\frac{2 r_0\lambda}{e\beta^2\gamma^3 a^2(s)}\,,\quad \text{with}\quad r_0 = \frac{e^2}{4\pi\epsilon_0 m c^2}\,.
  \end{align}
  The additional tune term, the tune shift, is calculated from the coefficients in Hills equation as
  \begin{align}
      \Delta Q = \frac{1}{4\pi} \oint K_y^\text{sc}(s) \beta_y(s)\,ds = -\frac{1}{4\pi} \oint \frac{2 r_0\lambda\beta_y(s)}{e\beta^2\gamma^3 a^2(s)}\,ds = -\frac{r_0 R\lambda}{e\beta^2\gamma^3}\,\left< \frac{\beta_y(s)}{a^2(s)} \right>\,.
  \end{align}
  Here 
  \[a(s) = \sqrt{\frac{\beta_{x, y}(s)\hat{\varepsilon}_{x, y}^n}{\beta\gamma}}\,,\]
  and $\hat{\varepsilon}_{x, y}^n$ is the normalized emittance including all particles. With this, the tune shift of a particle subject to the direct space charge fields of a uniform charge distribution is given by
  \begin{align}\label{eq:tuneshift_direct}
      \Delta Q_{x, y} &= -\frac{r_0 R\lambda}{e\beta\gamma^2\hat{\varepsilon}_{x, y}^n}\,.
  \end{align}
  Eq.~(\ref{eq:tuneshift_direct}) gives the direct space charge induced tune shift, at the same time highlighting a few fundamentals of direct space charge effects. The direct space charge induces tune shift:
  \begin{itemize}
      \item is negative, because space charge transversely always defocuses,
      \item is proportional to the line density and thus to the number of particles in the beam,
      \item decreases with energy like $1/(\beta\gamma^2)$ (when expressed in terms of normalized emittance) and therefore vanishes in the ultra-relativistic limit,
      \item does not depend on the local beta functions or beam sizes but is inversely proportional to the normalized emittance (remember, here, the emittance includes all particles).
  \end{itemize}
  
\begin{figure}[htbp]
      \centering
      \includegraphics[width=\linewidth]{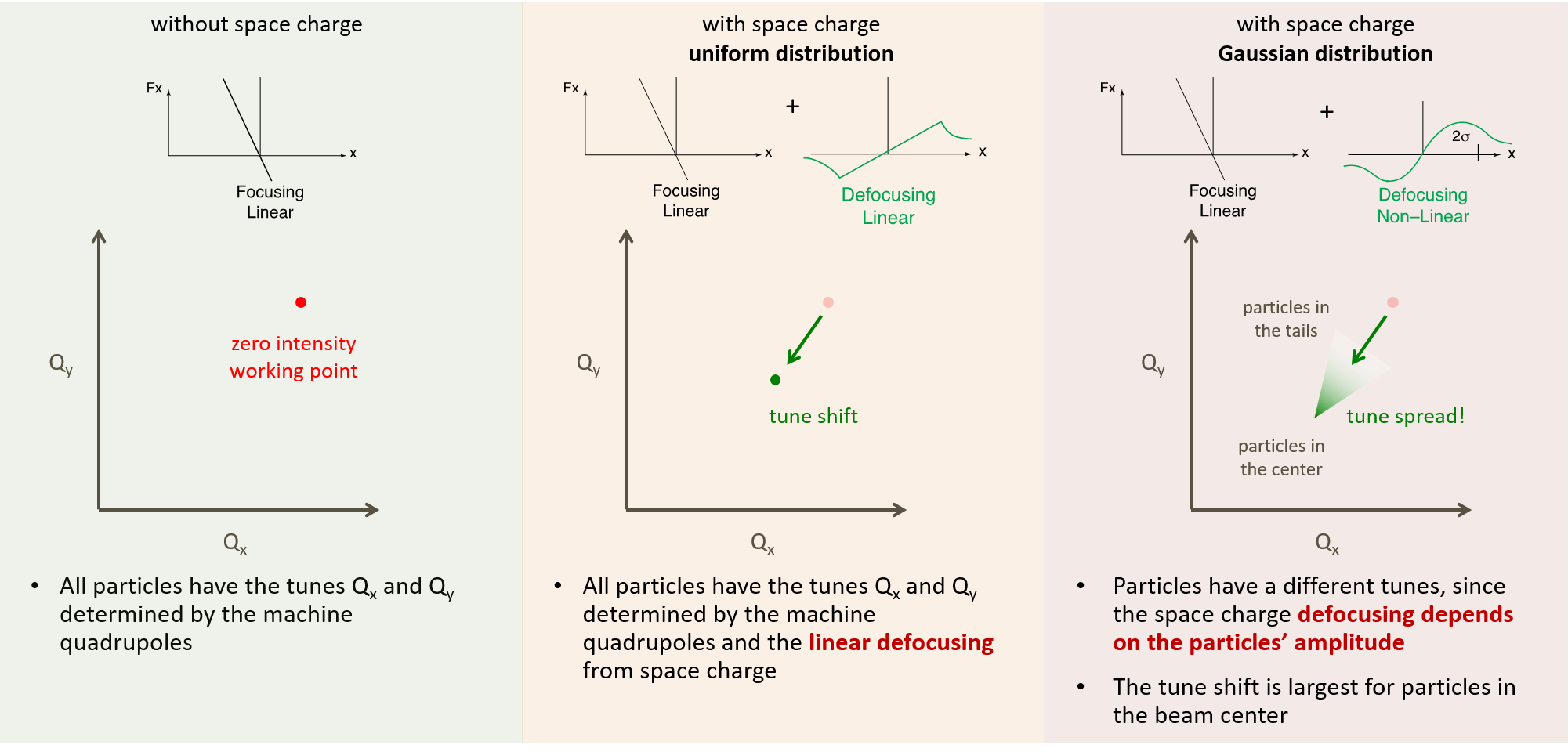}
      \caption{Different configurations of a charged particle subject to different types of direct space charge forces; the resulting detuning of particles is shown for the cases without space charge forces (green), with space charge forces within a uniformly charged beam (yellow) and with space charge forces within a Gaussian distributed beam (red).}
      \label{fig:spacecharge_direct_schema}
  \end{figure}

  Next, one can do yet another step of generalization, where one can move from considering a transversely uniform charged coasting beam to a transversely Gaussian distributed charged coasting beam. At this stage, we will no longer treat this strictly formally. We know from the above considerations that direct space charge induces a tune shift for all particles within a beam. Inspecting the situations discussed above, namely transversely uniform or Gaussian distributed charged coasting beams, we can compose a schema as depicted in Fig.~\ref{fig:spacecharge_direct_schema}.
  The figure shows the detuning to be expected for particles subject to different direct space charge forces. For the case of an individual particle within a machine, the tune will be the machine tune and corresponds to the zero intensity working point. A particle traveling within a uniformly charged coasting beam will experience a tuneshift as derived in Eq.~(\ref{eq:tuneshift_direct}). In fact, due to the uniformity of the charge distribution, all particles within the coasting beam will actually experience the same amount of tune shift. Therefore, all particles will occupy a single point in tune space which is shifted downwards from the zero intensity working point. Finally, a particle travelling within a Gaussian distributed coasting beam will also experience direct space charge forces which will lead to a tune shift. However, as opposed to the previous case where all particles experienced the same tune shift, for the Gaussian distributed case, due to the non-linearities of the resulting forces within the beam, particles at different transverse locations within the beam will experience a different amount of tune shift. Particles towards to core of the beam will actually exhibit the largest tune shift, whereas particle towards the beam edges will experience just very small tune shifts \footnote{The maximum space charge tune shift in a Gaussian beam distribution is stronger compared to the detuning in a uniform beam distribution when considering similar beam sizes due to the higher particle density in the beam}. The collection of particles within the beam will span a triangular area in tune space; this is called the tune footprint of the beam.

\begin{figure}[htbp]
      \centering
      \includegraphics[width=\linewidth]{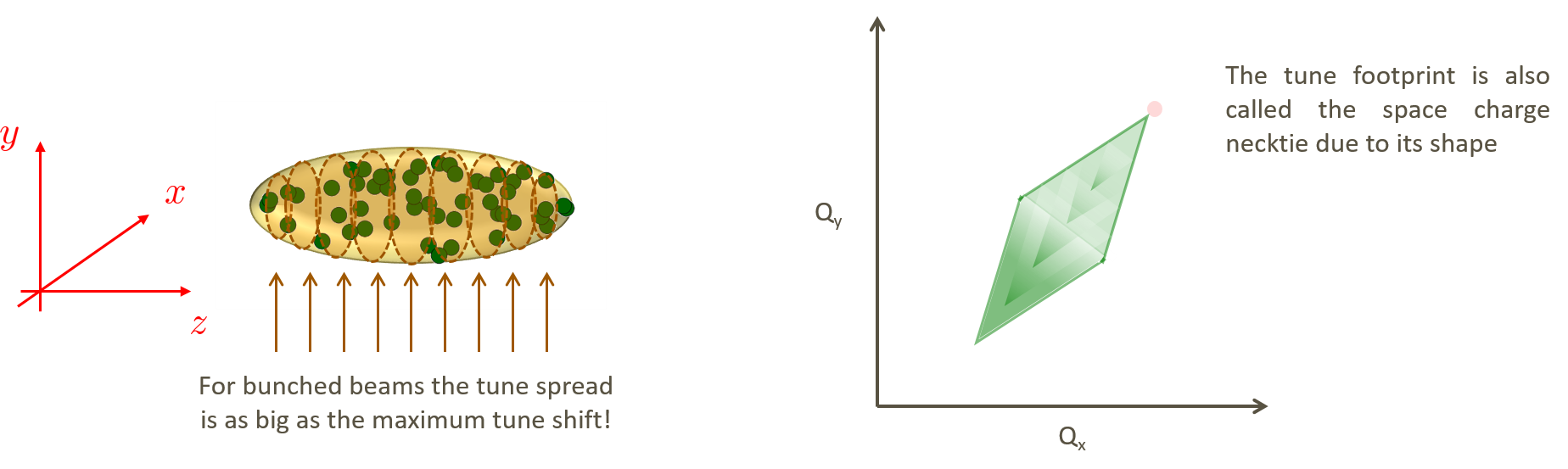}
      \caption{Characteristic space charge necktie tune footprint of a Gaussian bunched beam. Particles in regions with a large line charge density form larger triangles in tune space, particles it lower density regions form smaller triangles. All triangles superposed show the resulting total tune footprint of the bunch in tune space shown on the right hand side.}
      \label{fig:spacecharge_direct_footprint}
  \end{figure}

  In a last step of generalization, we will use the insights gained above for the Gaussian coasting beam to move to Gaussian bunched beams. We will again look at the impact rather qualitatively. Above in Fig.~\ref{fig:spacecharge_direct_schema} we have seen that a Gaussian coasting beam produces a tune footprint of triangular shape. Since we are dealing with a coasting beam, there is a translational invariance in the longitudinal direction within the beam; thus, it is irrelevant where particles are located longitudinally and there will be no impact on the tune shift and the resulting triangular shaped tune footprint from this parameter. This changes for a bunched beam. Depending on the local line charge density, the tune shifts will be more or less pronounced. This will reflect on the maximum offset as well as the size of the triangular shaped tune footprint. Figure~\ref{fig:spacecharge_direct_footprint} illustrates this on the example of a Gaussian bunched beam. In the center of the bunch, the line charge density is the largest, resulting in particles within this region to form in tune space a triangular area with a large offset and a wide spread. For particles towards the bunch edges, these triangles move closer to the zero intensity working point and become narrower. With all these triangles, formed by the particles in the different regions along the bunch, superposed, the resulting total tune footprint takes on the characteristic necktie shape depicted in the right hand side of Fig.~\ref{fig:spacecharge_direct_footprint}.  

  The fact that particle bunches occupy a finite region in tune space can have an important impact on the machine performance. From the scaling laws in Eq.~(\ref{eq:tuneshift_direct}) derived for coasting beams, and which hold similarly also for bunched beams, it becomes clear that the total occupied area in tune space is influenced by means of different parameters. In a real machine where errors and non-linearities will create resonances, an important criterion for defining the working point is the absence of significant resonance lines in the vicinity of the working point. It is also important, then, to keep all particles within the bunch as close as possible to the machine defined working point and to minimize the bunch's tune spread in order to prevent particles from crossing resonance lines. Dipole errors in the machine excite the integer resonances $(Q=n)$, quadrupole errors excite the half integer resonances $(Q=n+1/2)$; higher order resonances can be excited due to sextupoles and multipole errors.
  
\begin{figure}[htbp]
      \centering
      \subfigure[Initial tune footprint]{\includegraphics[width=0.45\linewidth]{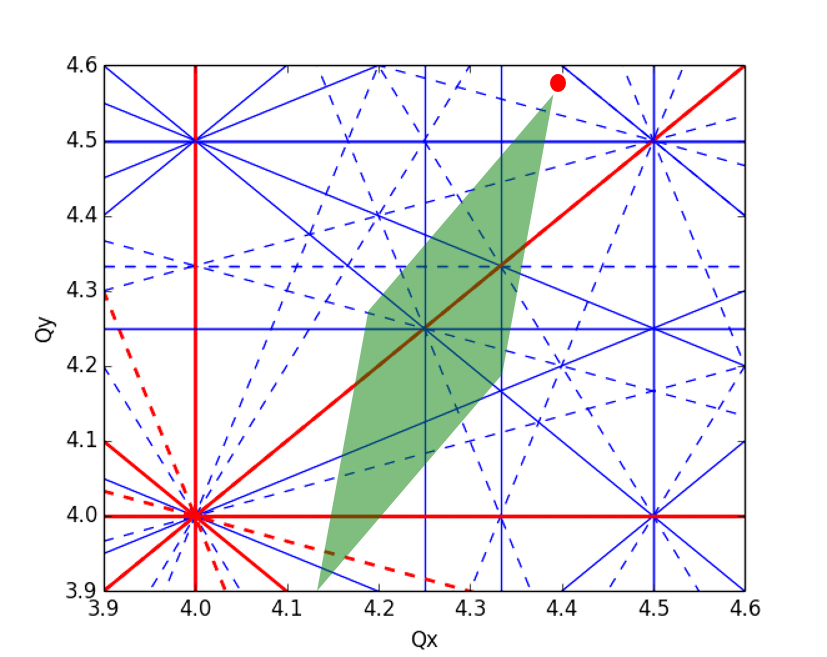}}
      \subfigure[Integer resonance crossing of beam core $\rightarrow$ blow-up]{\includegraphics[width=0.45\linewidth]{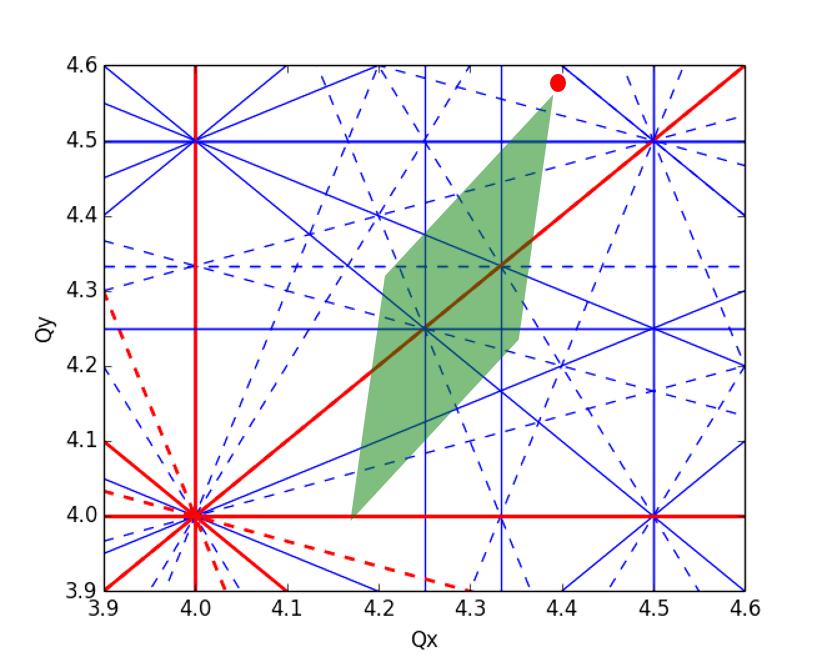}}
      \subfigure[Half-integer crossing of halo particles $\rightarrow$ losses]{\includegraphics[width=0.45\linewidth]{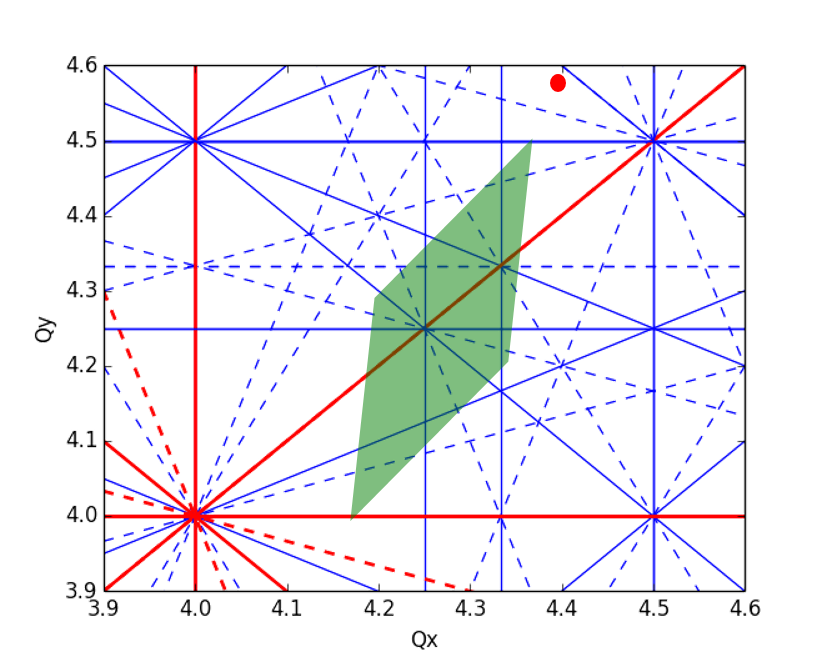}}
      \caption{Space charge necktie in a machine with resonances. The initial direct space charge induced tune footprint is large and covers both the integer as well as the half-integer resonance lines (a); core particles are pushed outward leading to an emittance blow-up of the beam (b); halo particles are driven into the aperture and lead to particle loss (c).}
      \label{fig:spacecharge_footprint_1}
  \end{figure}

  Direct space charge effects will lead to a certain tune footprint in tune space. If this tune footprint is sufficiently large, this inevitably leads to a crossing of resonance lines accompanied by a significant loss of beam quality in the machine. If we consider for instance a situation as depicted in Fig.~\ref{fig:spacecharge_footprint_1} we can see a beam with pronounced direct space charge injected into a non-linear machine with a variety of resonance lines. The resulting space charge necktie features a tune spread of more than 0.5 and thus covers several of these resonance lines including the integer and the half-integer resonance. Core particles crossing the integer resonance are driven towards the outside edges of the beam, leading to an emittance blow-up. As a result, the beam brightness decreases and the tune footprint shrinks together with the maximum tune shift. Particles located near the edges of the beam cross the half-integer resonance and are driven into the machine aperture leading to particle losses. The region around the half-integer resonance gets depleted and the tune footprint shrinks farther. Overall, because of the impact of the direct space charge effects in combination with machine errors and non-linearities, the beam has significantly lost in brightness and quality as the beam size has increased and the intensity has dropped as result of the particle losses.

\subsection{Direct space charge – mitigation techniques}

  It has become clear from the subsection above, that direct space charge effects can lead to a serious loss of beam quality. It is therefore important to find ways to minimize the impact of direct space charge effects. To identify effective measures it is useful to look back at Eq.~(\ref{eq:tuneshift_direct}) which gave an indication of the scaling of the direct space charge effects. There are several knobs that can be attacked to reduce the direct space charge forces.

\subsubsection{Line density}
\begin{figure}[htbp]
      \centering
      \subfigure[Single harmonic RF system]{\includegraphics[width=0.475\linewidth]{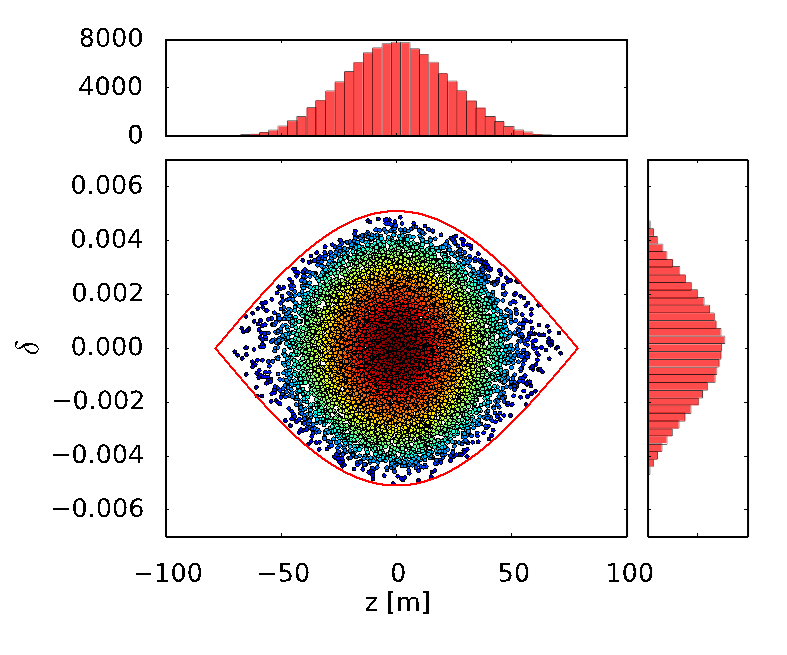}}
      \subfigure[Double harmonic RF system in bunch lengthening mode]{\includegraphics[width=0.475\linewidth]{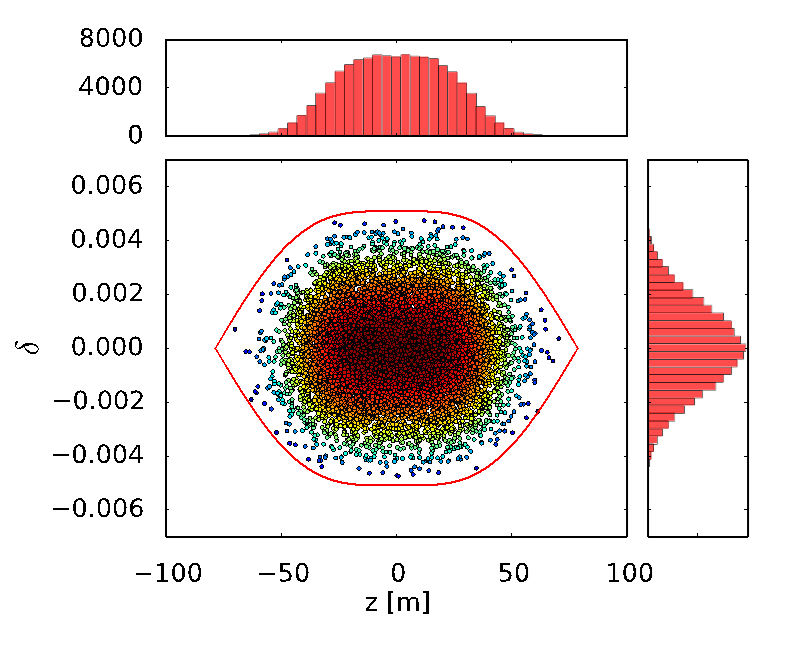}}
      \caption{Through the use of a double harmonic RF system in bunch lengthening mode, the longitudinal bunch profile can be flattened to reduce the peak line charge density while maintaining the total charge of the bunch.}
      \label{fig:spacecharge_footprint}
    \end{figure}

  Reducing the peak line charge density of bunches immediately leads to a reduction of the maximum tune shift and thus the extension of the space charge necktie in tune space. There are different ways to reduce the peak line charge density while maintaining the overall charge contained in a bunch. These range from the use of multi-harmonic RF systems to linearize the RF focusing forces and thus flatten the longitudinal bunch profile up to complicated RF gymnastics to produce "hollow" bunches which also feature a strongly homogenized longitudinal charge distribution. An example simulating the use of multiple harmonics to create flat bunch profiles is shown in Fig.~\ref{fig:spacecharge_footprint}

\subsubsection{Energy and ring geometry}
\begin{figure}[htbp]
    \centering
    \includegraphics[width=0.70\linewidth]{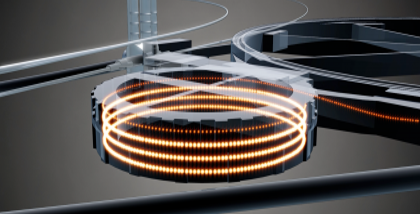}\vspace{4mm}
    \includegraphics[width=0.80\linewidth]{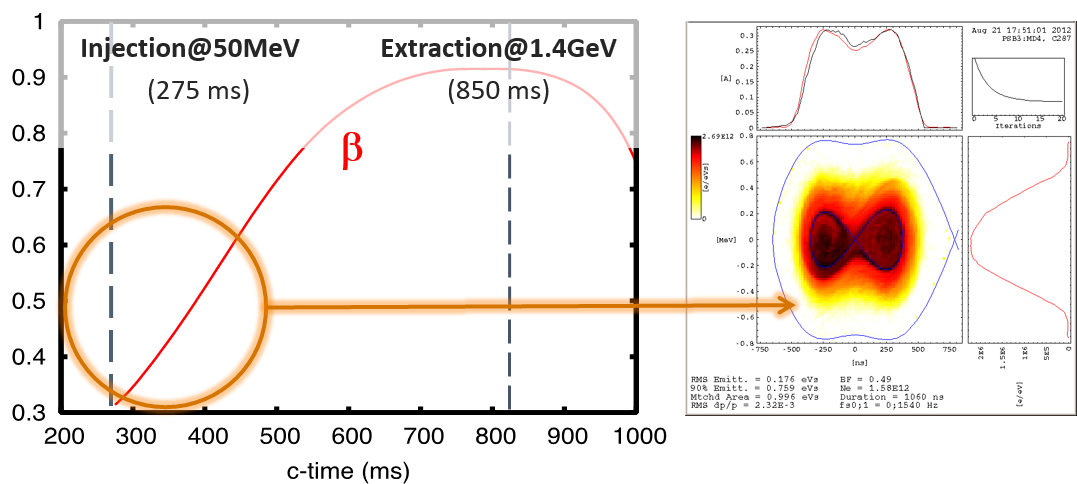}
    \caption{A schematic of the PS Booster with its four vertically stacked rings along with the magnetic cycle during pre-LS2 up to 2018.}
    \label{fig:psb-schematic}
  \end{figure}

  Rewriting Eq.~(\ref{eq:tuneshift_direct}) for cirlular Gaussian beams in a slightly modified manner as
  \[\Delta Q = -\frac{r_0 C \hat{\lambda}}{2\pi e \beta\gamma^2} \frac{1}{2\varepsilon^n_{x, y}}\,,\]
  it becomes evident that a large beam energy clearly helps to reduce direct space charge effects; but it also becomes clear that, in fact, a small ring circumference is beneficial as well. This is because the space charge induced detuning is actually an integrated effect and the integration is over one ring circumference. The Proton Synchrotron Booster (PSB) is a machine that makes use of several of the aforementioned space charge mitigation techniques simultaneously (see Fig.~\ref{fig:psb-schematic}).
    
  The PSB is actually the first synchrotron in the CERN accelerator chain which reaches up to the LHC. Consequently, the injected beam still has a very low energy and is specifically susceptible to direct space charge effects. It is therefore essential that the PSB has adequate space charge mitigation methods built-in in order to guarantee a decent beam quality of bunches which are ultimately sent to the LHC. Thus, the booster accelerates bunches rapidly from 50~MeV to 1.4~GeV over 530~ms. It is also composed of four small rings stacked on top of each other, rather than one large ring. This effectively reduces the machine circumference by a factor four (and thus the integrated detuning effect) while, nevertheless, allowing for long bunch trains to be injected into the downstream machine (Protons Synchrotron or PS) by means of a sophisticated distribution system. Moreover, in the first part of the cycle, where space charge it particularly important, bunches are flattened through the use of a second harmonic RF system. The future upgrade as part of the LIU-Project targets to further reduce space charge effects and to ultimately allow for a higher beam brightness in the machine by increasing the injection energy to 160~MeV.

\subsection{Indirect space charge}
  
  As mentioned already during the introduction of this section, a key feature of indirect space charge is the absence of direct interactions of particles among each other (this is exclusively specific to direct space charge) and the requirement of, instead, some surrounding material to act as mediator for indirect space charge effects. As an immediate consequence, this usually breaks the characteristic free space symmetry of the considered system configuration. As such - while direct space charge leads to purely incoherent effects - indirect space charge can also induce coherent effects. 

  We have seen above that direct space charge leads to a purely incoherent tune shift and a tune spread which produces the characteristic tune footprint with its typical necktie shape for bunched beams. We will now look at the effect of indirect space charge. We will briefly look at the different sources of indirect space charge, that it can lead to both incoherent as well as coherent tuneshifts, and how these tune shifts are usually parameterized by means of the Laslett coefficients.

\subsubsection{Image charges and image currents}

\begin{figure}[htbp]
      \centering
      \includegraphics[width=0.60\linewidth]{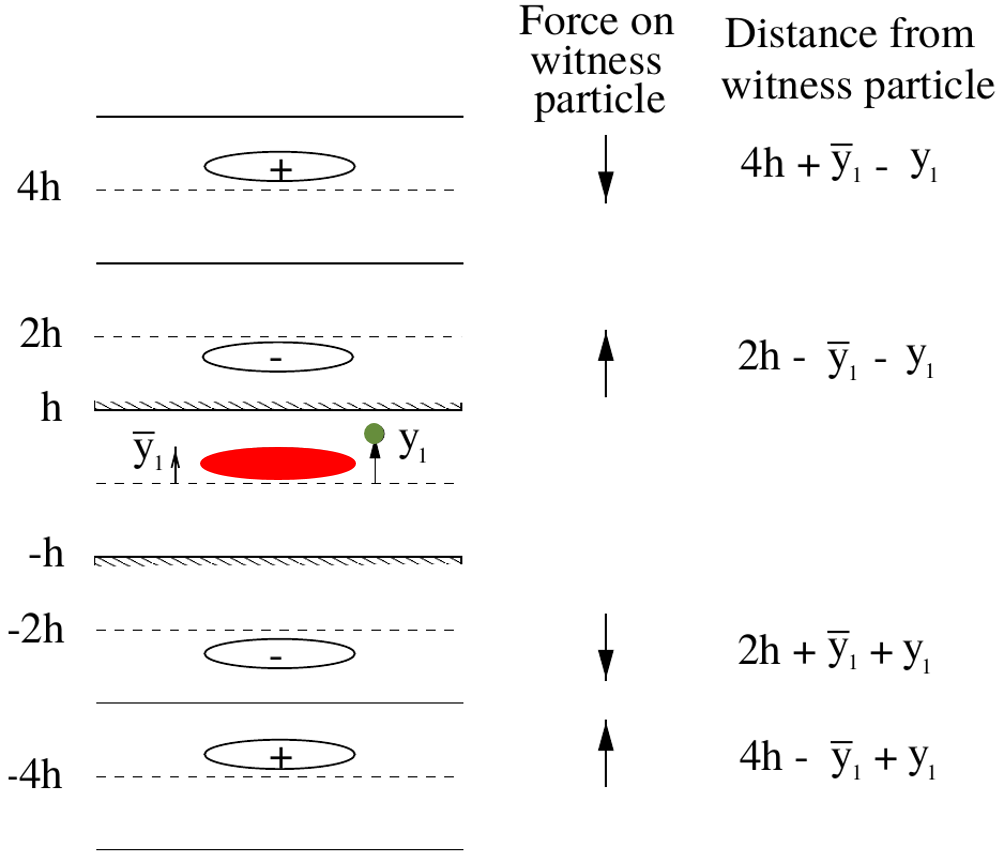}
        \caption{Configuration of source (red) and witness (green) charges in between two perfectly conducting parallel plates.}
      \label{fig:parallel_plates_iteration}
    \end{figure}

  While direct space charge treats exclusively the direct interaction of particles among each other in open space, indirect space charge relies on some surrounding material in which it induces image charges which then indirectly affect the particle system encompassed within. To get some quantitative insight in this effects, we start by considering a configuration indicated in Fig.~\ref{fig:parallel_plates_iteration}.
  
  Here, clearly, the free space symmetry is broken by the geometrical arrangement of the two perfectly conducting parallel plates; this gives a net impact on the centroid motion. Consequentially this will lead to both incoherent and coherent tune shifts. The electro- and magneto-static forces can be computed with the classical method of image charges \cite{Jackson:490457} (and derived in a very similar manner also in \cite{Ng:1012829}). We will consider the source charges as a beam of infinite length and a fixed line charge density of $\lambda$. This charge distribution will induce image charges on both of the perfectly conducting plates. The induced image charges in one plate will in turn induce further image charges in the opposite plate and vice versa in order to satisfy the boundary conditions, that is, to eliminate any parallel components of the electrostatic fields on the surface of either of the two plates.
  
  The electrostatic field at any given distance $d$ from a line charge $\lambda$ is easily computed as
  \begin{align}
      E_y = \frac{\lambda}{2\pi\epsilon_0} \frac{1}{d}\,.
  \end{align}

  We take to distance between the two parallel plates to be $2h$. Furthermore, we indicate the displacements from the center between the two parallel plates as $\bar{y}$ for the mean position of the source charge distribution and $y$ for the position of the witness particle. With $\bar{y}, y \ll h$, we can iteratively compute the total resulting field at the location of the witness particle as
  \begin{align}\label{eq:infinite_sum}\nonumber
      E_y(y, \bar{y}) = \frac{\lambda}{2\pi\epsilon_0} \biggl[
      & +\frac{1}{2h - \bar{y} - y} - \frac{1}{2h + \bar{y} + y} + \frac{1}{6h - \bar{y} - y} - \frac{1}{6h + \bar{y} + y} + \ldots\\
      & -\frac{1}{4h + \bar{y} - y} + \frac{1}{4h - \bar{y} + y} - \frac{1}{8h + \bar{y} - y} + \frac{1}{8h - \bar{y} + y} + \ldots
      \biggr]\,.
  \end{align}
  Equation~(\ref{eq:infinite_sum}) fortunately converges to
  \begin{align}
      E(y,\bar{y}) &= \frac{\lambda}{\pi\epsilon_0 h^2} \Bigl[(\bar{y} + y)\frac{\pi^2}{32} + (\bar{y} - y)\frac{\pi^2}{96}\Bigr] = \frac{F_y}{e}\,.
  \end{align}
  
  The resulting indirect space charge forces on a witness particle now depend on the positions of both the witness particle as well as the mean position of the full source charge distribution with respect to an external reference. This will give rise also to coherent effects. Before looking at these more closely, we will spend a few more words on the different types and phenomena of electromagnetic images.
  
  The previously depicted configuration is an example of stationary charges within some surrounding structure. These induce electrostatic image charges in this structure which results in the generation of electrostatic fields and forces. In this situation all electric fields vanish within a conductor while all static magnetic fields usually pass through (except for cases of very high permeability). Typically, we are dealing with moving charges within surrounding structures. Instead of image charges, these then rather induce image currents. The resulting induced fields become time varying and, in consequence, can penetrate into the surrounding material within a region up to the skin depth $\delta_w$. The skin depth $\delta_w$ depends on the properties of the surrounding material and on the frequency of the induced electromagnetic fields. Fields pass through the conductor wall if the skin depth is larger than the wall thickness $\Delta_w$. This happens at low frequencies. At higher frequencies and for a good conductor we have instead $\delta_w \ll \Delta_w$, and both electric and magnetic fields vanish within the wall.
  
\begin{figure}[htbp]
      \centering
      \subfigure[Electrostatic forces induces by image charges]{\includegraphics[width=0.45\linewidth]{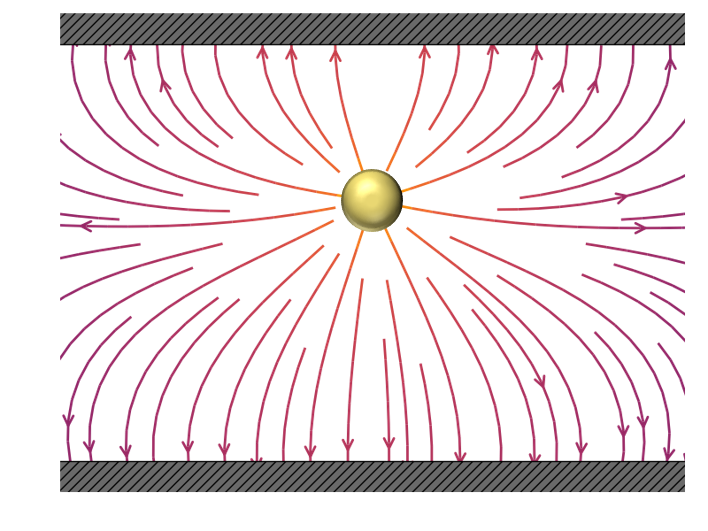}}\\
      \subfigure[Magnetic AC forces induced by image currents]{\includegraphics[width=0.45\linewidth]{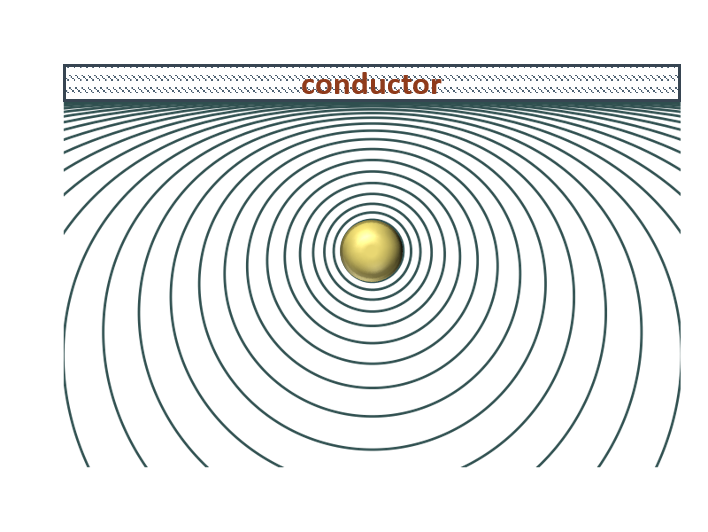}}
      \subfigure[Magnetic DC forces induced by image currents]{\includegraphics[width=0.45\linewidth]{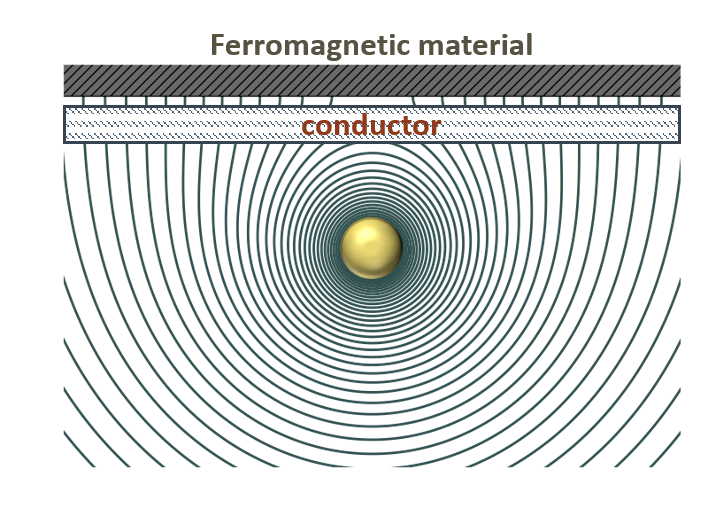}}
      \caption{Summary of indirect space charge induced image charge and current configurations for charged particle distributions travelling close to conducting surfaces: (a) charges at rest induce image charges and electrostatic fields - field lines are perpendicular to the surface; (b) if the skin depth is very small (rapidly varying fields), magnetic fields do not penetrate and the field lines are tangent to the surface; (c) at low frequencies the skin depth becomes larger and magnetic fields penetrate and pass through the vacuum chamber where they can interact with bodies behind the chamber.}
      \label{fig:image_configurations}
    \end{figure}

  Hence, whereas electric image charges induce electrostatic fields and forces, there are also electric image currents which can induce magneto-static fields and forces; here, we typically differentiated between AC and DC forces. Figure~\ref{fig:image_configurations} summarizes the different situations that can emerge as charged particle distributions are resting or travelling in between two parallel plates.

  In a similar fashion as done for the electrostatic case, we can again deploy the method of image currents (see again \cite{Jackson:490457}) to solve for the magneto-static fields and forces. We then obtain forces of the form:
  \begin{align}
      \frac{F_y}{e} &= -\frac{\lambda \beta^2}{\pi\epsilon_0 h^2} \Bigl[(\bar{y} + y)\frac{\pi^2}{32} + (\bar{y} - y)\frac{\pi^2}{96}\Bigr]\\
      \frac{F_y}{e} &= +\frac{\lambda \beta^2}{\pi\epsilon_0 h^2} \Bigl[(\bar{y} + y)\frac{\pi^2}{32} - (\bar{y} - y)\frac{\pi^2}{96}\Bigr]\,.
  \end{align}
  
\subsubsection{Tune shifts and Laslett coefficients}

  The aforementioned electric image charge, and magnetic ac and dc image current forces can all be brought into a general similar form as
  \begin{align}
      F_x \propto \frac{e\lambda}{\pi\epsilon_0 h^2}\,f(\bar{x}, x)\,,\quad F_y \propto \frac{e\lambda}{\pi\epsilon_0 h^2}\,f(\bar{y}, y)\,.
  \end{align}

  These forces can lead to both incoherent as well as coherent tune shifts. From these force expressions, the tune shifts can easily be computed. In fact, it turns out that the tune shifts can be parameterized via the Laslett coefficients \cite{Laslett:1963zz}. For example, for electric image charge forces between two perfectly conducting parallel plates separated by the distance $2h$, the incoherent and coherent tune shifts can be expressed as:
  \begin{align}
      \Delta Q^\textrm{inc}_{x,y} = -\frac{2\langle \beta_{x,y} \rangle r_0 R}{e\beta^2\gamma}\, \frac{\varepsilon^{x,y}_1}{h^2}\,,\quad 
      \Delta Q^\textrm{coh}_{x,y} = -\frac{2\langle \beta_{x,y} \rangle r_0 R}{e\beta^2\gamma}\, \frac{\xi^{x,y}_1}{h^2}\,.
  \end{align}

\begin{figure}[htbp]
      \centering
      \includegraphics[width=0.60\linewidth]{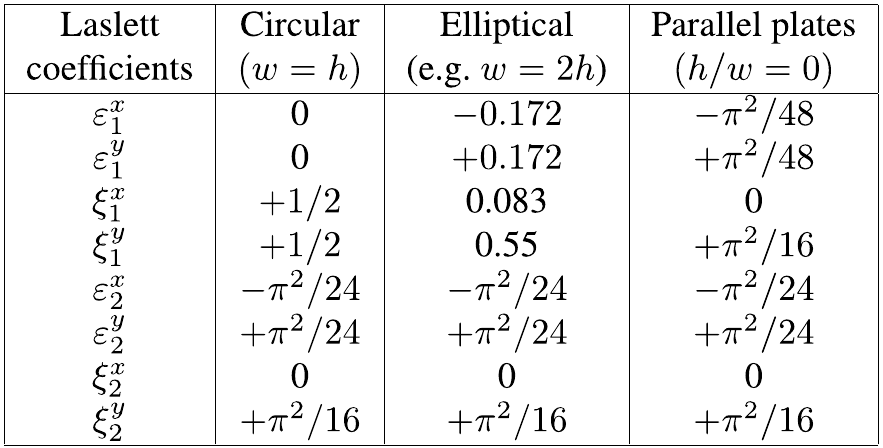}
        \caption{Laslett coefficients for different geometries; they are always defined with respect to the vertical half gap $h$ or $g$. For the ferro-magnetic boundary, parallel plates are always exclusively assumed.}
      \label{fig:laslett_coefficients}
    \end{figure}

  The Laslett coefficients can be evaluated for different geometries and are classified in incoherent and coherent tune shifts for electric, magnetic AC and magnetic DC image charges and currents. Table~\ref{fig:laslett_coefficients} shows some of the Laslett coefficients most frequently used.
  
\subsubsection{Example of indirect space charge effects}
  
\begin{figure}[htbp]
      \centering
      \includegraphics[width=\linewidth]{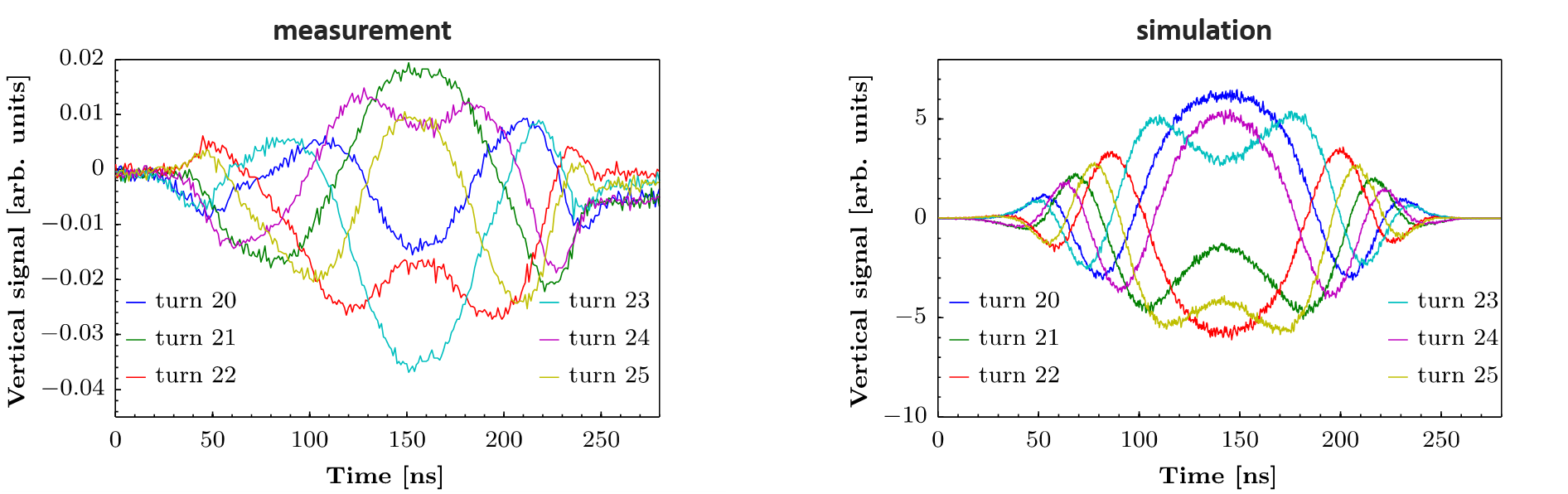}
        \caption{The coherent intra-bunch vertical motion along the longitudinal extension of the bunch measured turn-by-turn is shown on the left. Macroparticle simulations render the same coherent intra-bunch motion once indirect space charge effects have been included, as shown on the right.}
      \label{fig:ps_injection_position}
    \end{figure}

  Indirect space charge effects can have different impacts the beam charge and the surrounding material. The effects can be of incoherent or of coherent nature. An example of a coherent indirect space charge effect was observed at the CERN Proton Synchrotron (PS) in 2014. Here, a strong intra-bunch motion combined with slow losses developed for a few 100~\textmu s after injection into the machine. The intra-bunch pattern could not be explained until macroparticle simulations were conducted which explicitly took into account the indirect space charge induced fields. With these fields integrated into the simulations, the measurement results could be reproduced with a remarkable precision as highlighted by Fig.~\ref{fig:ps_injection_position}.

\begin{figure}[htbp]
      \centering
      \includegraphics[width=\linewidth]{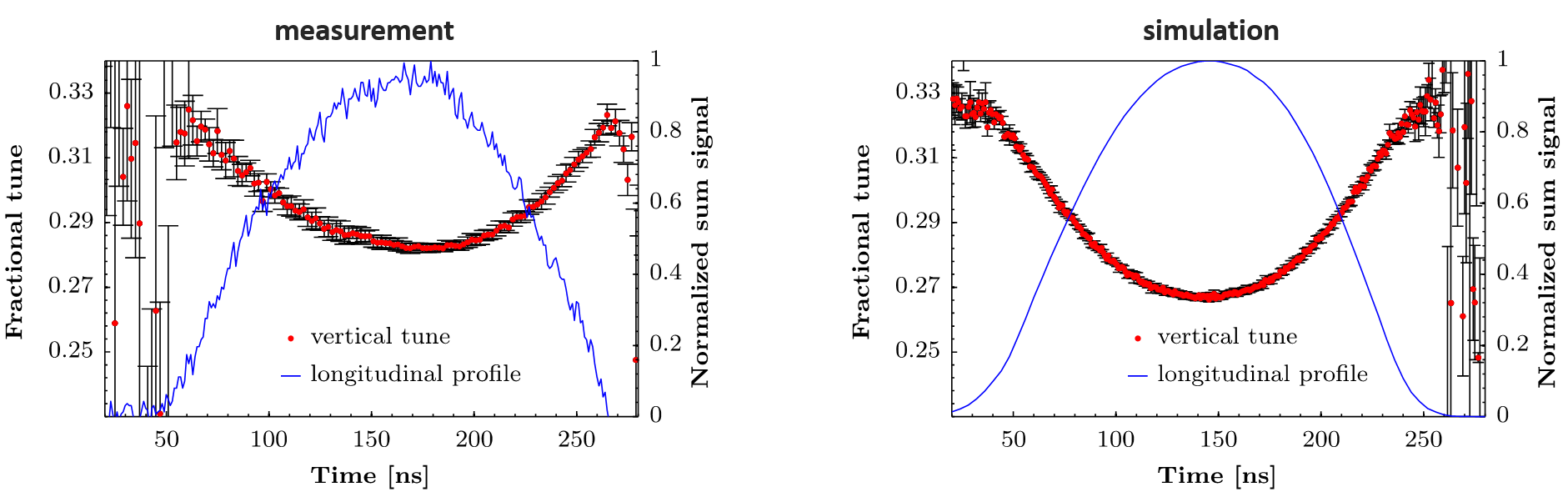}
        \caption{The coherent vertical tune shift along the longitudinal extension of the bunch as measured is shown on the left. Macroparticle simulations render the same coherent vertical tune shift once indirect space charge effects have been included, as shown on the right.}
      \label{fig:ps_injection_tune}
    \end{figure}

  Moreover, the coherent tune shift induced by the indirect space charge forces was measured along the longitudinal extension of the bunch. The same analysis was done again in macroparticle simulations. With the inclusion of indirect space charge effects into the simulations, the coherent tune shifts along the bunch, which were obtained in measurements, could be very well reproduced by the simulations as highlighted by Fig.~\ref{fig:ps_injection_tune}.
    
  Thus, it was understood from simulations that the observed intra-bunch motion was induced by the beam injected off-center in combination with the indirect space charge effect, which causes a tune shift along the bunch proportional to the local charge density.
  
  To conclude on the overview and comparison of direct and indirect space charge effects, we summarize below the main features of each of the two kinds of space charge effects:
  \begin{itemize}
      \item Direct space charge
      \begin{itemize}
          \item interaction of the bunch particles with the self induced electromagnetic fields in free space
          \item results in an incoherent tune shift (or spread)
          \item the space charge force along a bunch is modulated with the local line density along the bunch and this results in an additional tune spread
          \item decreases with energy like $\beta^{-1}\gamma^{-2}$
          \item is a typical performance limitation for low energy machines
      \end{itemize}
      
      \item Indirect space charge
      \begin{itemize}
          \item interaction with image charges and currents induced in perfect conducting walls and ferromagnetic materials close to the beam pipe
          \item results in incoherent and coherent tune shifts (or spreads), some of which are proportional to the average line density (dc) and others to the peak line density (ac)
          \item the contributions to the coherent and incoherent tune shifts for different standard geometries are expressed in terms of Laslett coefficients
          \item decreases with energy like $\beta^{-2}\gamma^{-1}$
      \end{itemize}
  \end{itemize}

  So far, we have introduced direct and indirect space charge as collective effects. The corresponding forces were not externally given but dependent on the actual particle distribution within the beam (remember, we looked  at  single  particles  as  well  as  uniform  and  Gaussian distributions). The forces led to incoherent and coherent tune shifts.

  We will now go a step further and investigate more complicated structures. We will try to find a smart way to deal with these structures. In the course of this, we will generalize and extend the direct and indirect space charge effects towards the concept of wake fields and impedances.

\subsection{From indirect space charge to (resistive) wall wakes}

\begin{figure}[htbp]
      \centering
      \subfigure[Electrostatic forces induces by image charges]{\includegraphics[width=0.6\linewidth]{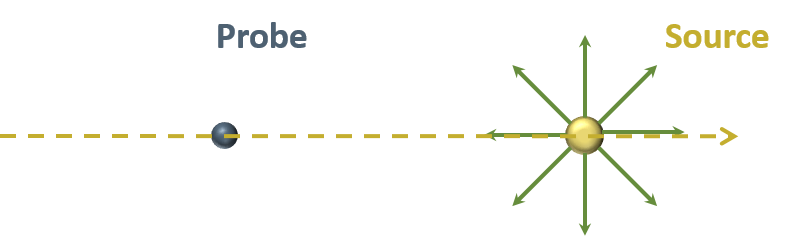}}\vspace{8mm}
      \subfigure[Electrostatic forces induces by image charges]{\includegraphics[width=0.6\linewidth]{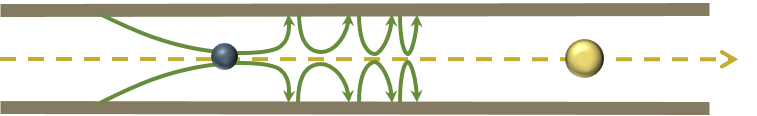}}\vspace{8mm}
      \caption{Configuration of charged particles travelling, (a) in free space or, (b) within an encompassing structure; on the top figure, probe particles are effected directly by the source particles via the Lorentz force; on the bottom figure, which shows an encompassing structure with smooth boundaries, probe particles are effected indirectly by the source particles’ induced image charges and currents.}
      \label{fig:surrounding_chamber_reswall}
  \end{figure}

  Up to now, this section has been dealing with direct and indirect space charge effects. The first treats forces exerted in free space, the latter describes forces induced via image charges and currents within surrounding material. It is this latter type of forces that we will now further categorize. For this, we take a close look at Fig.~\ref{fig:surrounding_chamber_reswall}. The top image shows the situation for direct space charge forces in free space. The bottom image illustrates the situation for indirect space charge forces, where we consider a smooth surrounding structure. As we have seen earlier, the charged particle distribution travelling through this smooth structure induces image charges and currents within the structure surface. These image charges and currents can exert forces onto the charged particles themselves. Up to know, we have always assumed perfectly conducting surfaces such that the image charges always traveled perfectly in sync with the charged particles. If we now assume instead smooth structures with some finite resistivity, the image charges will start lagging behind as they travel slower through the surface material compared to the charged particle distribution, which travels in vacuum. The resulting effect is illustrated in the bottom image of Fig.~\ref{fig:surrounding_chamber_reswall}. The slower travelling image charges induce electromagnetic fields behind the source particles. We call these trailing electromagnetic fields, which are left behind by the leading charged particles, the wake fields.

\begin{figure}[htbp]
      \centering
      \subfigure[Cylindrical structure]{\includegraphics[width=0.45\linewidth]{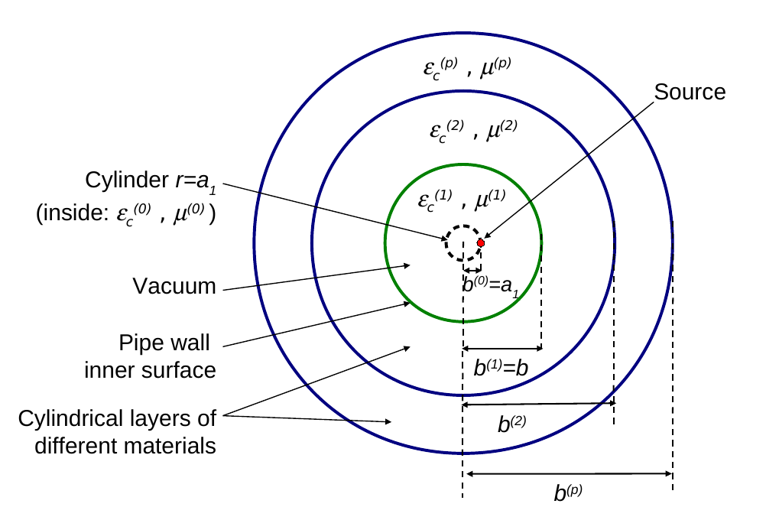}}
      \subfigure[Flat structure]{\includegraphics[width=0.45\linewidth]{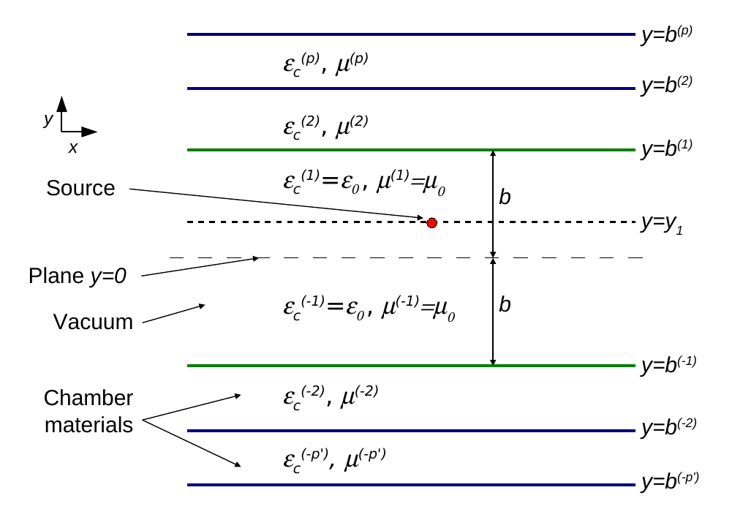}}
  \caption{Examples of smooth multi-layer structures, longitudinally translation invariant with regular cross-sections; a charged particle placed within these structures will induce electromagnetic fields. These fields can be computed semi-analytically by solving Maxwell's equations using the field matching technique.}
      \label{fig:multilayer}
  \end{figure}

\begin{figure}[htbp]
    \centering
    \subfigure[Direct space charge]{
      \includegraphics[height=0.26\linewidth]{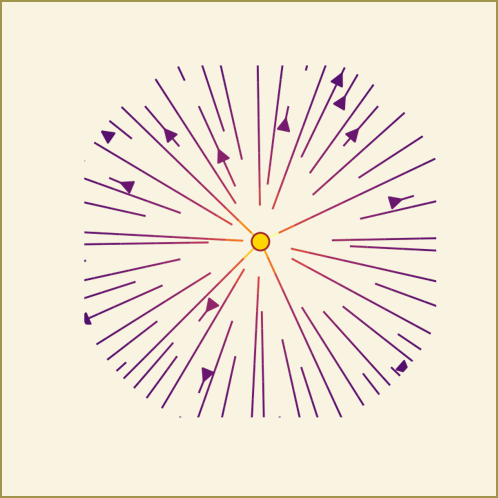}
    }
    \subfigure[Indirect space charge]{
      \includegraphics[height=0.26\linewidth]{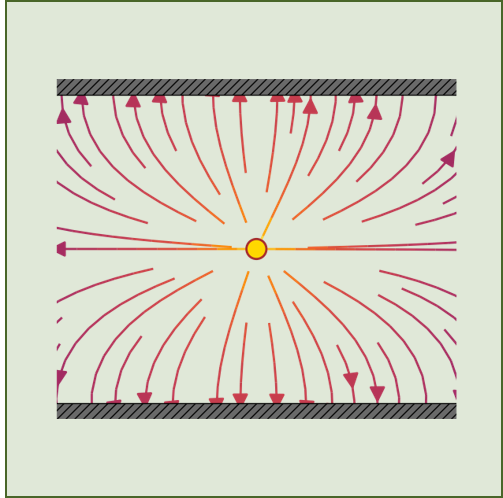}
    }
    \subfigure[Resistive wall wakes]{
      \includegraphics[height=0.26\linewidth]{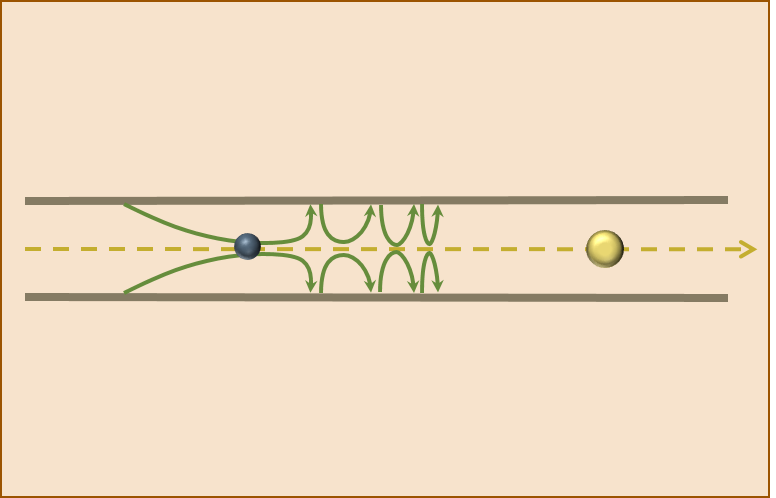}
    }\\\vspace{8mm}
    \includegraphics[width=0.45\linewidth]{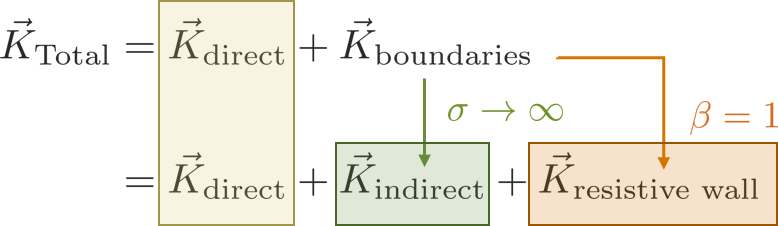}
    \caption{(a) Component independent of the surrounding boundaries $\rightarrow$ direct space charge in free space; (b) component which is independent of the surrounding material properties and purely dependent on the surrounding geometry $\rightarrow$ indirect space charge in smooth perfectly conducting structures; (c) component which is dependent on the surrounding material electromagnetic properties $\rightarrow$ resistive wall wakes.}
    \label{fig:towards_wakes}
  \end{figure}

  These resistive wall wake fields can be computed analytically or semi-analytically for smooth, infinitely long structures. A code which has been developed to perform these calculations for multi-layer structures is ImpedanceWake2D \cite{Mounet:1451296}. For these types of calculations, we typically consider a smooth multi-layer structure, longitudinally translation invariant and transversely bounded. We let a charged point particle travel through this smooth multilayered structure. For this configuration and for certain geometries, the induced electromagnetic fields can be computed by means of Maxwell’s equations (longitudinal and transverse electromagnetic fields) with field matching at the boundaries. Two examples of such geometries are shown in Fig.~\ref{fig:multilayer}. Upon evaluation of the induced fields for such structures, it turns out, that the resulting electromagnetic fields can be decomposed into three components:
  \begin{itemize}
      \item a component which is entirely independent of the surrounding boundaries; this is the direct space charge field present in free space
      \item a component which is independent of the surrounding material properties but purely dependent on the surrounding geometry; this is the indirect space charge field present in smooth perfectly conducting structures
      \item a component which is dependent on the surrounding material electromagnetic properties; this turns out to be what we have previously called the resistive wall wakes
  \end{itemize}
  Typically, the electromagnetic fields which result from the boundary conditions, contain both components, the indirect space charge fields for perfectly conducting walls, and the resistive wall wakes, which are very often modeled only for ultra-relativistic beams. For the limits of infinite conductivity, these boundary fields (or wall wakes) reduce to the pure indirect space charge fields; for the case of ultra-relativistic source particle distributions, the wall wakes turn into the commonly used resistive wall wakes for $\beta = 1$. This is illustrated in Fig.~\ref{fig:towards_wakes}. We will not get into this any further at this point. The calculation of wake fields is a commonly performed task and analytical expressions exist for various types and geometries with examples shown in \cite{Chao:246480, Ng:1012829}, for instance.

\begin{figure}[htbp]
    \centering
    \includegraphics[width=0.6\linewidth]{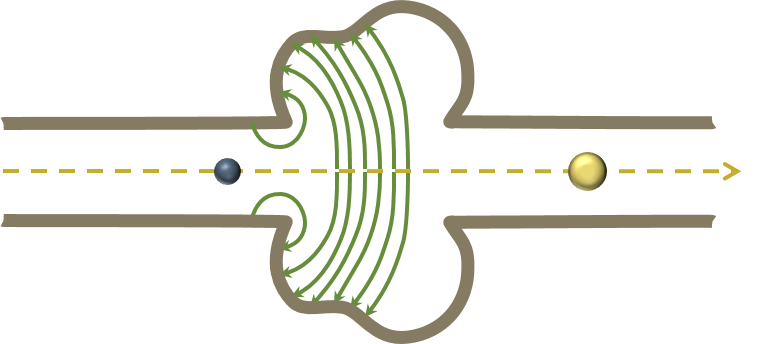}
    \caption{Configuration of charged particles travelling through a resonating structure with discontinuities in boundaries; fields are excited by the leading source particles and can keep ringing, thus effecting trailing probe particles.}
    \label{fig:surrounding_chamber_resonator}
  \end{figure}

  In a final step of generalization, we consider now a generic geometry of the kind shown in Fig.~\ref{fig:surrounding_chamber_resonator}. These geometries feature non-smooth, sometimes discontinuous boundaries. Charged particles traversing these structures naturally also generate wake fields. However, these wake fields are usually no longer analytically or semi-analytically computable and one needs to revert to numerical simulations and time-domain or frequency-domain Maxwell solvers (FDTD, FEM, etc.), which can be numerically very expensive.
  
  To self-consistently simulate the interaction of charged particle systems with their environment by means of wake fields, in principle, one would need to solve the full set of Maxwell’s equations at every time step in order to obtain the electromagnetic fields at every location within the structure and to evaluate the forces at the probe particle’s locations. This becomes very tedious and virtually impossible for a 27~km ring such as the LHC. The question naturally arises, how one can treat these phenomena more effectively in our models

  For this, we normally use a set of assumptions to simplify the problem:
  \begin{enumerate}
    \item Rigid beam approximation: the beam traverses the discontinuity of the vacuum chamber rigidly
    \item Impulse approximation: what the beam really cares about is the integrated impulse as it completes the traversal of the discontinuity
  \end{enumerate}

  We will see in the next section, how one can put these assumptions to use to obtain a formalism by which an effective treatment of this interplay between charged particle systems and their surrounding environment even within complex structures can be made possible. For this, we will introduce the concept of wake fields and impedances. We will also study the effect of these on both the machine and on the beam.
\clearpage\pagebreak
\section{Wake fields and impedances}
\label{sec:wakesimpedances}

  In the previous sections we have studied multi-particle systems, often expressed as the single particle distribution function $\bm{\psi}$, in terms of their description and, in absence of charges, in terms of their collective behaviour. We got to see phenomena as Liouville's theorem, coherent- and incoherent motion and decoherence and filamentation. We then moved to the inspection of charged multi-particle systems where we analyzed direct and indirect space charge effects. These were first forms of collective effects in that they depended on the charges and the distributions of the multi-particle systems inducing these effects. Direct and indirect space charge effects lead to incoherent and coherent tune shifts and tune spreads which usually has a detrimental impact on the machine and on the charged particle beams. At the end of the previous section we learned how indirect space charge and resistive wall wakes are linked to each other and how the resistive wall wakes can be computed for simple geometries. We will now investigate the concept of wake fields and impedances in more depth to see how these are typically used to model complex geometries in large machines. 

  We will discuss at first the idea and the definition of the wake function which will lead to the description of wake fields. We will look at the specific cases of longitudinal and transverse wake functions and introduce the notion of the impedance. We will then analyze the impact that the interaction of charged multi-particle systems with wake fields and impedances can have both onto the machine as well as the charged multi-particle systems themselves. This will lead us to phenomena such as beam induced heating or coherent beam instabilities.

\subsection{Concept of wake fields}

  We have understood that obtaining all electromagnetic fields and forces induced by charged multi-particle systems within their surrounding environment by solving the full electrodynamics usually becomes a huge simulation effort. These types of studies would be virtually impossible to be carried out for large accelerators.

  It is for this reason, that one reverts to the concept of the wake function. The wake function is essentially the electromagnetic impulse response of any given structure. From this one obtains wake fields and impedances, which can be used to formally study the electromagnetic interaction of structures with a passing beam.

  \begin{figure}[htbp]
    \centering
    \includegraphics[width=0.95\linewidth]{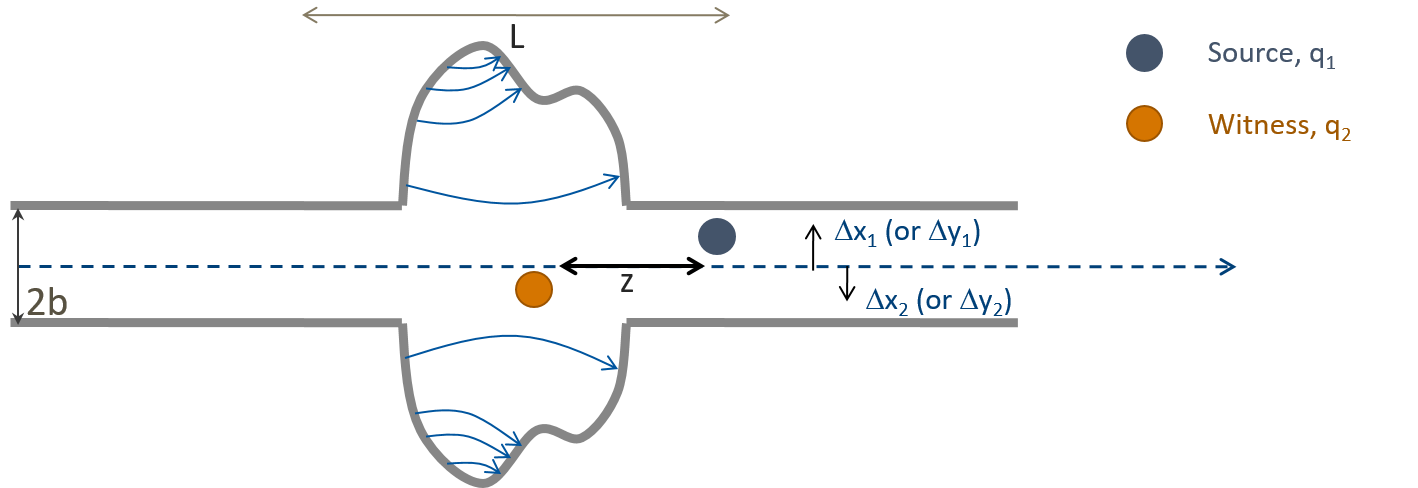}
    \caption{Illustration of a leading source charged particle and a trailing witness charged particle traversing a surrounding structure; the source particle induces electromagnetic fields and forces that are then picked up by the witness particle.}
    \label{fig:wake_function}
  \end{figure}

  To introduce the concept of the wake function, we consider a configuration as shown in Fig.~\ref{fig:wake_function} where we essentially have a leading source charged particle and a trailing witness charged particle traversing a surrounding structure (we have already looked at this type of configuration when introducing space charge effects). We denote the charge of the source particle as $q_1$ and the charge of the witness particle as $q_2$. We can, most generally, write down the force exerted by the source particle onto the witness particle via the surrounding structure as
  \begin{align}
    \vec{F}_\textrm{wake} = \vec{F}_\textrm{wake}(x_1,y_1,s_1,x_2,y_2,s_2,t)\,,
  \end{align}
  where $(x_1,y_1,s_1)$ and $(x_2,y_2,s_2)$ are the source and the witness particle's positions within the surrounding structure at any given time $t$. Of course, $\vec{F}_\textrm{wake}$ is really just the Lorentz force exerted by the source particle onto the witness particle
  \begin{align}
      \vec{F}_\textrm{wake} = q_2\,\Bigl[ \vec{E}(x_1,y_1,s_1,x_2,y_2,s_2,t) + v_z\vec{e}_z \times \vec{B}(x_1,y_1,s_1,x_2,y_2,s_2,t) \Bigr]\,,
  \end{align}
  where $\vec{E}$ and $\vec{B}$, however, have been fully self-consistently calculated taking into account all charged particles and surrounding materials properties. To become computationally more effective, we now make the two aforementioned assumptions or approximations:
  \begin{itemize}
      \item the rigid beam approximation,
      \item the impulse approximation.
  \end{itemize}
  
  The first approximation says that all charged particles traverse the structure rigidly, i.e., they do not change their relative positions while moving through the structure. The second item expresses the fact that, ultimately, what we are interested in, is not so much the exact forces at every location and at every time withing the given structure. But, instead, we are rather interested in the net integrated effect after full passage of any given structure. That is, we would like to characterize a structure by the integrated force that a source charged particle will have exerted on a witness charged particle after passage of the given structure. This can be expressed formally; we will restrict ourselves for now to just the horizontal plane. Furthermore, we denote $s_1$ as $s$ and define $z = s_2 - s_1$; due to the rigid beam approximation, $z$ will stay constant as the source and the witness particles traverse the structure. We can then express the integrated force (which results in a change of energy of the witness particle) as
  \begin{align}\label{eq:wake_function}
      \Delta E_2 = \int \vec{F} (x_1, x_2, z; s) \,ds = -q_1q_2\,w(x_1, x_2, z)\,.
  \end{align}
  Equation~(\ref{eq:wake_function}) defines the wake function $w(x_1, x_2, z)$ which is a function of the horizontal offsets of and the distance between the source and the witness particles. Thus, the wake function $w$ is proportional to the integrated forces felt by the witness particle. It is a key property of any structure when studying collective effects. Once the wake function is known, the integrated forces on any witness particle exerted by an entire charged particle distribution due to the superposition principle can be easily evaluated as
  \begin{align}\label{eq:wake_fields}
    \Delta E_2(z) = \sum_i q_i q_2\,w(x_i, x_2, z-z_i) \rightarrow \int \lambda_1(x_1, z_1)\,w(x_1, x_2, z-z_1)\,dx_1 dz_1\,.
  \end{align}
  We can see from Eq.~(\ref{eq:wake_fields}) that the forces now become dependent on the particle distribution function. Looking closely, we can even recognize that the wake function mathematically actually resembles a Green's function.
  
  We have thus introduced the concept of the wake function. These can simplify our handling of induced electromagnetic fields within complex structures. The wake function is the electromagnetic response of a structure and is in fact an intrinsic property of any such structure.
  
  In practice, we will never compute the full wake function but we will separate between longitudinal and transverse wake fields. We then treat these in an expansion which further simplifies our treatment. Complementary to the wake fields one can also move to frequency domain and use the impedance.

\subsection{Longitudinal and transverse wake fields and impedances}

  Starting from Eq.~(\ref{eq:wake_function}), we will first consider only the longitudinal component of the forces which will result in giving the longitudinal wake function. We also perform a Taylor expansion in the offset $\Delta x_1$ of the source and the offset $\Delta x_2$ of the witness particles. We obtain
  \begin{align}\label{eq:longitudinal_wake_function}
      \Delta E_2 = \int F_z(\Delta x_1, \Delta x_2, z; s) \approx -q_1 q_2 \Bigl[ 
      \underbrace{W_\parallel(z)}_\textrm{Zeroth order term} + 
      \underbrace{O(\Delta x_1) + O(\Delta x_2)}_\textrm{Higher order terms} \Bigr]\,.
  \end{align}
  
  In Eq.~(\ref{eq:longitudinal_wake_function}) the zeroth order term considers both the source and the witness particles traversing on-axis and at a fixed distance of $z$ the indicated beam pipe and cavity. This term is usually the dominant term responsible for longitudinal wake field induced effects. The higher order terms in the offsets of source and witness particles are neglected. Thus, it is this first term $W_\parallel(z)$ that is identified with the longitudinal wake function. Now, $W_\parallel(z)$ is a quantity that can be simulated or even measured for any given structure. Once $W_\parallel(z)$ is known, any interaction of a charged multi-particle distribution with this structure can be effectively evaluated.
  
  The value of the wake function in $z=0$ is related to the energy lost by the source particle in the creation of the wake field. This becomes clear if we consider
  \begin{align}
    W_\parallel(z) = -\frac{\Delta E_2}{q_1 q_2}\quad \xrightarrow[\substack{q_2\rightarrow q_1}]{\substack{z\rightarrow 0}}\quad W_\parallel(0) = -\frac{\Delta E_1}{q_1^2}\,.
  \end{align}
  In order to compute the total wake field effects induced by an entire charged particle distribution of line charge density $\lambda_1(z_1)$, Eq.~(\ref{eq:longitudinal_wake_function}) can be generalized to the convolution
  \begin{align}\label{eq:longitudinal_convolution}
      \Delta E_2(z) \propto \int \lambda_1(z_1)\,W_\parallel(z-z_1)\,dz_1\,.
  \end{align}
  We can see from Eq.~(\ref{eq:longitudinal_convolution}) that the wake function of an accelerator component is basically its Green's function in time domain (i.e., its response to a pulse excitation). This expression is particularly useful for macroparticle models and simulations, because it can be used to describe the driving terms in the single particle equations of motion. The convolution in Eq.~(\ref{eq:longitudinal_convolution}) can also be represented in frequency domain. This will then yield the longitudinal impedance of an accelerator component which is basically its transfer function in frequency domain. Thus, the formal definition of the longitudinal beam coupling impedance of a structure is given by means of its longitudinal wake function as
  \begin{align}\label{eq:longitudinal_impedance}
    Z_\parallel(\omega) = \int_{-\infty}^\infty W_\parallel(z) \exp\left(-\frac{i\omega z}{c}\right)\,\frac{dz}{c}\,.
  \end{align}

  Next, again starting from Eq.~(\ref{eq:wake_function}), we will consider the transverse components of the forces. This will results in the transverse wake functions. We will limit ourselves to the horizontal plane, without loss of generality. As for the longitudinal case, we perform a Taylor expansion in the offset $\Delta x_1$ of the source and the offset $\Delta x_2$ of the witness particles. We then obtain
  \begin{align}\label{eq:transverse_wake_function}
        \beta c\,\Delta p_{x_2} &= \int F_x(\Delta x_1, \Delta x_2, z; s) \approx -q_1 q_2 \Bigl[ 
        \underbrace{W_{C_x}(z)}_\textrm{Zeroth order term} + 
        \underbrace{W_{D_x}(z)\,\Delta x_1}_\textrm{Dipole wake term} +
        \underbrace{W_{Q_x}(z)\,\Delta x_2}_\textrm{Quadrupole wake term}  + \ldots\Bigr]\,.
  \end{align}
  
  It it becomes evident from Eq.~(\ref{eq:transverse_wake_function}) that in the transverse plane, in order to capture the impact on the transverse motion, we need to keep higher order terms in the transverse offsets, as there are otherwise no transverse dynamical terms left. Thus, we have a zeroth order term, which we call the constant wake function $W_{C_x}(z)$, which will lead to a constant closed orbit distortion along the bunch. The next term is the dipolar wake function $W_{D_x}(z)$ which leads to orbit kicks and detuning; if these kicks become resonant, the dipolar wake fields can drive coherent instabilities. The final term considered is the quadrupole wake function $W_{Q_x}(z)$. This term introduces a detuning along the bunch and thus leads to an overall tune spread. Again, each of these quantities $W_{C_x}$, $W_{D_x}$ and $W_{Q_x}$ can be simulated or even measured for any given structure. And, again, once $W_{C_x}$, $W_{D_x}$ and $W_{Q_x}$ are known, any interaction of a charged multi-particle distribution with this structure can be effectively evaluated. In general, the wake fields in Eq.~(\ref{eq:transverse_wake_function}) induce a transverse deflecting kick of the witness particle. These kicks are independent on the witness particle's position for the constant and (to first order) for the dipole wake fields which is why these terms lead to orbit kicks. They are proportional to the witness particle's position for the quadrupolar term and, consequentially, lead to a detuning. First order coupling terms between the $x$ and $y$ planes have been neglected, as have been all higher order terms including mixed higher order terms with products of the dipolar and quadrupolar offsets.
  
  We can rewrite the individual terms in Eq~.(\ref{eq:transverse_wake_function}) if we consider only one wake field at a time; thus, considering first the dipolar wake, we can derive
  \begin{align}\label{eq:dipolar_wake_field_kick}
      W_{D_x}(z) = -\frac{\beta^2 E_0}{q_1 q_2}\,\frac{\Delta x_2'}{\Delta x_1}\quad \xrightarrow[\substack{q_2\rightarrow q_1}]{\substack{z\rightarrow 0}}\quad W_{D_x}(0) = 0\,.
  \end{align}
  Looking closely at Eq.~(\ref{eq:dipolar_wake_field_kick}) we can deduce that
  \begin{itemize}
      \item the value of the transverse dipolar wake function in $z=0$ vanishes because source and witness particles are traveling parallel and they can only - mutually - interact through space charge, which is explicitly excluded in this framework
      \item $W_{D_x}(0^-) < 0$ since source particles induce image charges which are aligned towards the source particle's offset, thus, the trailing witness particles are always deflected towards the source particle's offset (i.e., $\Delta x_1$ and $\Delta x_2'$ have the same sign)
  \end{itemize}
  
\begin{figure}
      \centering
      \includegraphics[width=0.42\linewidth]{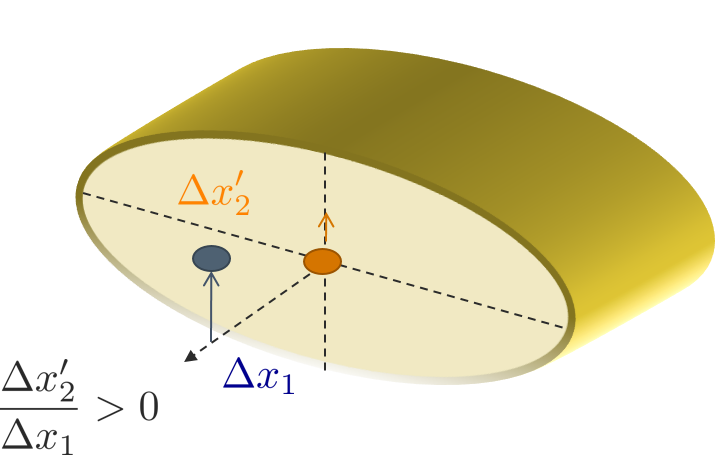}
      \includegraphics[width=0.56\linewidth]{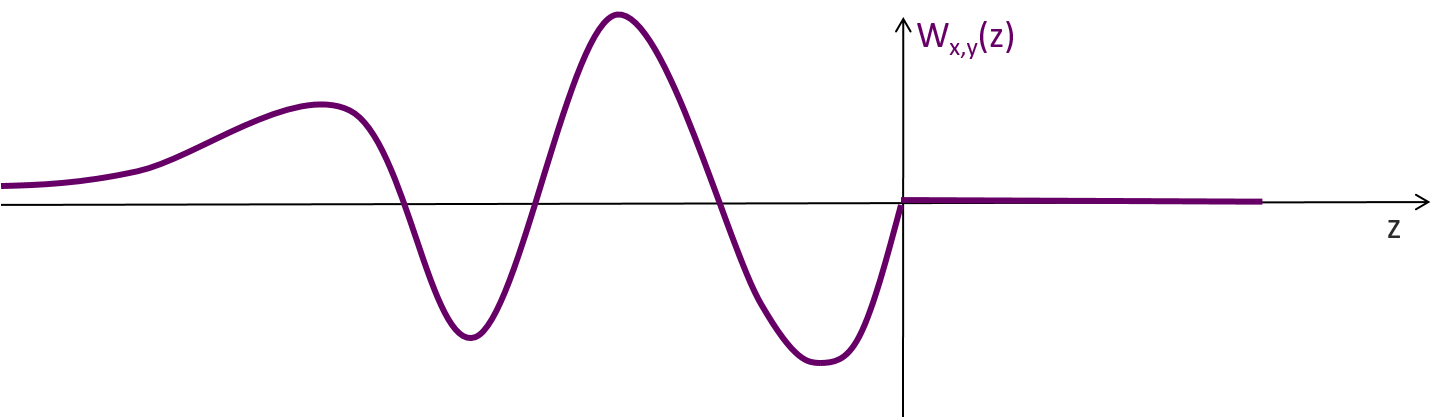}
      \caption{Configuration for the excitation of dipole wake fields; the leading particles induces image charges that attract the trailing particles towards the leading particles. Thus, consistent with its definition, the dipolar wake turns negative for $z \rightarrow 0^-$.}
      \label{fig:wakes_shape_dipole}
  \end{figure}

  Fig.~\ref{fig:wakes_shape_dipole} illustrates the situation of source and witness particles and their interaction with a dipolar wake field.
  
  Similarly, considering the quadrupolar wake, we can derive
  \begin{align}\label{eq:quadrupole_wake_field_kick}
      W_{Q_x}(z) &= -\frac{\beta^2 E_0}{q_1 q_2}\,\frac{\Delta x_2'}{\Delta x_2}\quad \xrightarrow[\substack{q_2\rightarrow q_1}]{\substack{z\rightarrow 0}}\quad W_{D_x}(0) = 0\,.
  \end{align}
  Looking now at Eq.~(\ref{eq:quadrupole_wake_field_kick}) we can deduce this time, that
  \begin{itemize}
    \item the value of the transverse dipolar wake function in $z=0$ vanishes because source and witness particles are traveling parallel and they can only - mutually - interact through space charge, which is explicitly excluded in this framework
    \item $W_{Q_x}(0^-)$ can be of either sign since source particles are considered to travel through the surrounding geometry on-axis and thus will induce image charges depending on the geometry and boundary conditions; thus, trailing witness particles can be either attracted or deflected yet further off axis
  \end{itemize}
  
\begin{figure}
      \centering
      \includegraphics[width=0.38\linewidth]{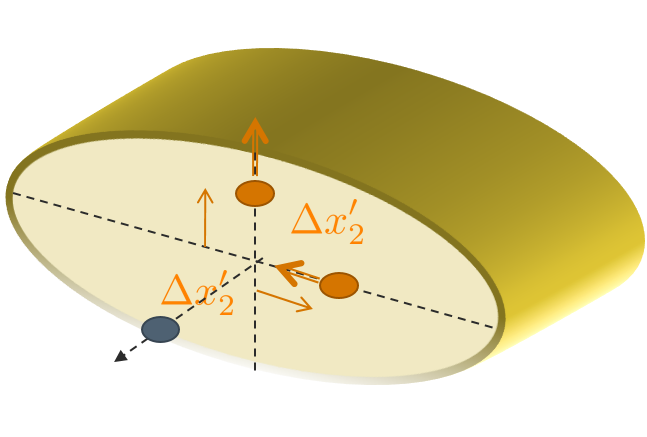}
      \includegraphics[width=0.60\linewidth]{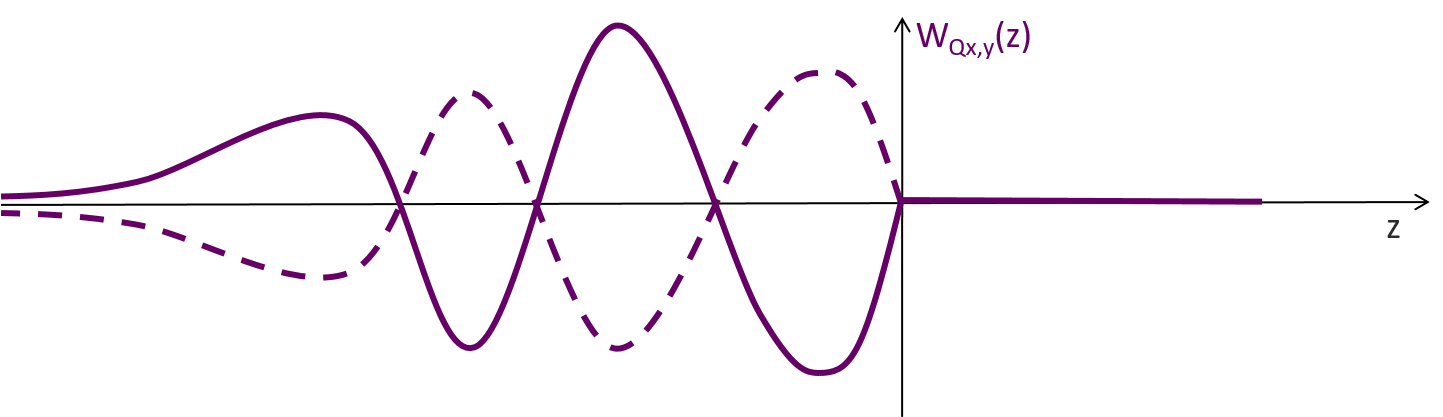}
      \caption{Configuration for the excitation of quadupole wake fields; the leading particles travel through the surrounding structure on-axis and induces image charges according to the structure's geometry. These can then either attract or deflect trailing particles from the axis. Thus, consistent with its definition, the quadrupolar wake turns either negative or positive for $z \rightarrow 0^-$.}
      \label{fig:wakes_shape_quadrupole}
    \end{figure}

  Fig.~\ref{fig:wakes_shape_quadrupole} illustrates the situation of source and witness particles and their interaction with a quadrupolar wake field.
  
  Again, in order to compute the total wake field effects induced by an entire charged particle distribution of line charge density $\lambda_1(z_1)$, Eq.~(\ref{eq:transverse_wake_function}) can be generalized to the convolution
  \begin{align}\label{eq:transverse_convolution}
      \Delta x_2'(z) \propto \int \bigl[ \langle x_1 \rangle\lambda_1 \bigr](z_1)\,W_{D_x}(z-z_1)\,dz_1\,.
  \end{align}
  We can see again from Eq.~(\ref{eq:transverse_convolution}) that also the transverse wake function of an accelerator component is basically its Green's function in time domain (i.e., its response to a pulse excitation). This expression also in the transverse plane is particularly useful for macroparticle models and simulations, because it can be used to describe the driving terms in the single particle equations of motion. The convolution in Eq.~(\ref{eq:transverse_convolution}) can again also be represented in frequency domain. This will then yield the transverse impedance of an accelerator component which is basically again just its transfer function in frequency domain. Thus, the formal definition of the transverse beam coupling impedance of a structure is given by means of its transverse wake function as
  \begin{align}\label{eq:transverse_impedance}
    Z_{D_x}(\omega) &= i\int_{-\infty}^\infty W_{D_x}(z) \exp\left(-\frac{i\omega z}{c}\right)\,\frac{dz}{c}\,,\\
    Z_{Q_x}(\omega) &= i\int_{-\infty}^\infty W_{Q_x}(z) \exp\left(-\frac{i\omega z}{c}\right)\,\frac{dz}{c}
  \end{align}

\subsection{Impact of wake fields and impedance on the accelerator environment}

  We have defined the wake function, longitudinal and transverse wake fields and impedances and have hopefully understood why the concept is beneficial and how it can be put to use to make computations of collective effects more effective. We will now look into a few phenomena linked to wake fields and impedances. We will start by exploring the impact of impedances onto the machine. This will bear the problem of beam induced heating.
  
\begin{figure}
      \centering
      \includegraphics[width=0.9\linewidth]{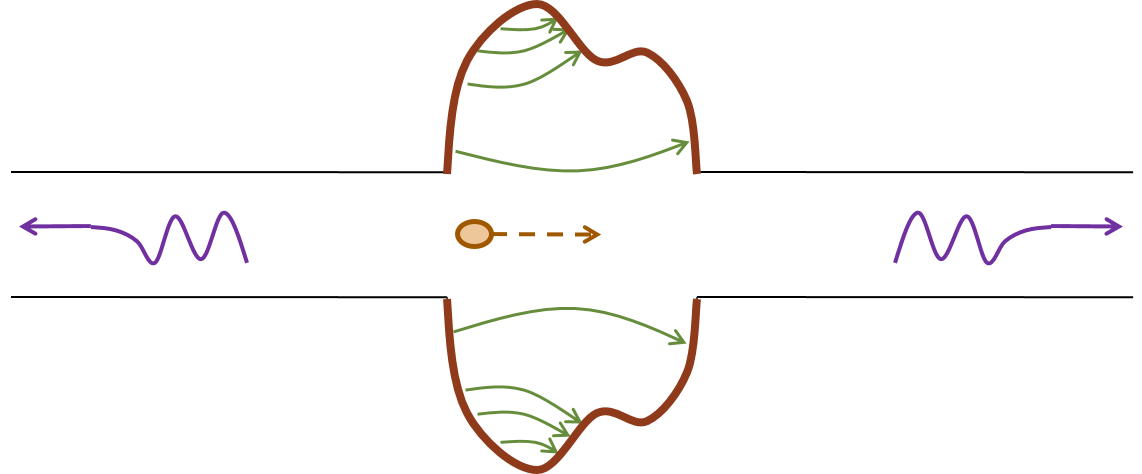}
      \caption{Energy balance for power dissipation of the beam into the machine; energy can be trapped (green), dissipated into the walls (red), transferred to the beam (orange) or travel down the beam pipe to other locations (purple).}
      \label{fig:energy_balance}
  \end{figure}

  As a charged multi-particle distribution traverses a device that has a certain impedance, we have seen above, that this distribution induces and generates wake fields depending on the device's impedance. Of course, due to one of the fundamental law of nature, namely the conservation of energy, the energy for the generation of these fields needs to come from somewhere and it also needs to dissipate to somewhere as these fields decay. This leads us to investigate the global energy balance when considering a charge multi-particle distribution traversing a device with a finite impedance. The source of the energy is actually the RF power that is absorbed by the beam. Ideally all of this power is stored as beam energy. Unfortunately, some of this energy is dissipated by the beam. To understand this dissipation better we look at Fig.~\ref{fig:energy_balance}.
  
  As indicated by Fig.~\ref{fig:energy_balance}, the energy lost by the source splits into:
  \begin{itemize}
      \item Electromagnetic energy of the modes that remain trapped in the object (green):
      \begin{itemize}
          \item Partly dissipated on lossy walls or into purposely designed inserts or HOM absorbers (red)
          \item Partly transferred to following particles (or the same particle over successive turns), possibly feeding into an instability (orange)
      \end{itemize}
      \item Electromagnetic energy of modes that propagate down the beam chamber (above  cut-off), eventually lost on surrounding lossy materials (purple)
  \end{itemize}
  This means, that the energy loss of a particle bunch
  \begin{itemize}
      \item causes beam induced heating of the machine elements (damage, outgasing),
      \item feeds into both longitudinal and transverse instabilities through the associated EM fields,
      \item is compensated by the RF system determining a stable phase shift.
  \end{itemize}
  
  The computation of the energy loss can be carried out explicitly and is done for instance in \cite{Ng:1012829}. It turns out, that the amount of coupling of a beam to a surrounding structure is quantified by the overlap of the beam's power spectrum with the structure's beam coupling impedance. Thus, to compute the energy lost by the beam into the surrounding structure, one simply needs to evaluate the beam power spectrum and use the structure impedance in order to then compute the specific overlap integral.
  
\subsubsection{Example: SPS extraction kickers}

  An example where the beam induced heating became a real problem was in the CERN Super Proton Synchtrotron (SPS). The SPS extraction kicker MKE are used to deflect the beam in the SPS at flat top into the transfer lines to the LHC upon fast extraction. Throughout the entire SPS cycle the beam keeps circulating in the machine, traversing the MKE kickers structure at a rate of 43~kHz. During this time the MKE is an entirely passive element with all of its components (ferrite, conductors) exposed to the beam. It turned out that using the beam for LHC filling (4 x 200~ns spaced trains of 72 x 25~ns spaced bunches) led to an unacceptable heating of these elements. The temperature was raised above the Curie temperature which lead to Ferrite degradation and inhibited the extraction of the beam. Moreover, this heating caused out-gassing and strong pressure rise in the kicker sector, with consequent beam interlocking due to the poor vacuum.
  
  As mentioned just two paragraphs above, the induced heat is a function of the beam power spectrum and the device impedance. The beam that is used by the LHC experiments is clearly specified and it would have been detrimental for the luminosity to step back in bunch intensity or total number of bunches. The other point of action was the impedance of the extraction kicker. This could, in fact, be modified in a beneficial manner by the use of serigraph printing.
  
\begin{figure}[htbp]
      \centering
      \subfigure[Original MKE without serigraphy]{\hspace{6mm}
      \includegraphics[height=0.3\linewidth,valign=b]{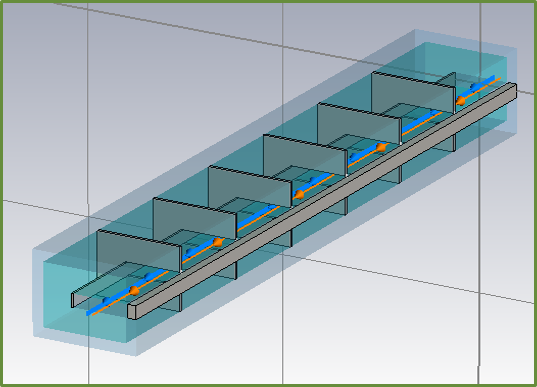}}\hspace{6mm}
      \subfigure[New MKE with serigraphy]{\includegraphics[height=0.3\linewidth,valign=b]{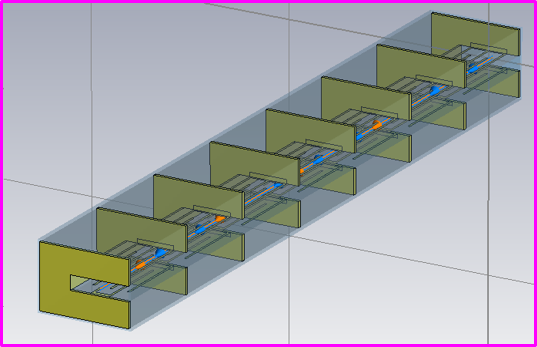}}

      \subfigure[The original MKE broadband impedance (green)]{\includegraphics[width=0.495\linewidth,trim=1mm 1mm 7mm 1mm,clip]{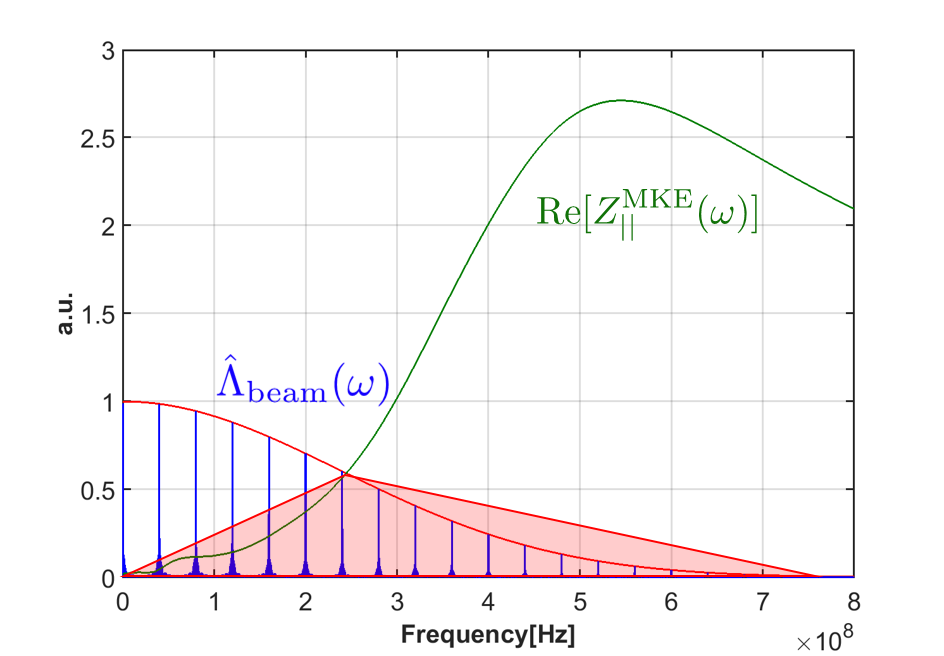}}\hfill
      \subfigure[The serigraphed MKE narrowband impedance (pink)]{\includegraphics[width=0.495\linewidth,trim=1mm 1mm 7mm 1mm,clip]{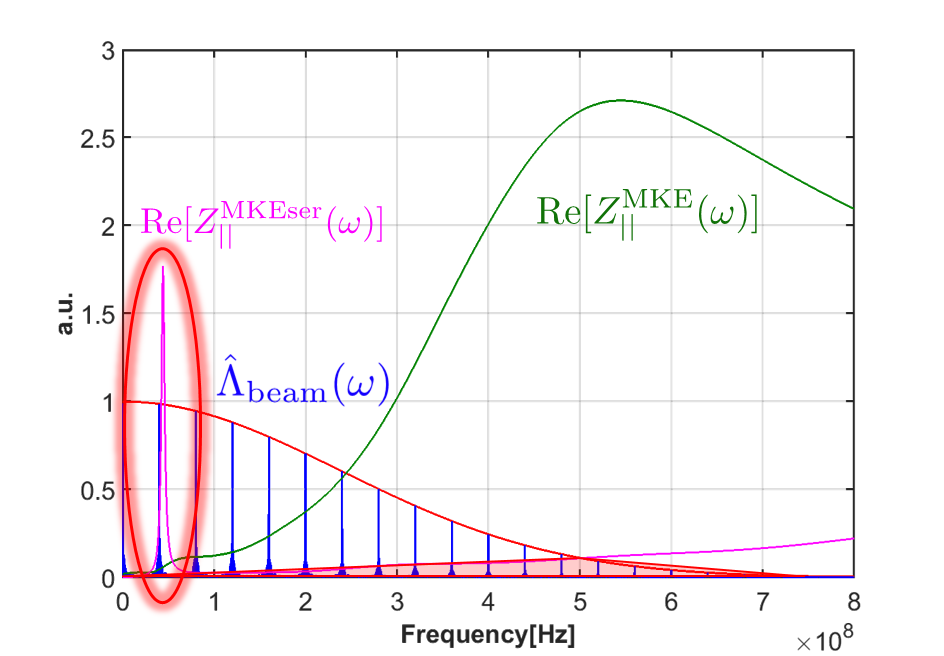}}
      \caption{The left hand side shows the original MKE broadband impedance (green) covering a wide range of frequencies well across the beam power spectrum (blue comb); the red shaded triangular area indicates the overlap. The right hand side is the serigraphed MKE narrowband impedance (pink) covering just a very narrow region of frequencies; the red shaded triangular area indicates the overlap}
      \label{fig:mke_serigraphy}
  \end{figure}

  Figure~\ref{fig:mke_serigraphy} shows a comparison between the original MKE structure and an improved serigraphed structure. The original kicker structure is composed of several modules separated by conductor stripes (segmentation) with bare ferrite blocks, fed by an inner and an outer conductor. The new design was conceived very much like the original, but now the modules have ‘serigraphed’ ferrite blocks (i.e., with patterns of silver paste screen printed on the ferrite surface exposed to the beam). This serigraphy has an important impact on the structure's impedance. 
  
  The left hand side in Fig.~\ref{fig:mke_serigraphy} shows the original structure without serigraphy with its typical broad-band  behaviour. The wake fields excited in the kicker structure decay fast. The energy loss of the beam to the kicker structure is given by the overlap of the impedance in green with the beam power beam spectrum in blue. It is indicated by the red shaded triangular area. The kicker impedance already becomes significant at frequencies for which the beam spectrum has not fully decayed, which causes the undesired heating. The kicker impedance is dominated by the losses in the ferrite blocks. In order to lower the kicker impedance one thus needs to introduce a ferrite shielding. This can be done by printing a striped pattern of good conductor onto the ferrite (serigraphy).
  
  The right hand side in Fig.~\ref{fig:mke_serigraphy} shows the serigraphed kicker. This structure exhibits a strong ringing due to the electromagnetic trapping along the serigraphy fingers. As a result, the impedance is almost entirely suppressed over the full beam power spectrum and only features a narrow peak at a low frequency. The overlap of the impedance, in pink, with the beam power beam spectrum, in blue, becomes very minor, provided the impedance spike doesn't exactly hit a spike in the beam power spectrum (blue comb). This effectively suppresses the kicker heating.
  
\begin{figure}[htbp]
      \centering
      \includegraphics[width=0.8\linewidth]{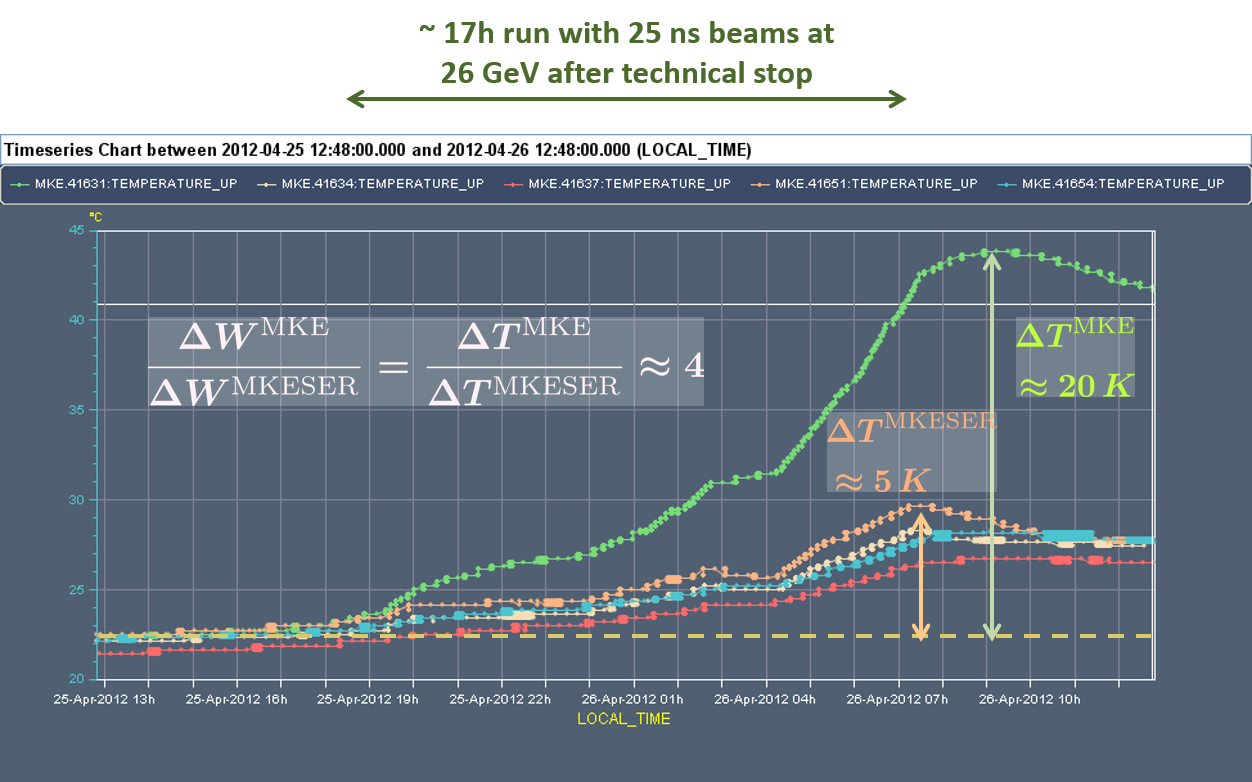}
      \caption{Original kicker heating (green) vs. serigraphed kicker heating (orange) over a time span of about 17~h running with 25~ns beams in the SPS. An improvement of roughly a factor of 4 in kicker heating could be achieved thanks to the serigraphy.}
      \label{fig:mke_temparatures}
  \end{figure}

  In an experiment in the SPS, the machine was operated during a 17 hour run with 25~ns beams at 26~GeV after technical stop with both a normal as well as a serigraphied extraction kicker structure installed. The improvement in kicker heating can be extracted from Fig.~\ref{fig:mke_temparatures}, highlighting that the serigraphy gave a factor~4 improvement in the kicker heating.
  
  This demonstrates that the computation of the impedance of any element present or to be installed in the machine is an absolutely crucial step. Impedance reduction measures and campaigns can significantly help to increase the stability and lifetime of machine components and to improve the overall machine performance. In order to compute the impedance of a given structure or device there are different approaches:
  \begin{itemize}
    \item Analytical or semi-analytical approach, when the geometry is simple (or simplified)
    \begin{itemize}
      \item Solve Maxwell’s equations with the correct source terms, geometries and boundary conditions up to an advanced stage (e.g., resistive wall for axisymmetric chambers)
      \item Find closed expressions or execute the last steps numerically to derive wakes and impedances
    \end{itemize}
    \item Numerical approach
    \begin{itemize}
      \item Different codes have been developed over the years to solve numerically Maxwell’s equations in arbitrarily complicated structures
      \item Examples are CST Studio Suite (Particle Studio, Microwave Studio), ABCI, GdFidL, HFSS, ECHO2(3)D.\footnote{An exhaustive list can be found from the program of the ICFA mini-Workshop on “Electromagnetic wake fields and impedances in particle accelerators”, Erice, Sicily, 23-28 April, 2014}
    \end{itemize}
    \item Bench measurements based on transmission/reflection measurements with stretched wires
    \begin{itemize}
        \item Seldom used independently to assess impedances; usefulness mainly lies in that they can be used for validating 3D EM models for simulations
    \end{itemize}
  \end{itemize}

\subsection{Description of a coherent beam instability and the instability loop}

  In the previous subsection, we have seen how the impedance of a device can have an impact on the machine environment and cause, for example, beam induced heating. This can lead to outgassing or damage of the device. Therefore, devices need to be carefully designed in order to minimize their impedance. 
  
  We will now see, that impedances can also have a direct impact on the passing beam itself. This can lead to impedance induced coherent beam instabilities. Before studying coherent beam instabilities in more depth in the next section, we will first understand the basic concepts and mechanisms behind these.
  
  As a first question one may ask, why the study of coherent beam instabilities is actually so important. It became clear earlier, that badly dimensioned impedances, causing beam induced heating, can be a severe problem for devices installed in the machine, and also for the machine in general. Beam induced heating will lead either to outgassing, deterioration or even destruction of devices or parts of the machine. Impedances effecting the beam, however, can be similarly dangerous. A beam passing into a coherent instability can start performing oscillations at a fast growing amplitude. This normally leads to significant beam degradation due to emittance blow-up or particle losses. In the worst case, the oscillations become so large that the entire beam is lost by impacting into the machine walls. Depending on the deposited beam power this can lead to serious machine damage (vacuum leaks, magnet quenches) and high radiation doses along with a high machine activation. It is thus crucial to remain under the instability threshold intensities or to have adequate instability mitigation measures in place. This often puts a hard limit on the maximum attainable beam intensity and brightness and thus on the overall machine performance. Understanding the instability and its underlying mechanism is thus essential in order to identify the instability sources along with possible mitigation measures and cures. It also permits to suppress the instability by allowing to dimension adequate active feedback systems.

\subsubsection{The instability mechanisms}

\begin{figure}[htbp]
    \centering
    \subfigure[Broadband resonator; left: bunch profile and wake kicks, right: wake function and impedance]{\includegraphics[width=\linewidth]{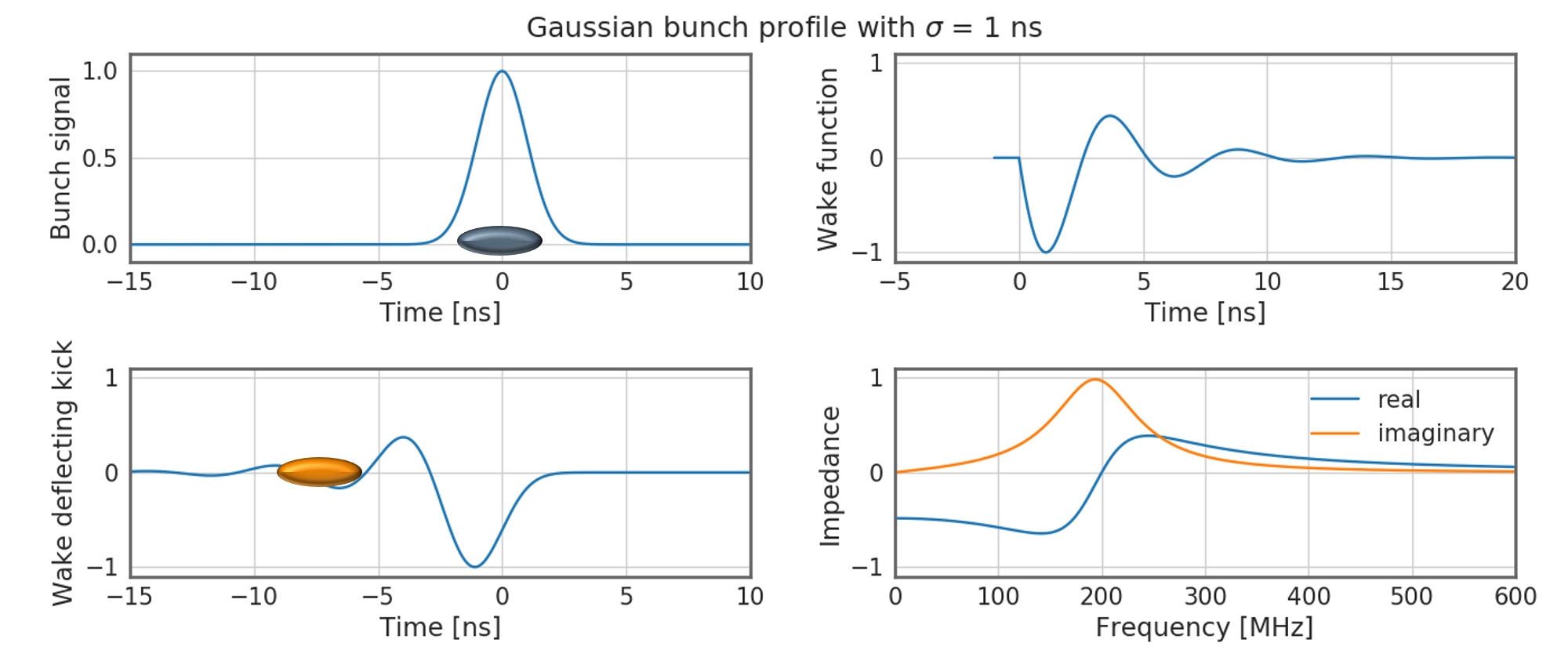}}
    \subfigure[Narrowband resonator; left: bunch profile and wake kicks, right: wake function and impedance]{\includegraphics[width=\linewidth]{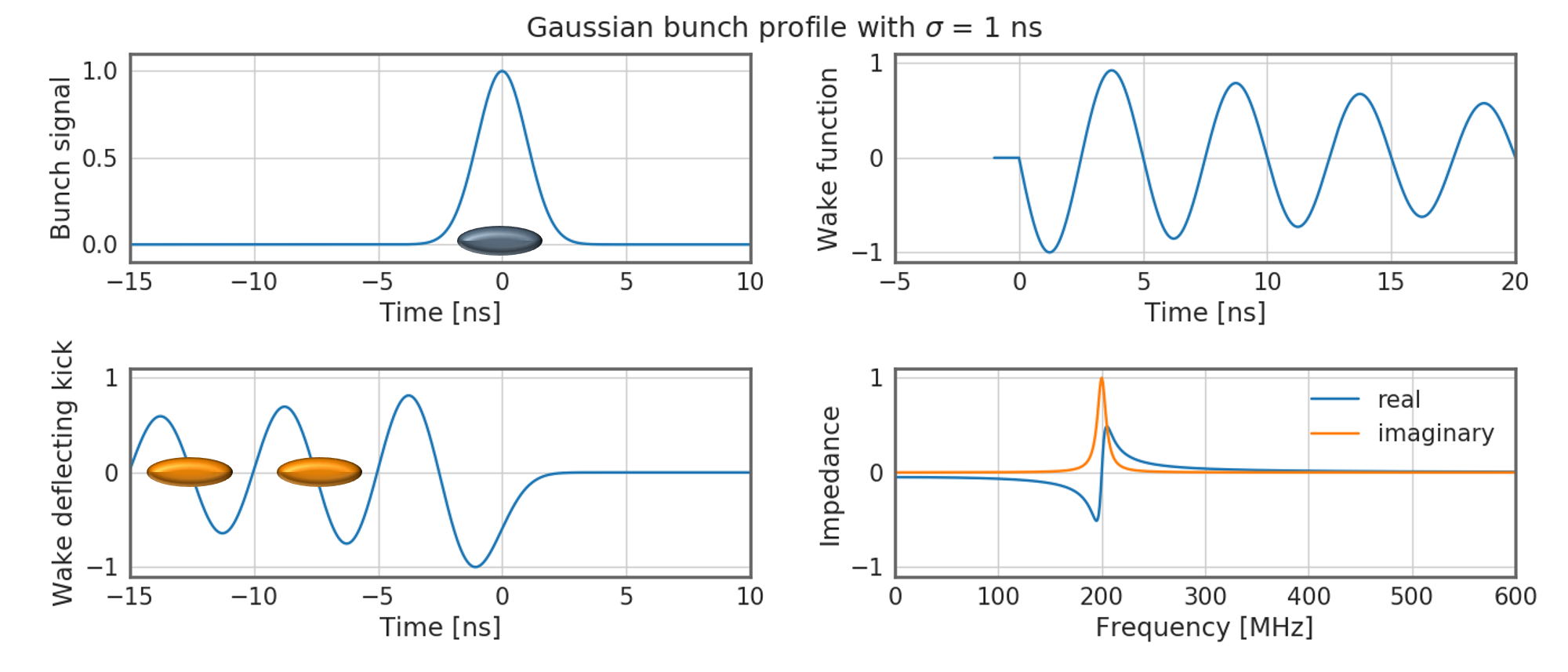}}
    \caption{Broadband (a) and narrowband (b) resonator impedances with wake function and impedance shown on the right and line charge density and wake kicks shown on the left hand side.}
    \label{fig:wakes_bunch_interaction}
  \end{figure}

  We will try to grasp the origins and the development of a coherent beam instability in an intuitive manner. If we look at the plots in Fig.~\ref{fig:wakes_bunch_interaction}, we can see illustrated, the interaction of a charged multi-particle system with the wake function or the impedance of a broadband (a) and a narrowband (b) resonator, respectively. In general, we obtain both wake functions and impedances by using Maxwell’s equations to compute the impulse response for a given structure either in time domain or in frequency domain, respectively. These wake functions and impedances specifically of resonating devices, can be approximated analytically quite will using a resonator model; this is what is shown in Fig.~\ref{fig:wakes_bunch_interaction}.
  
  The right hand side of each figure shows the wake function (impulse response) on the top and the imaginary and the real part of the impedance on the bottom. The left hand side shows the charge distribution with its line charge density and the wake kick along the bunch computed from the convolution in Eq.~(\ref{eq:transverse_convolution}). It is clearly visible how the broadband resonator has fields decaying very fast in terms of bunch length. Thus, broadband resonators wake will drive mainly intra-bunch motion and single bunch instabilities. The narrowband resonator, on the other had, has fields that keep ringing across several subsequent bunches. Thus, narrowband resonators are able to effectively drive coupled bunch instabilities.

\begin{figure}[htbp]
    \centering
    \includegraphics[width=0.45\linewidth]{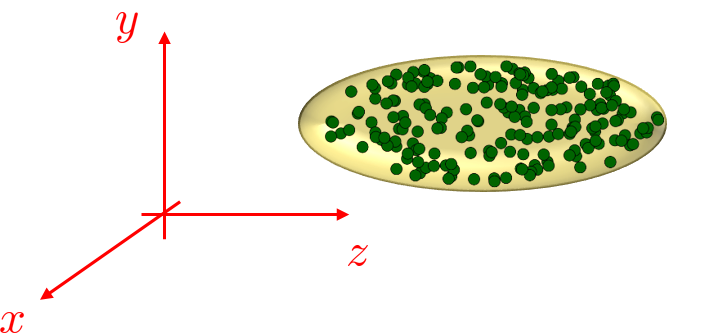}
    \includegraphics[width=0.50\linewidth]{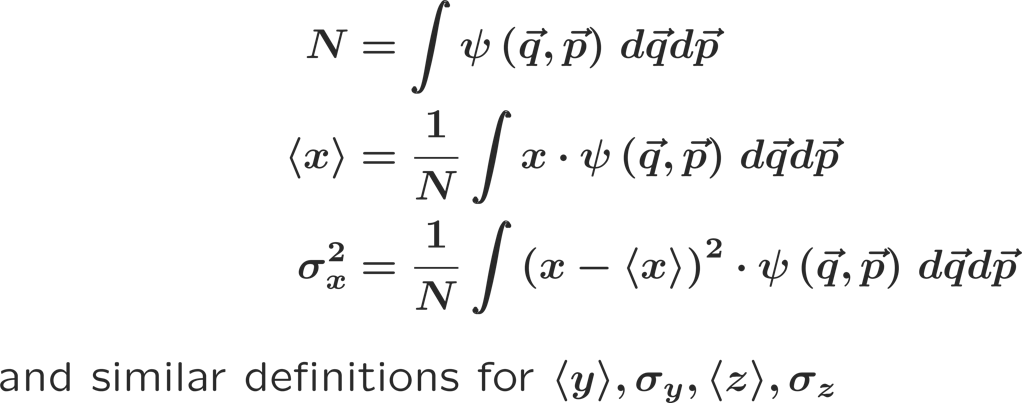}
    \caption{A coherent beam instability is the occurrence of an exponential growth of one of the statistical moments of the charged multi-particle distribution.}
    \label{fig:instability_moments}
  \end{figure}

  In the multi-particle system (described by $\bm{\psi}$) and state (described as $\bm{\left(\vec{p},\vec{q}\right)_N}$) picture, a coherent instability can be regarded as the exponential growth of any of the moments of a charged multi-particle distribution (e.g. mean positions, standard deviations, etc.); this is illustrated in Fig.~\ref{fig:instability_moments}. Such a growth of moments can occur if the betatron and the synchrotron motion and the wake fields manage to synchronize with one another such that they get into constructive resonance. In this case, a distinct bunch oscillation pattern will be excited – a so called bunch coherent mode. For fast decaying fields which are short-ranged, this will be a single bunch mode; for slowly decaying fields which are long-ranged this will be a coupled bunch mode. Whether intra-bunch or coupled bunch motion are excited is determined by the resonator frequency, the bunch length and the bunch spacing. The coherent bunch, or beam signal will then grow exponentially. This can result in beam loss or emittance growth.

\begin{figure}[htbp]
    \centering
    \includegraphics[width=\linewidth]{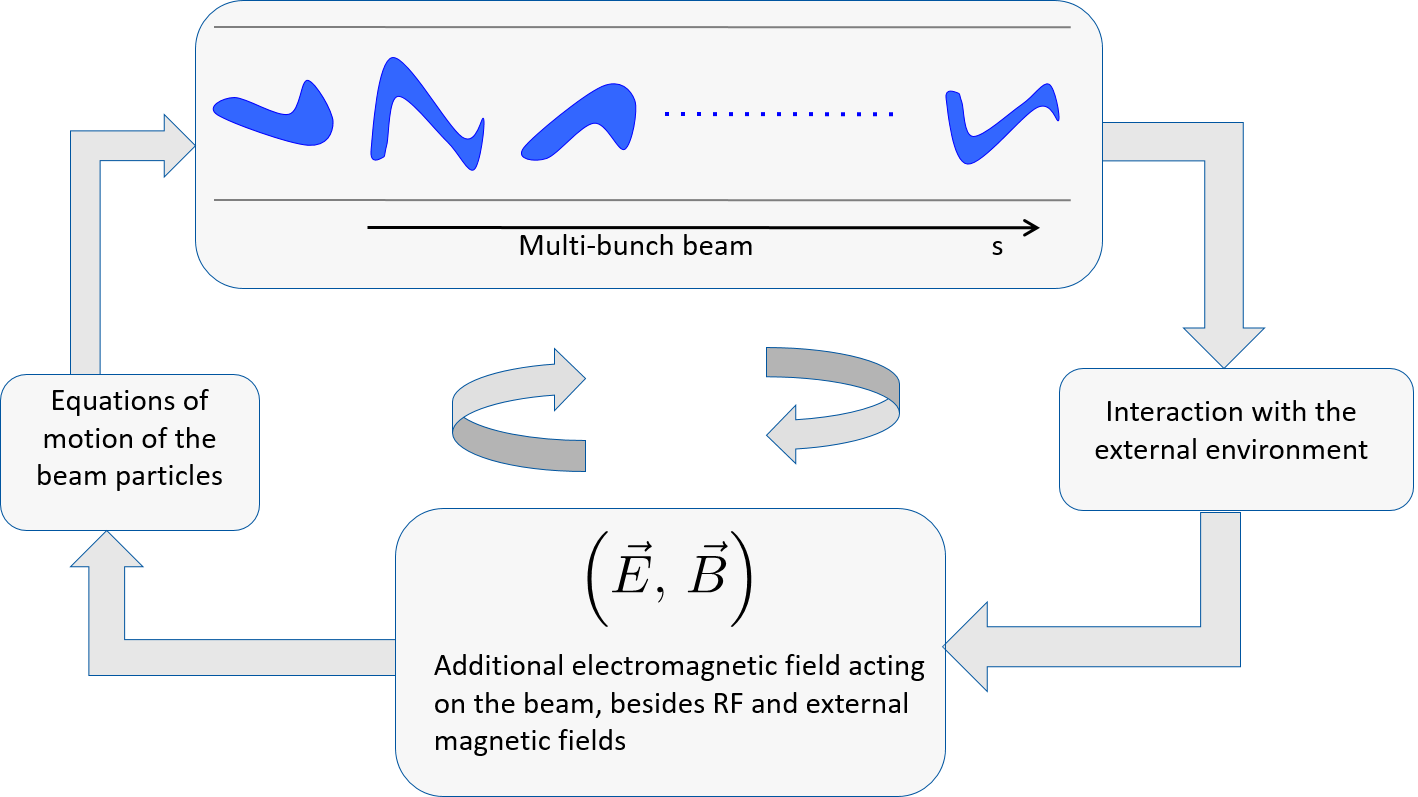}
    \caption{The instability loop illustrating the dependencies within the interaction of a charged multi-particle system or a charged particle beam with the external machine environment.}
    \label{fig:instability_loop}
  \end{figure}

  We will at last look at the instability loop depicted in Fig.~\ref{fig:instability_loop}. Looking at a charged multi-particle system $\bm{\psi}$ as a charged particle beam circulating within an accelerator under the influence of both external fields as well as the self-induced fields from devices with a finite impedance, we can identify several classes and scenarios of action. The charged multi-particle system or the charged particle beam will interact with the external environment to produce additional electromagnetic fields. These additional fields will enter the equations of motion and change the dynamics of the charged particle beam. If this loop is closed, as indicated by Fig.~\ref{fig:instability_loop} the charged particle beam will find a new equilibrium solution. This new solution can feature a new regular pattern which can either be stable or it can exhibit an exponential growth, either across several bunches for a coupled bunch instability, or within a single bunch, for a single bunch instability.
  
  Finally, instability mitigation techniques exist that will act on different parts of the instability loop in Fig.~\ref{fig:instability_loop}. For instance, one can impinge on the right side of the loop by optimizing the interaction with the external environment. This is done by optimizing the impedance or launching dedicated impedance reduction campaigns for the entire machine. One can also try and act on the right hand side by modifying the equations of motion. This is done by introducing additional terms with the help of sextupoles or octupoles, leading to effects such as chromaticity or Landau damping to help suppressing instabilities. And finally, one can act on the instability loop by adding yet another loop which operates in parallel to counter-act the instability growth. Such an external loop can be realized as an active feedback system to ensure beam stability. 

  We have seen how the concept of wake fields and impedances can be used to help effectively evaluating the interaction of charged particle beams with the accelerator environment as collective effects. We have seen that these collective effects can have a detrimental machine, on the example f beam induced heating, but also on the charged particle beam itself, as coherent beam instabilites. Both effects can be devastating to the machine and are serious performance limitations. We looked at the instability loop and used this to identify several mitigation methods.
  
  In the last section, we will more closely investigate the phenomena of coherent beam instabilities.


\clearpage\pagebreak
\section{Coherent beam instabilities}
\label{sec:instabilities}

In this last part of our introductory lecture on collective effects in beam dynamics we will focus on the impact and phenomena of collective effects exerted on the beam. Namely, we will look into more detail at the different forms and some examples of coherent beam instabilities. We have already conceptually seen in the previous section how coherent beam instabilities can develop when studying the instability loop. We have also seen a few approaches for mitigation techniques when inspecting the instability loop. 




As already mentioned a few times, coherent beam instabilities are a serious limitation for any accelerator approaching its performance limits in terms of intensity and brightness. Induced by collective effects, these instabilities typically become more pronounced and vicious as the beam intensity increases; this effectively provides an intensity limitation for the respective accelerators with implications on accelerator specific performance parameters, such as the maximum attainable brightness or luminosity, for example. It is thus important to understand the mechanism that leads to coherent instabilities. It is also important to understand the characteristics and the signatures of these instabilities in order to understand their occurrence and to find adequate means of mitigation in order to push the performance of the accelerator beyond its limits. 

In this section, we will first look at a few real world examples of coherent beam instabilities encountered at some of the CERN accelerators. We will then try to classify, in the transverse plane, a few of the most commonly occurring instabilities. We will learn, roughly, how these can be described theoretically. Finally, we will spend a few words on coherent instabilities encountered in the longitudinal plane.

\subsection{Examples of coherent beam instabilities at CERN}

A very nice showcase for different types of transverse coherent beam instabilities was provided during the years 2015 until about 2018 during the Run~2 of the LHC in the CERN Super Proton Synchrotron (SPS). The SPS acts as the final injector for the LHC. Proton beams composed of four batches of 72 bunches (288 bunches) are accelerated from 26~GeV to an extraction energy of 450~GeV. The LHC Injectors Upgrade Project (LIU) \cite{} targets an increase of the beam brightness by roughly doubling the intensity and reducing the transverse emittance in order to prepare the CERN injectors complex to deliver beams for the High Luminosity LHC (HL-LHC) upgrade \cite{BéjarAlonso:2749422} which is planned to start in 2025, with the goal to increase the LHC luminosity by one order of magnitude. 

In order to understand and explore the present limitations of the SPS, dedicated studies were done with high intensity beams injected and circulated in the machine. In these studies different types of instabilities occurred which limited the maximum attainable intensities at the time.

\begin{figure}[htbp]
    \centering
    \subfigure[Bunch-by-bunch horizontal positions snapshot]{\includegraphics[width=0.495\linewidth]{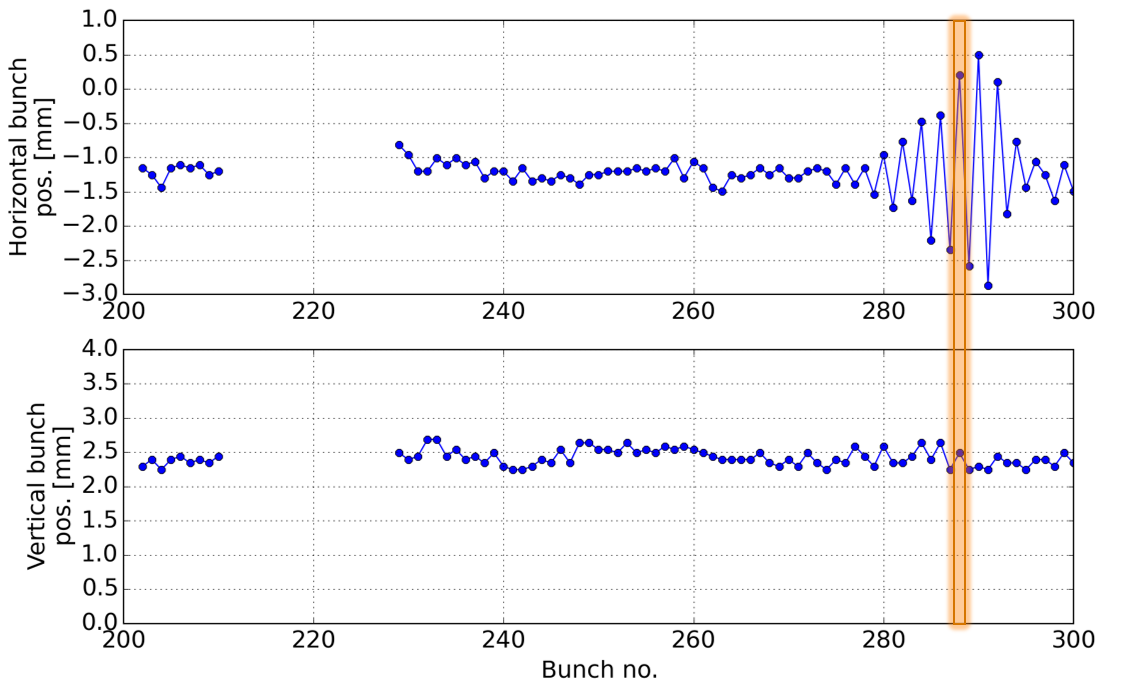}}
    \subfigure[Turn-by-turn horizontal position for an individual bunch]{\includegraphics[width=0.425\linewidth]{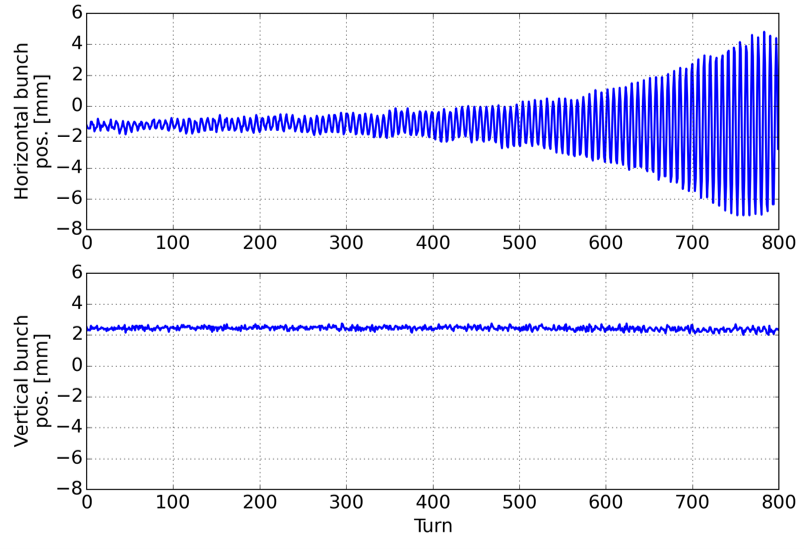}}
    \caption{Horizontal coupled bunch instability observed during the 2015 scrubbing run in the SPS.}
    \label{fig:coupled_bunch}
\end{figure}

An initial instability was observed during the scrubbing run in 2015\footnote{Scrubbing runs are performed regularly in the SPS, in particular after a longer stop; the goal of these runs is to operate the machine at a high duty cycle with high intensity beams in order to deliver high doses of electron cloud to condition the surfaces of the inner vacuum chamber; this ultimately reduces electron cloud effects and improves the final beam quality}. This was a coupled bunch instability in the horizontal plane. An image with a typical signature of such a coupled bunch instability is shown in Fig.~\ref{fig:coupled_bunch}. The top figure displays the bunch-by-bunch pattern for the last part of the beam. It is clearly visible, how in the horizontal plane there is a strong correlated oscillation of bunches towards the end of the last batch of 72 bunches. In fact, bunches are strongly oscillating at a fixed phase relation of 180 degrees from one bunch to the next. Isolating a single bunch and observing its time evolution reveals an exponential growth of the oscillation amplitude. This is shown on the bottom plot of Fig.~\ref{fig:coupled_bunch} for the bunch in slot 287 (indicated by the orange bar). This type of instability is induced by the long range wake fields coming from resistive wall impedances, for example.

With a bunch spacing of 25~ns, the mode frequency of the observed coherent instability becomes 40~MHz. In order to damp this type of instability, it is sufficient to employ a transverse feedback system with a frequency reach of up to 20~MHz. The SPS transverse damper normally has enough bandwidth for this, but had not initially been set up to provide sufficient gain in this frequency range. Hence, the horizontal coupled bunch instability emerged. In one of the following runs, the transverse damper was adjusted to provide gain along its full bandwidth. After this, the instability was successfully suppressed and was not longer observed. Instead, however, a new type of instability arose.

\begin{figure}[htbp]
    \centering
    \includegraphics[width=\linewidth]{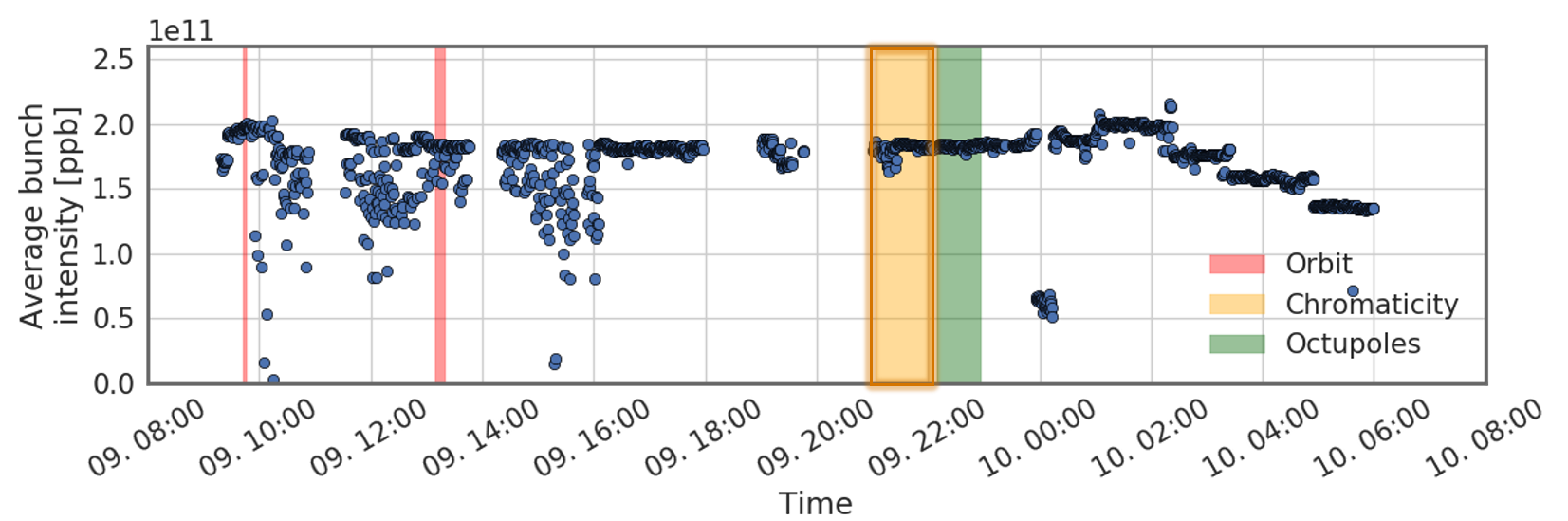}
    \caption{High intensity tests during dedicated MDs in 2017 in the SPS.}
    \label{fig:high_int_mds}
\end{figure}

\begin{figure}[htbp]
    \centering
    \subfigure[Chromaticity $\xi = 0.2$]{\includegraphics[width=0.495\linewidth]{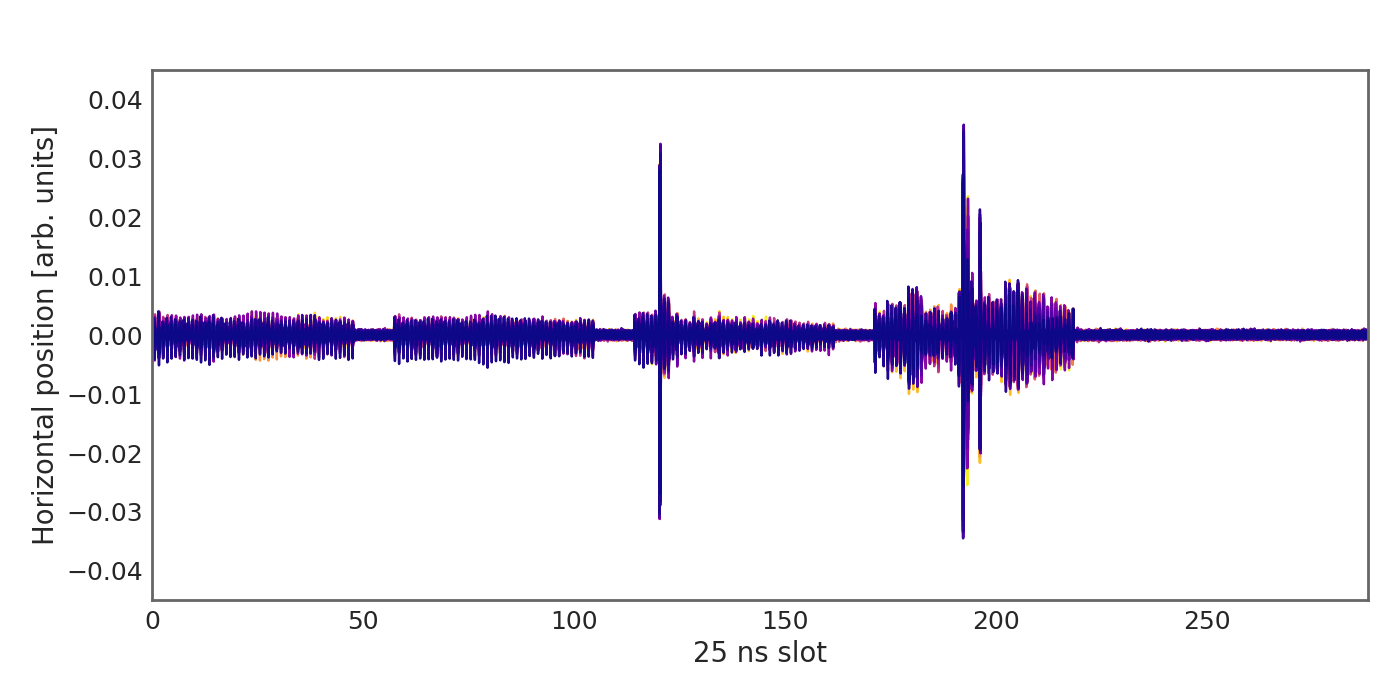}}
    \subfigure[Chromaticity $\xi = 0.4$]{\includegraphics[width=0.495\linewidth]{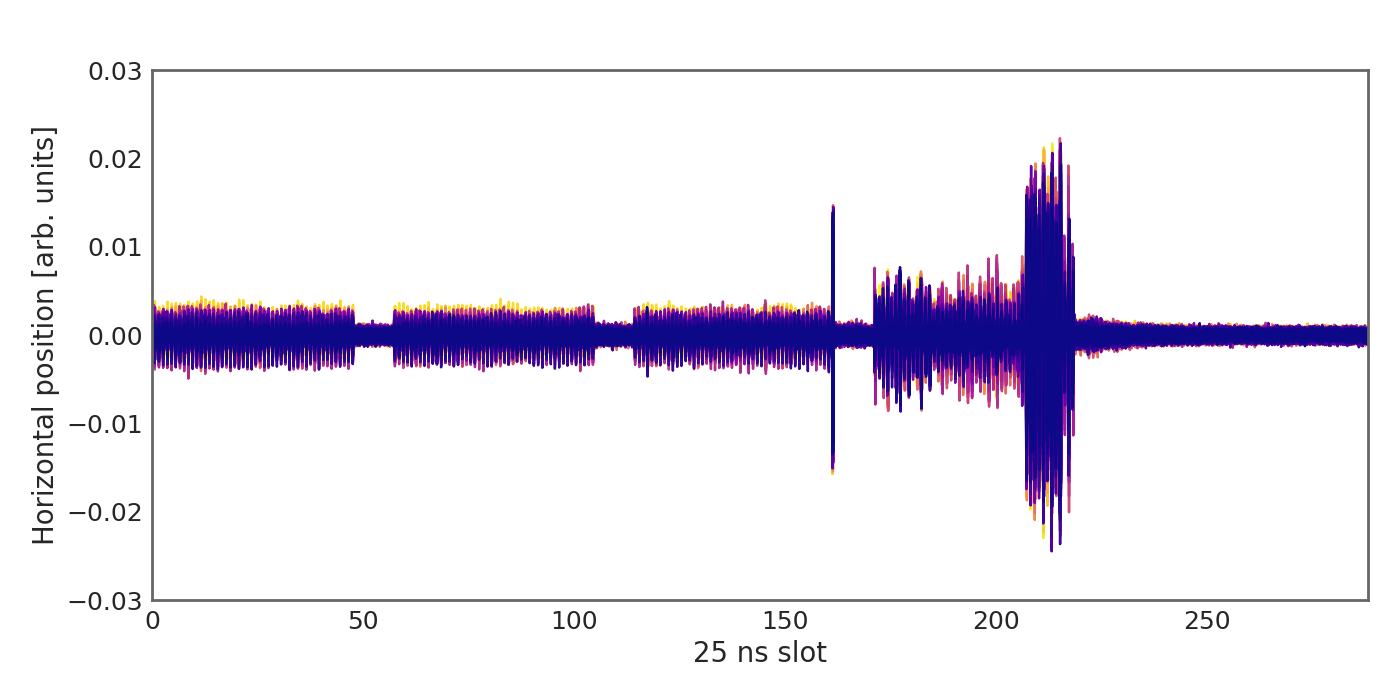}}
    \subfigure[Chromaticity $\xi = 0.6$]{\includegraphics[width=0.500\linewidth]{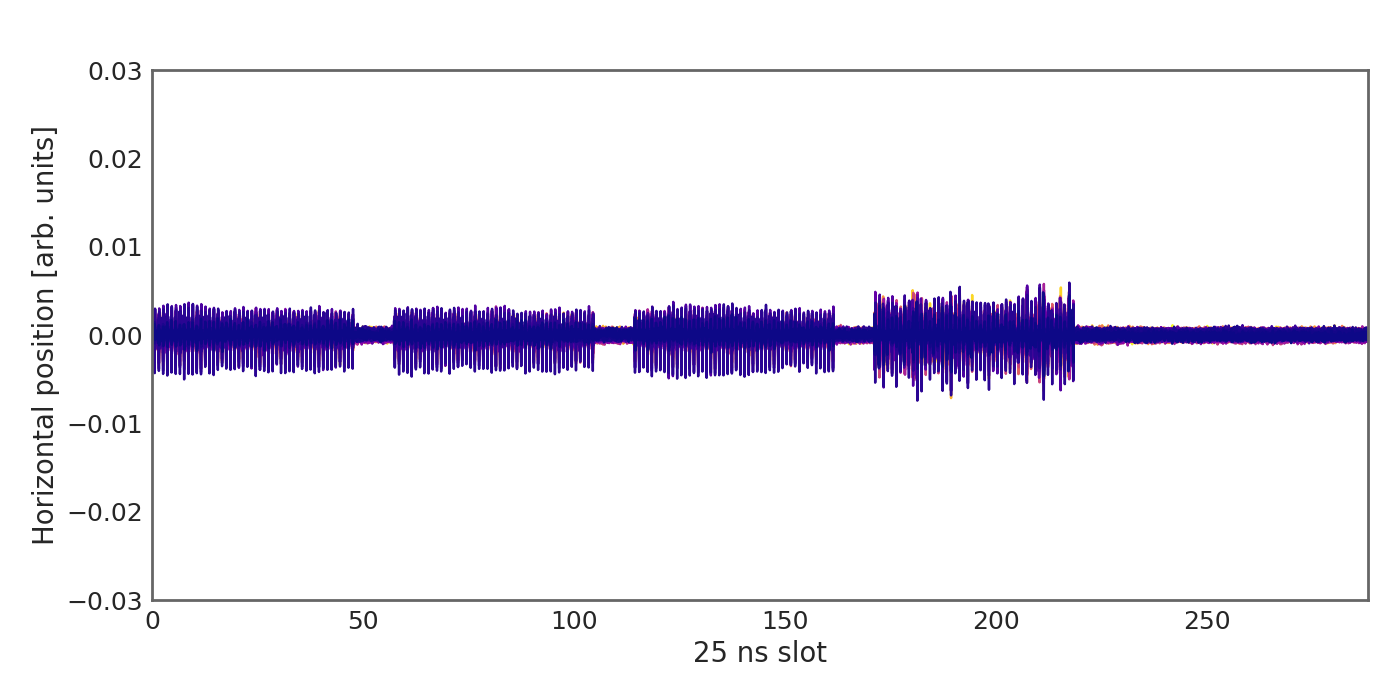}}
    \caption{High intensity tests during dedicated MDs in 2017 in the SPS.}
    \label{fig:high_int_headtail_modes}
\end{figure}

Figure~\ref{fig:high_int_mds} shows an evolution plot of single bunch intensities during dedicated high intensity studies at injection energy in the SPS. During these studies, high intensity beams of up to 192 bunches (4 batches of 48 bunches) were injected into the SPS. As mentioned above, the transverse damper was set up to provide gain across its full bandwidth, thus, successfully mitigating the horizontal coupled bunch instability. Nevertheless, individual bunches within the later batches could be observed becoming unstable. During a certain period, indicated by the yellow region in Fig.~\ref{fig:high_int_mds}, a chromaticity scan was done. Single bunch instabilities were observed for lower chromaticity values, and they disappeared for high chromaticities beyond 0.5. Figure~\ref{fig:high_int_headtail_modes} depicts the turn-by-turn, bunch-by-bunch signals obtained by the headtail monitor for the different chromaticities. The headtail monitor records the turn-by-turn traces of the full beam at a high sampling rate allowing to resolve the transverse turn-by-turn motion of individual bunches. The plots in Fig.~\ref{fig:high_int_headtail_modes} show the turn-by-turn acquisitions of the headtail monitor superposed one turn after the other. A single trace corresponding to one turn shows the horizontal oscillation amplitude as a function of the 25~ns bunch slots. Violent single bunch instabilities are clearly observed for chromaticity values of around $\xi=0.2$. The instabilities persist also for higher chromaticities around $\xi=0.4$. At chromaticities above $\xi=0.5$ the beam starts to settle down as instabilities are suppressed.

\begin{figure}[htbp]
    \centering
    \subfigure[Headtail mode 1 for chromaticity $\xi=0.2$]{\includegraphics[width=0.465\linewidth]{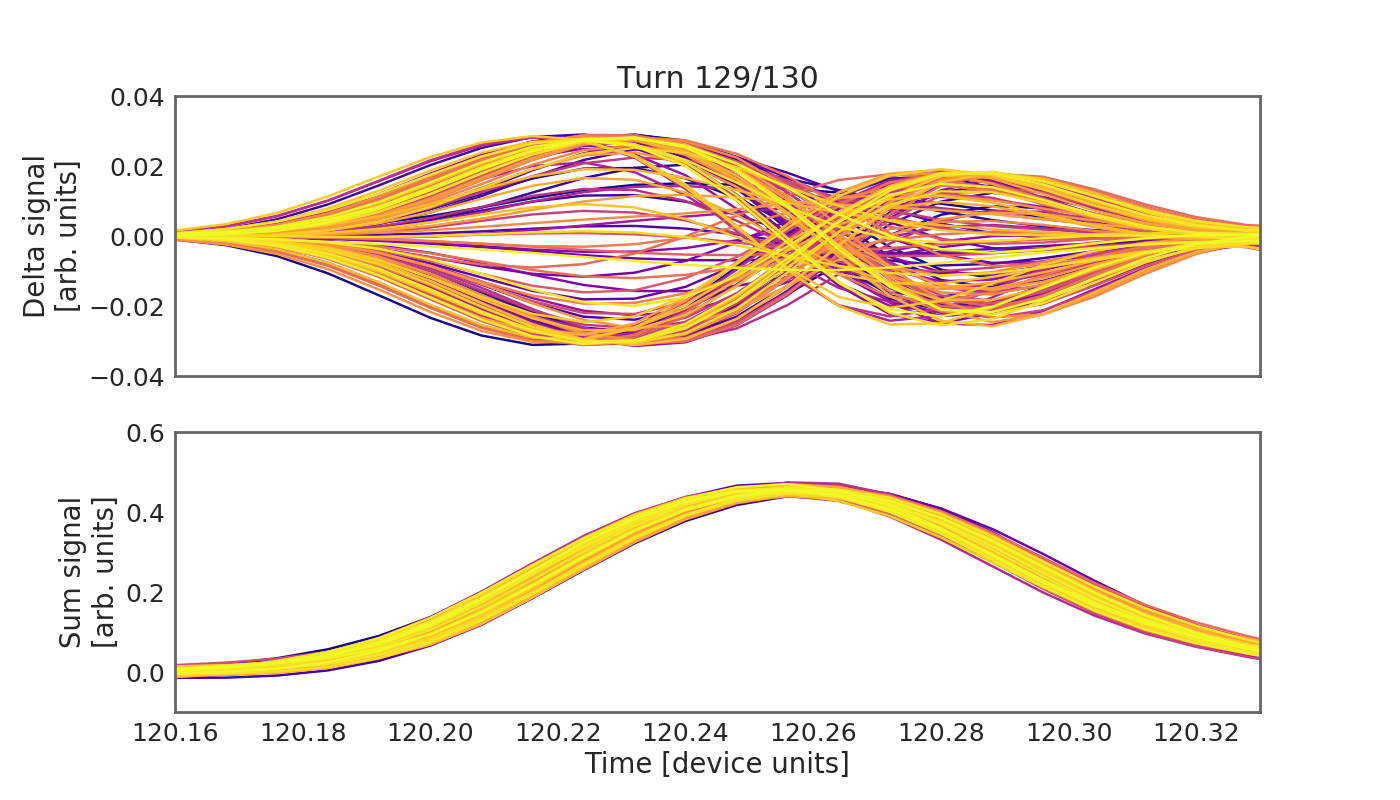}}
    \subfigure[Headtail mode 2 for chromaticity $\xi=0.4$]{\includegraphics[width=0.465\linewidth]{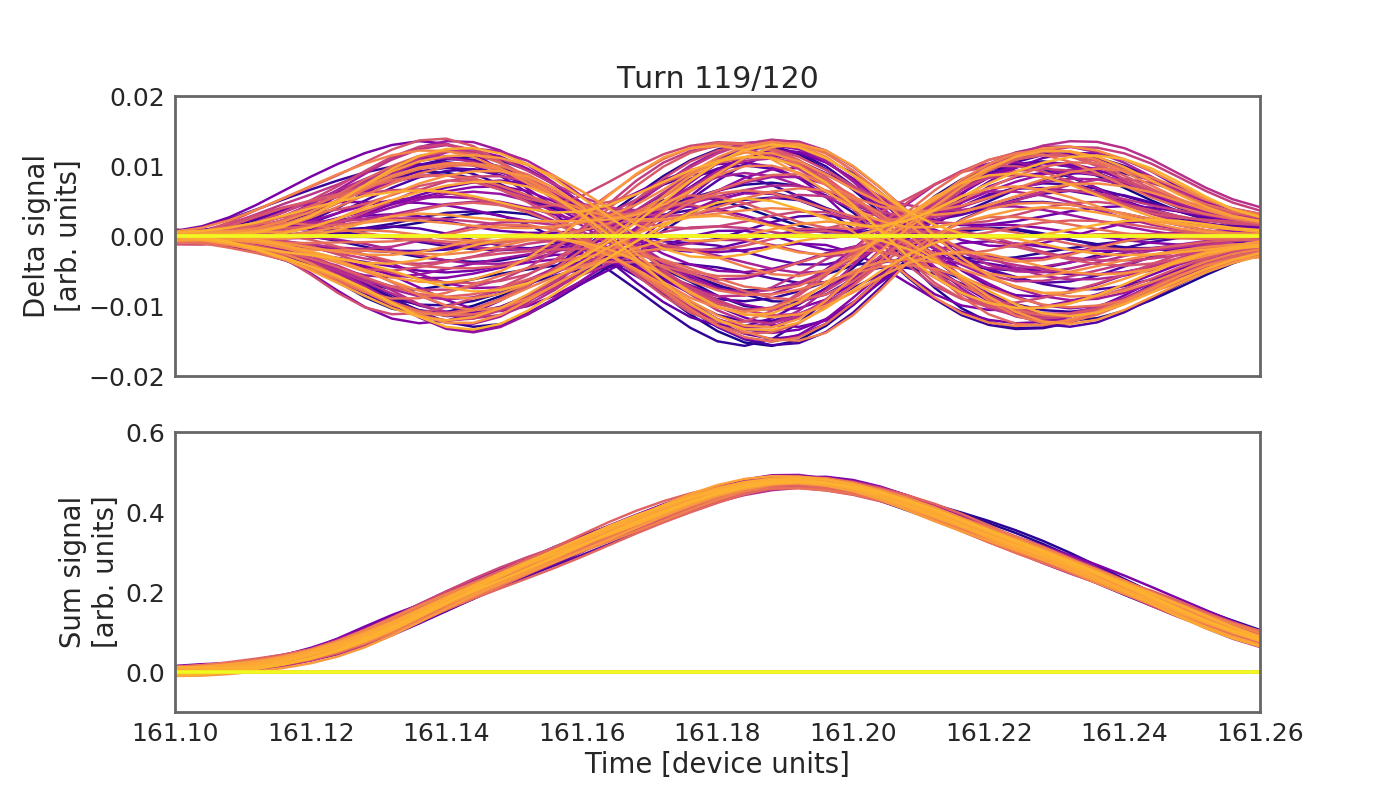}}
    \caption{A zoom into some of the traces in the figure above reveals strongly correlated intra-bunch motion as the typical signature of headtail modes.}
    \label{fig:high_int_headtail_modes_zoom}
\end{figure}

It is interesting now to zoom into a single bunch instability signal and to inspect the acquired bunch motion. Such a close-up is shown in Fig.~\ref{fig:high_int_headtail_modes_zoom}; the left hand side shows the instability signature obtained for a chromaticity of $\xi=0.2$, the right hand side shows the instability signature for a chromaticity of $\xi=0.4$. It becomes evident that the single bunch motion follows a coherent, regular and stationary pattern with a high frequency content which manifests itself as a pronounced, stationary intra-bunch motion. The fixed phase relation is now between longitudinal slices along the bunch rather than between neighbouring bunches along the beam. The frequency content of this motion for the short bunches around 3~ns or less ranges from 500~MHz up to more than 1 GHz which is far beyond the bandwidth of the transverse damper. This explains why the transverse feedback system is not able to damp this type of instability.

The instabilities observed in Fig.~\ref{fig:high_int_headtail_modes_zoom} are a well known phenomenon of transverse collective effects and are know as (slow) headtail instabilities. We will study this effect in a phenomenological fashion in the next subsection.

\subsection{Slow headtail instability at finite chromaticity}

\begin{figure}[htbp]
    \centering
    \includegraphics[width=0.950\linewidth]{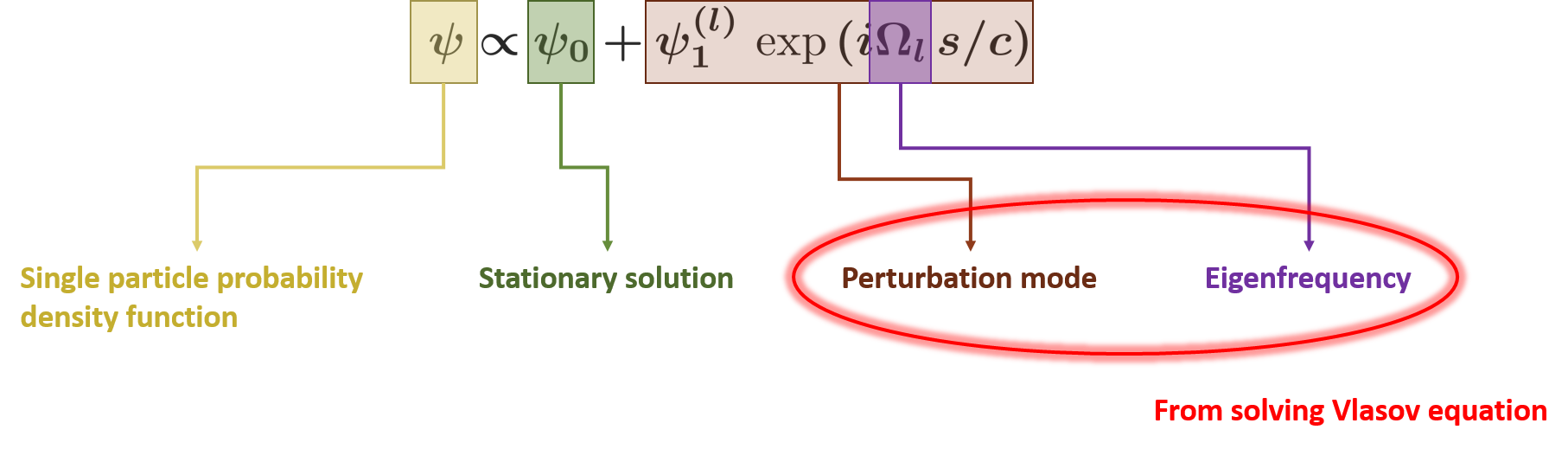}
    \label{fig:headtail_vlasov}
    \caption{Schematic illustration of relationship between stationary solution, headtail modes and Vlasov equation for the description of headtail modes.}
\end{figure}

Headtail modes are one of the typical manifestations of single bunch collective effects. They are induced by short range wakefields. Headtail modes formally emerge from the Vlasov equation in the presence of impedances, where they evolve as eigenmodes of the coupled accelerator-beam system. The eigenvalue of these modes is the complex tune $\bm{\Omega}$. This number $\bm{\Omega}$ fully characterizes a mode and thus, also the associated instability. Fig.~\ref{fig:headtail_vlasov} schematically illustrates how the stationary solution, the headtail mode, with its complex tune $\bm{\Omega}$, and the Vlasov equation relate to each other. The single particle probability density function $\bm{\psi}$ is written as the superposition of the stationary solution $\bm{\psi}_0$, which is the single particle probability density function in absence of collective effects, and a stationary perturbation $\bm{\psi}_1$ which evolves from the solution of the Vlasov equation\footnote{Eigenmodes are the eigenfunctions of the time evolution operator $\frac{\partial}{\partial t}\,\bm{\psi}_1$; the eigenvalues are the complex tune $\bm{\Omega}$.}. The real part of the complex tune $\operatorname{Re}(\bm{\Omega})$ gives the coherent tune shift $\Delta Q_l$ due to the impedance. The imaginary part of the complex tune $\operatorname{Im}(\bm{\Omega})$ gives the instability growth rate $\tau_l^{-1}$. Both are key parameters to characterize an instability.

In general, writing down the full Vlasov equation with the beam coupling impedance and finding a solution in terms of eigenmodes and eigenfrequencies is a non-trivial task. In 1972, Frank Sacherer wrote down an approximate model and solutions for the headtail modes, which very well matched observations in the CERN Proton Synchrotron (PS) \cite{Sacherer:322545}. These were written as
\begin{align}\label{eq:sacherer_modes}
    p_l(z) &= \left\{
      \begin{matrix}
          \cos \left(\left(l+1\right) \pi \dfrac{z}{\hat{z}}\right),\, l=0,2,4,\ldots\\\\
          \sin \left(\left(l+1\right) \pi \dfrac{z}{\hat{z}}\right),\, l=1,3,5,\ldots
      \end{matrix}
    \right.
\end{align}
with $z$ the longitudinal position along the bunch and $\hat{z}$ the bunch length; $l$ is the headtail mode number.

\begin{figure}[htbp]
    \centering
    \subfigure[Headtail modes 0 -- 3, chromaticity $\xi = 0$]{\includegraphics[width=0.900\linewidth]{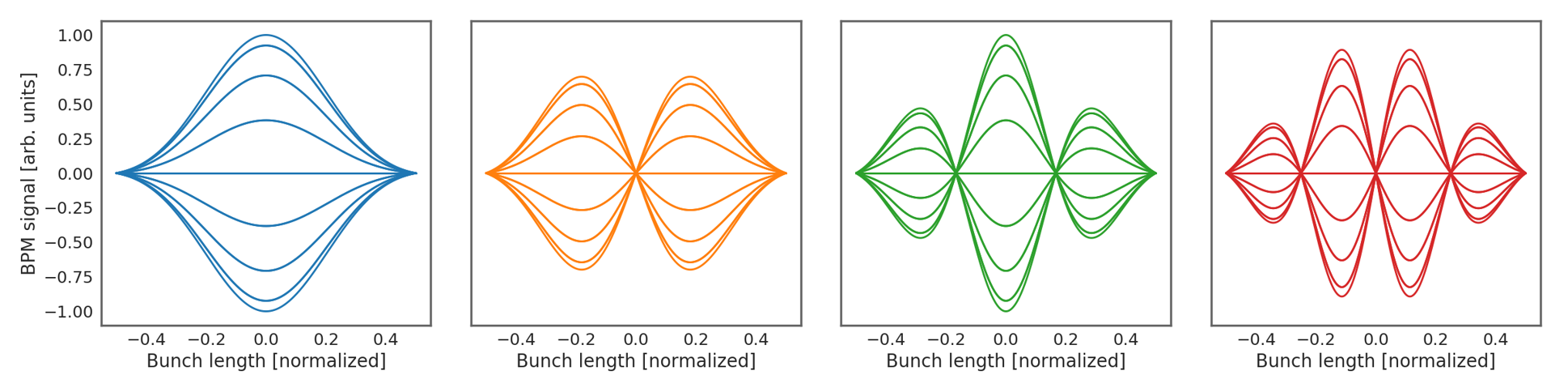}}
    \subfigure[Headtail modes 0 -- 3, chromaticity $\xi > 0$]{\includegraphics[width=0.900\linewidth]{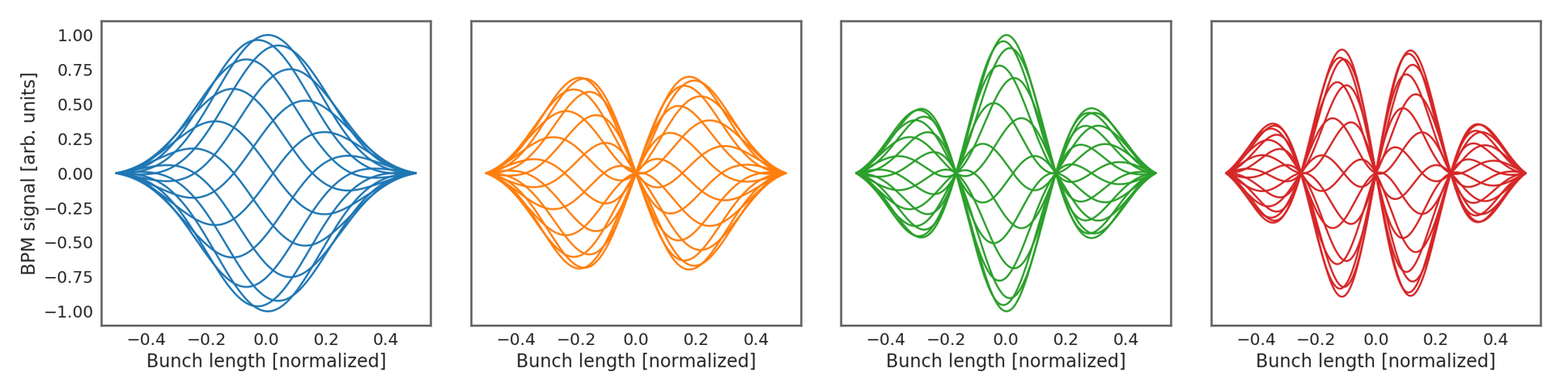}}
    \caption{Headtail modes reconstructed from Schaerer's approximate model, following Eq.~(\ref{eq:sacherer_pickups}), for zero chromaticity (top) and for finite chromaticity (bottom).}
    \label{fig:headtail_chroma}
\end{figure}

Sacherer found in \cite{Sacherer:864422} that the observed signal at a (wideband) pickup for a given mode $l$ with this can well be described as
\begin{align}\label{eq:sacherer_pickups}
    S &\propto p_l(z)\,\exp\left(
    -2\pi i \left( k Q_x + \frac{\xi Q_x \omega_0}{2\pi\eta\beta c}\,z \right)\,.
    \right)
\end{align}
Here, $k$ is the turn number, $Q_x$ the horizontal tune, $\xi$ the normalized chromaticity, $\eta$ the slippage factor and $\omega_0$ the revolution frequency. It is to be noted, that these bunch modes are always latently present, but are not usually excited. It requires a beam coupling impedance to establish an energy transfer together with chromaticity to generate a synchronization of the bunch motion with the wake fields kicks in order to drive a given bunch mode into resonance. Figure~\ref{fig:headtail_chroma} shows the first four headtail modes constructed from Eq.~(\ref{eq:sacherer_pickups}) at zero chromaticity (top) as well as at some finite chromaticity (bottom).

Chromaticity in general, plays a critical role in the excitation of headtail modes. Chromaticity leads to a de-phasing of the betatron oscillations and can bring about a synchronicity condition between betatron and synchrotron motion and the associated wake kicks so as to excite a given headtail mode and to provide a channel for the energy transfer between the impedance sources and the beam. Thus, chromaticity is a key parameter in determining which headtail mode will be excited. It turns out, that the growth rate of a given headtail mode $l$ roughly follows
\begin{align}\label{eq:chroma_growth_rates}
    \frac{1}{\tau} &\propto (1 + l) ^{-1}\,.
\end{align}
One can see from Eq.(\ref{eq:chroma_growth_rates}), that the lower order modes are the fastest growing ones and are thus the most violent. Hence, chromaticity must be well controlled in order to maintain stability. At a chromaticity of exactly zero, all headtail modes are equally damped and excited and there is no rise of headtail modes to be expected. Operationally, it is difficult to run at exactly zero chromaticity. One thus tunes chromaticity to ensure suppression of the lower order modes in favour of the less violently growing higher order modes. These are then mitigated by other means (i.e., Landau damping). As a matter of fact, the lowest order mode - mode 0 - needs to be treated in a specifically particular manner. When below transition, mode 0 is normally excited for positive chromaticity values. The situation inverts when the machine is operated above transition; here, mode 0 is excited for negative chromaticity values. As lowest order headtail mode, mode 0 is the fastest growing headtail instability and can be rather violent. Thus, one tunes the chromaticity to always suppress mode 0. This means below transition a machine must be operated with negative chromaticity whereas above transition the chromaticity must always be slightly positive; these crucial modes of operating a synchrotron are summarized in Tab.~\ref{tab:chromaticity}.

\begin{table}[htbp]
    \centering
    \begin{tabular}{rll}
    \toprule
         & Below transition & Above transition \\
    \midrule
      Chromaticity negative & \textbf{\textcolor{Green}{l = 0 stable; l > 0 unstable}} & \textbf{\textcolor{Crimson}{l = 0 unstable; l > 0 stable}}\\
      Chromaticity positive & \textbf{\textcolor{Crimson}{l = 0 unstable; l > 0 stable}} & \textbf{\textcolor{Green}{l = 0 stable; l > 0 unstable}}\\
    \bottomrule
    \end{tabular}
    \caption{Overview over stable operation regimes for machines running below or above transition.}
    \label{tab:chromaticity}
\end{table}

This concludes our studies on the slow headtail modes. We have seen how these can be phenomenologically described as the Sacherer’s sinusoidal modes. We have also learned that it requires impedances and chromaticity for these modes to actually be resonantly excited; this happens as a synchronization between the betatron and synchrotron motion and the associated wake kicks so as to excite a given headtail mode along the bunch and to provide a channel for the energy transfer between the impedance sources and the beam. And we know now that machines must typically be operated in certain chromaticity regimes, depending on the state of transition, in order to suppress the lowest order, and thus, most violent headtail mode - mode 0.

\subsection{Fast headtail or transverse mode coupling instability}

If the machine is operated exactly at zero chromaticity, there are no headtail modes that are excited. In practice, zero chromaticity is not accurately attainable; thus the machine must be operated with a finite chromaticity. If a machine is operated with the right sign of chromaticity, mode 0 is usually successfully suppressed. Higher order headtail modes can often (but not always easily, as the example of the high intensity beams in the SPS has shown) be damped by other means, as already stated above. Nevertheless, there is yet another type of instability that can still occur even - and in fact preferably - at zero chromaticity and above a certain intensity threshold. This instability is exceptionally fast and violent and often provides a hard limit on the maximum attainable intensity in a machine. It emerges when two headtail modes couple. For this reason, this instability is called a Transverse Mode Coupling instability (TMCI). We will again phenomenologically give some background on the TMCI in this subsection. Good explanations of the TMCI with plenty of background information can be found in \cite{Chao:246480, Ng:1012829}.

A bunch circulating in a perfectly linear machine with no other external or collective effects performs betatron oscillations at a given tune $Q_x$, as derived from Hill's equation, according to
\[
y(s) = \sqrt{2 J_y\beta_y(s)}\,\cos\left(\frac{Q_y}{R}s\right)\,.
\]

\begin{figure}[htbp]
    \centering
    \subfigure[Pure betatron motion]{\includegraphics[width=0.490\linewidth]{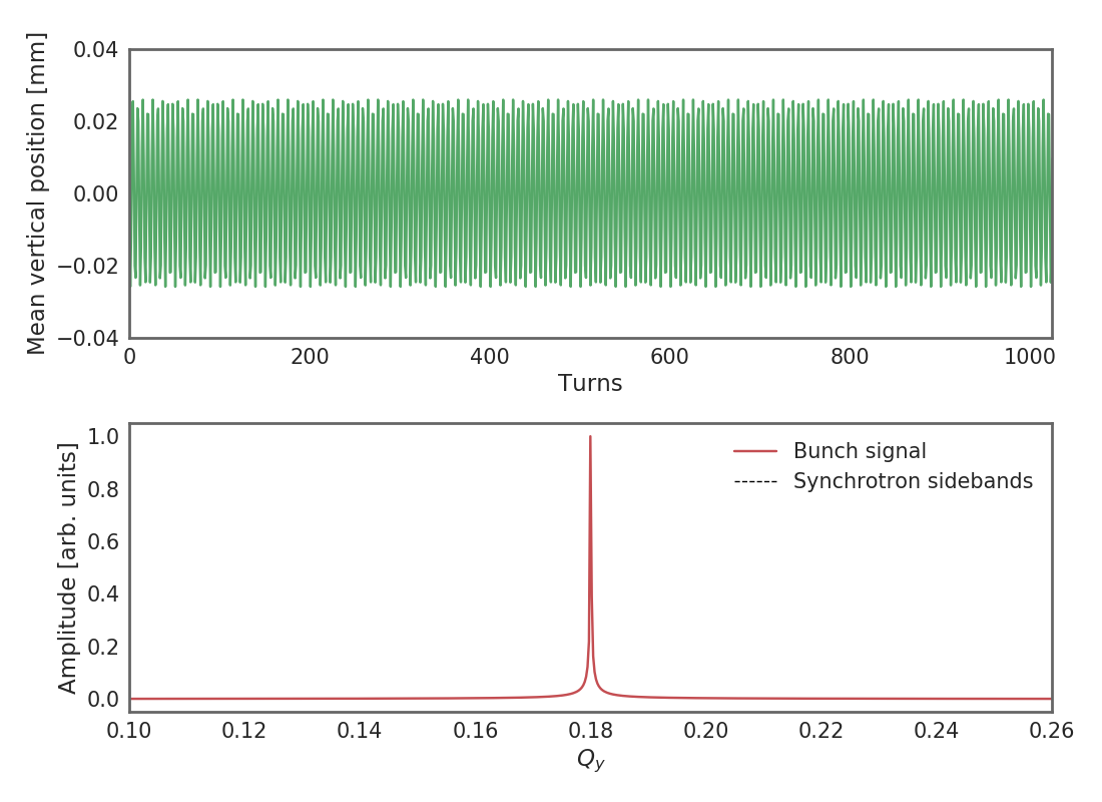}\label{fig:spectrum}}
    \subfigure[Coupled betatron and synchrotron motion]{\includegraphics[width=0.490\linewidth]{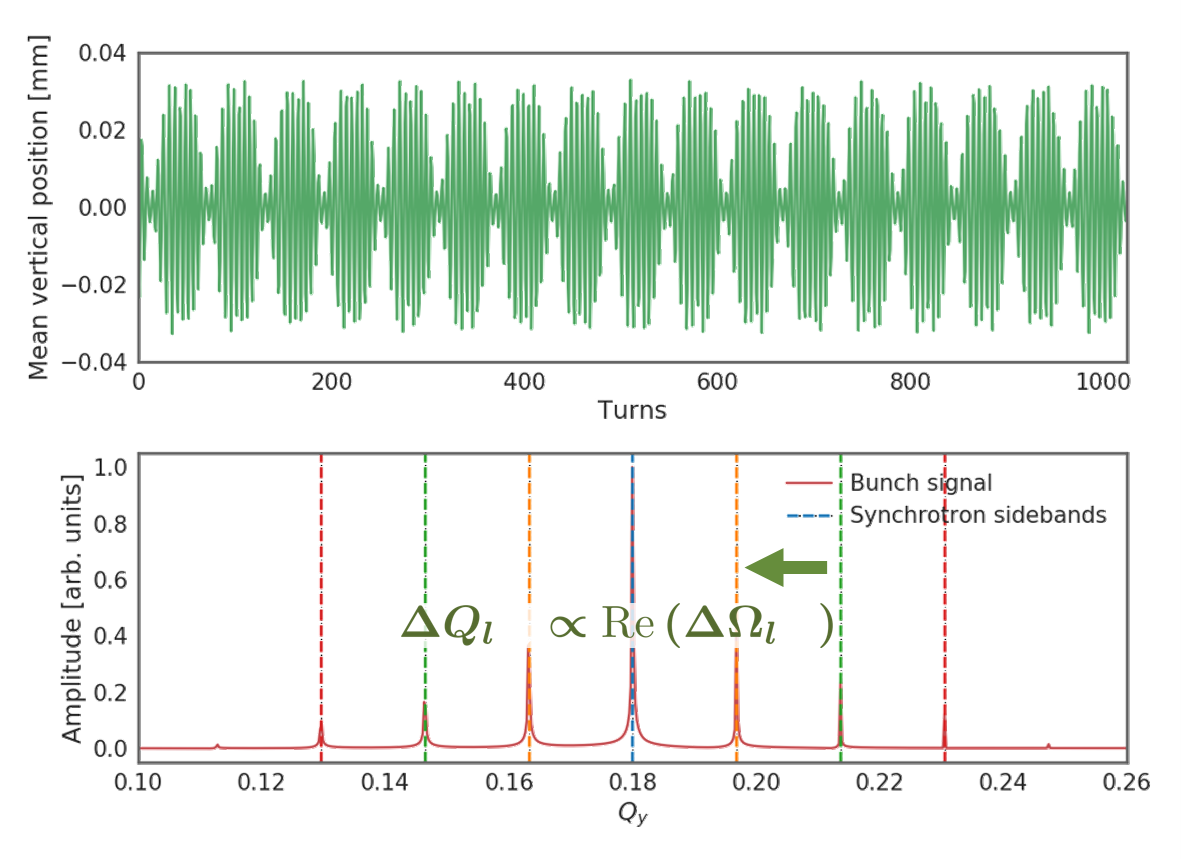}    \label{fig:spectrum_chroma}}
    \caption{Spectrum of a bunch performing a pure betatron motion in a perfectly linear machine; there is a single and clearly pronounced tune line.}
\end{figure}

The bunch spectrum for this type of motion can be obtained from a frequency analysis of the bunch centroid motion. This motion along with it's spectrum has been plotted in Fig.~\ref{fig:spectrum}. A single, clear line is visible at the tune. If instead, there exists a coupling between the transverse and the longitudinal motion, as is the case for chromaticity or in the presence wake fields, for instance, the betatron motion gets modulated with the synchrotron frequency. As a consequence, when doing a frequency analysis of the bunch centroid motion, one can observe not only the tune line, but also several (synchrotron) sidebands corresponding to additional bunch oscillation modes, who's degeneracy has been lifted by the coupling. This situation is clearly depicted in Fig.~\ref{fig:spectrum_chroma}. The separation between the individual lines is one synchrotron tune, for low intensity beams.

\begin{figure}[htbp]
    \centering
    \subfigure[Pure betatron motion]{\includegraphics[width=0.800\linewidth]{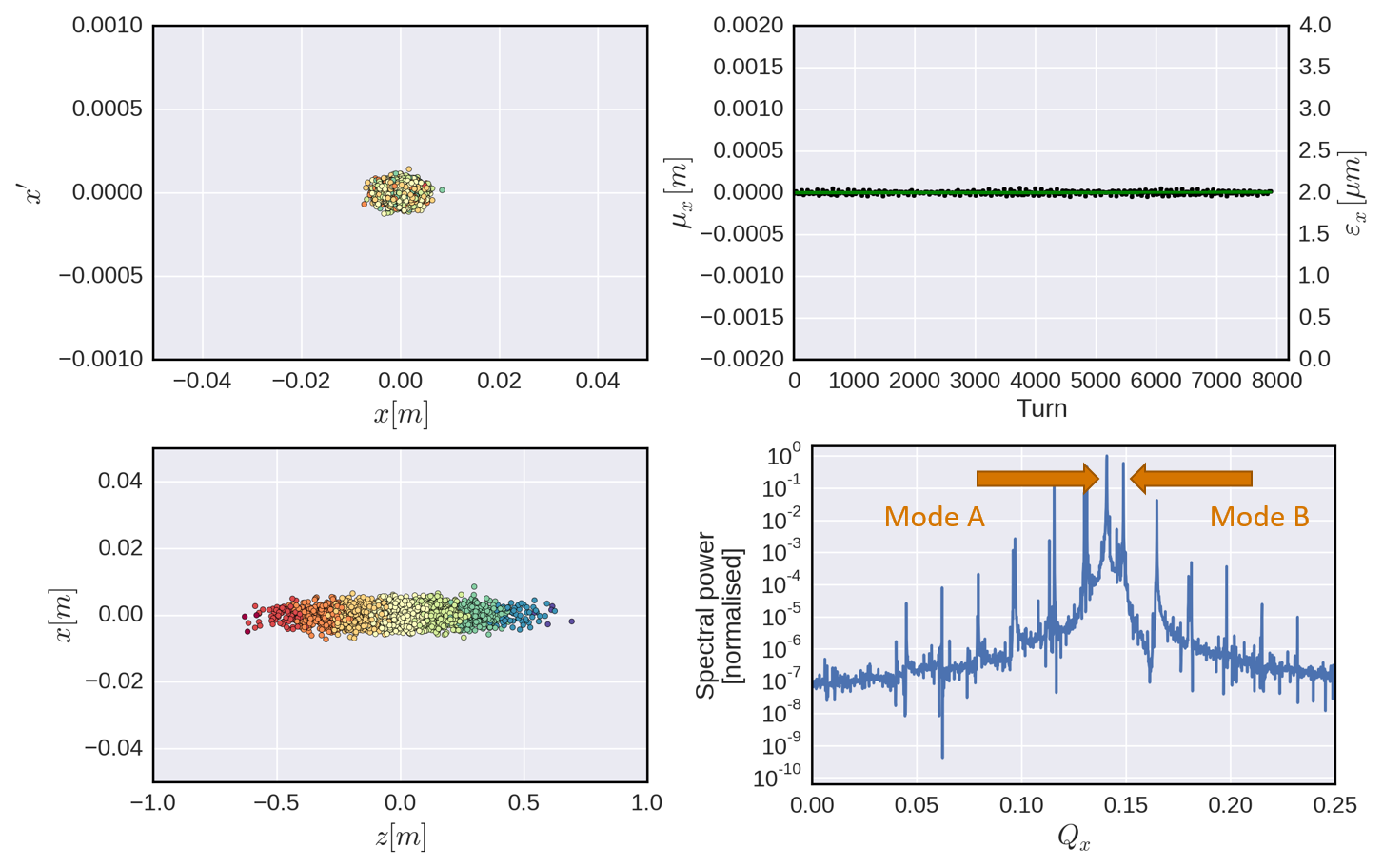}\label{fig:spectrum_simulation}}
    \subfigure[Coupled betatron and synchrotron motion]{\includegraphics[width=0.800\linewidth]{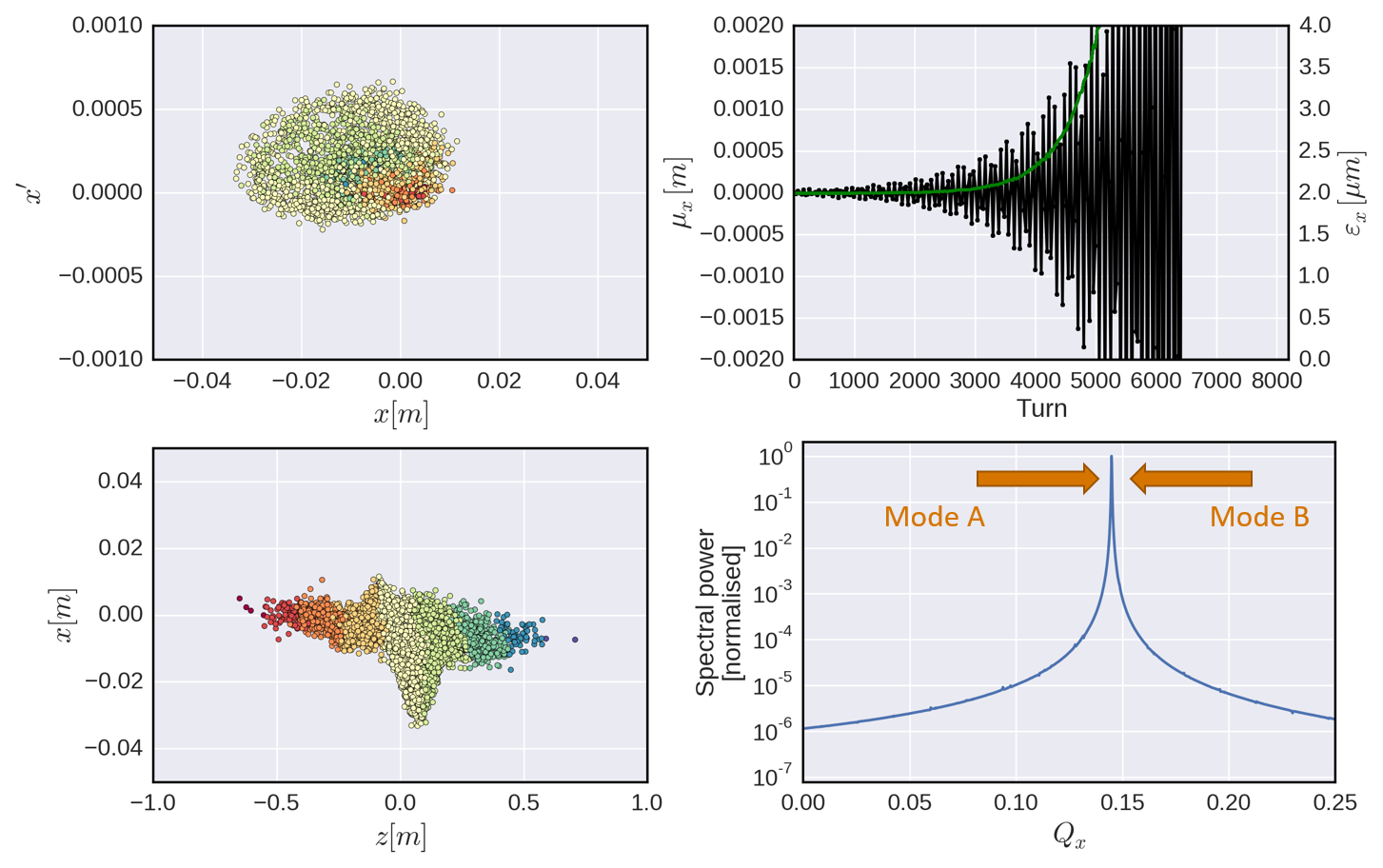}    \label{fig:spectrum_simulation_chroma}}
    \caption{Spectrum of a bunch performing a pure betatron motion in a perfectly linear machine; there is a single and clearly pronounced tune line.}
\end{figure}

Reasonably realistic scenarios can be evaluated using macro-particle simulation codes. These can treat multi-particle systems in realistic accelerator environments taking into account machine parameters and impedance models. Figures~\ref{fig:spectrum_simulation} and \ref{fig:spectrum_simulation_chroma} show the results of macro-particle simulations done for the SPS. On each of the figures, one can see the horizontal phase space (top-left), the full bunch view from the top (bottom-left), the centroid motion (top-right) and on the bottom right, the bunch spectrum obtained form the frequency analysis of the centroid motion. Figure~\ref{fig:spectrum_simulation} shows the situation below the threshold intensity using a broadband resonator impedance model for the SPS. In this situation, then bunch is inherently stable. However, due to the strongly coupled planes, especially via the wake fields, one can detect many coherent lines in the bunch spectrum shown on the bottom-right plot. These are coherent bunch modes which exist due to the presence of the wake fields. The peculiarity here is, that the individual lines shift differently in frequency as the bunch intensity is increased. Thus, it can happen that, as the intensity increases, two modes actually approach each other in frequency domain. This is indicated in Fig.~\ref{fig:spectrum_simulation} by the two arrows showing how modes $A$ and $B$ are pushing towards each other with increasing bunch intensity. Once the two lines fall onto each other the two modes become degenerate at the same frequency; they couple and this coupling leads to an exceptionally fast and potent instability. This is why this type of instability is called a Transverse Mode Coupling Instability, or TMCI. The situation of mode coupling is depicted in Fig.~\ref{fig:spectrum_simulation_chroma}. The coupled mode becomes the, by far, single dominant line. The bunch oscillates vastly with a strong intra-bunch motion and the bunch centroid motion grows exponentially at a very fast growth rate (< 1000 turns, in this case). This coupling of modes happens at a distinct bunch intensity. Below this intensity, the bunch remains stable, above this intensity, it becomes severely unstable. A TMCI is thus a threshold effect; the TMCI threshold poses a hard limit on the maximum attainable intensity of most accelerators.

Increasing the TMCI threshold of a machine is not always straightforward. The best and most sustainable solution is often to apply the cure directly at the source. Namely, this means keeping a tight and stringent impedance budget for a machine or carrying out dedicated impedance reduction campaigns for machines where the impedance budget has been exceeded. The TMCI threshold can also be increased by optimizing the optics of a machine. This has been done in recent years at the SPS \cite{Bartosik:1644761}. By moving to a lower integer tune of the machine, the dispersion in the dipoles is increased which leads to a lower transition energy of the machine. In consequence, the slippage factor is increased, which leads to a larger synchrotron tune and thus to a larger separation of the individual modes observed in Figs.~\ref{fig:spectrum_chroma} or \ref{fig:spectrum_simulation}.

\begin{figure}[htbp]
    \centering
    \subfigure[T][Transverse Mode Coupling]{\includegraphics[width=0.545\linewidth]{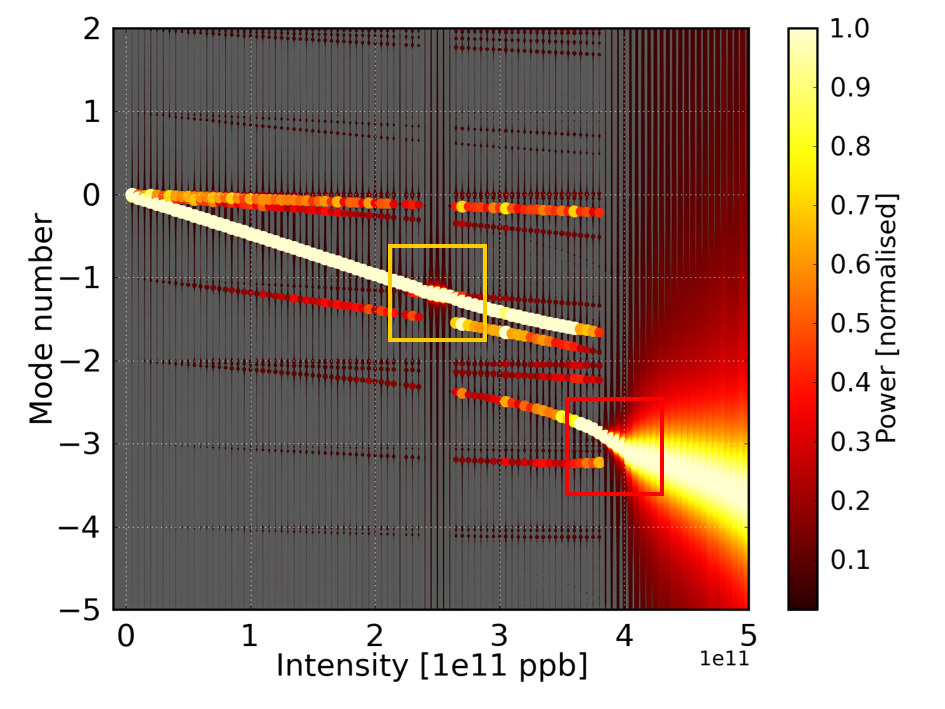}\label{fig:tmci_plot}}
    \subfigure[T][Intra-bunch motion for the slow and the fast TMCI]{\includegraphics[width=0.445\linewidth]{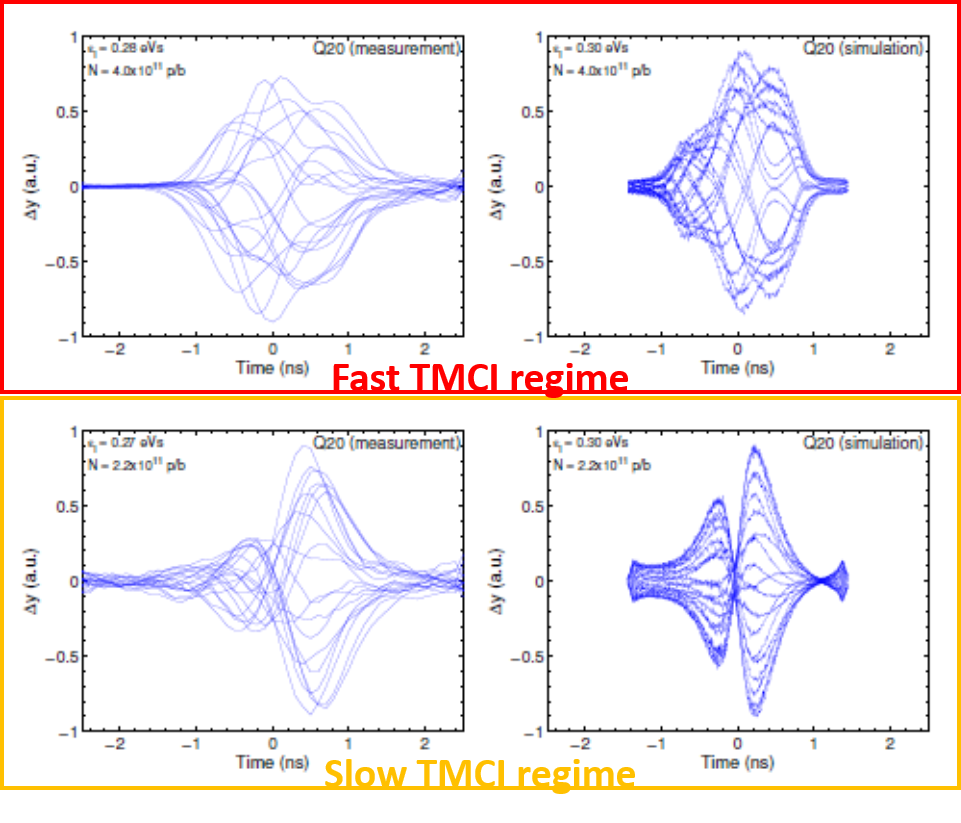}\label{fig:tmci_waveforms}}
    \caption{Graphical visualization of the Transverse Mode Coupling instability as simulated for the SPS using a broadband impedance model. A weak coupling can be observed for lower intensities (red box); the actual TMCI occurs at an intensity of roughly 4e11 ppb, which is the current ultimate intensity limit of the machine.}
\end{figure}

Figure~\ref{fig:tmci_plot} shows a typical plot used for visualizing the TMCI as function of intensity. At close to zero intensity we can find a clear mode 0 along with the different synchrotron sidebands yielding modes 1, 2, etc., each separated by the synchrotron tune. With rising bunch intensity, mode 0 shifts downwards, as do most of the other modes. However, the rate of the modes shifting is different and, consequently, the different mode lines cross. A first encounter of two modes occurs at about 2.5e11 ppb between modes 0 and -1. This is just a weak coupling and the modes separate as the intensity increases. The actual TMCI takes place at an intensity of about 4e11 as modes -2 and -3 couple strongly. This leads to a violent instability accompanied by a large loss of particles down to below the TMCI threshold. The strong coherent intra-bunch motion patterns are shown in Fig.~\ref{fig:tmci_waveforms} for the two different TMCI regimes both as simulated and as measured in the machine. It becomes clear from the inspection of Fig.~\ref{fig:tmci_plot} that, increasing the separation of the different synchrotron sidebands, or mode lines, by increasing the synchrotron tune will lead to a later coupling, in terms of intensity, between the modes, and thus, push the TMCI intensity threshold to larger values. This techniques has been successfully deployed in the SPS as one of the pre-requisites for the LIU Project.

A final, very recent potential mitigation of TMCI is via active feedback. As we can see from Fig.~\ref{fig:tmci_waveforms}, the TMCI generates a very strong intra-bunch motion which, depending on the bunch length, can feature very high frequency content (in the SPS, for instance, this frequency can reach into the GHz range). Due to its severity and also its mechanism, a TMCI is usually very hard, if not close to impossible, to effectively mitigate by passive means, such as chromaticity or Landau damping, for example. Conventional transverse feedback systems operating with short bunches are able to act on the bunch centroid motion, but don't usually have the required bandwidth to access the frequency regime of the intra-bunch motion. Recently, new kicker structures and very fast and powerful digital signal processing chains have been developed for a demonstrator prototype wideband feedback system to be suited for the mitigation of high frequency instabilities, in particular TMCI or electron cloud effects, in the SPS. Experiments using this novel system were carried out in 2017 \cite{Li:2640634}.

\begin{figure}[htbp]
    \centering
    \subfigure[TMCI threshold in the SPS]{\includegraphics[width=0.480\linewidth]{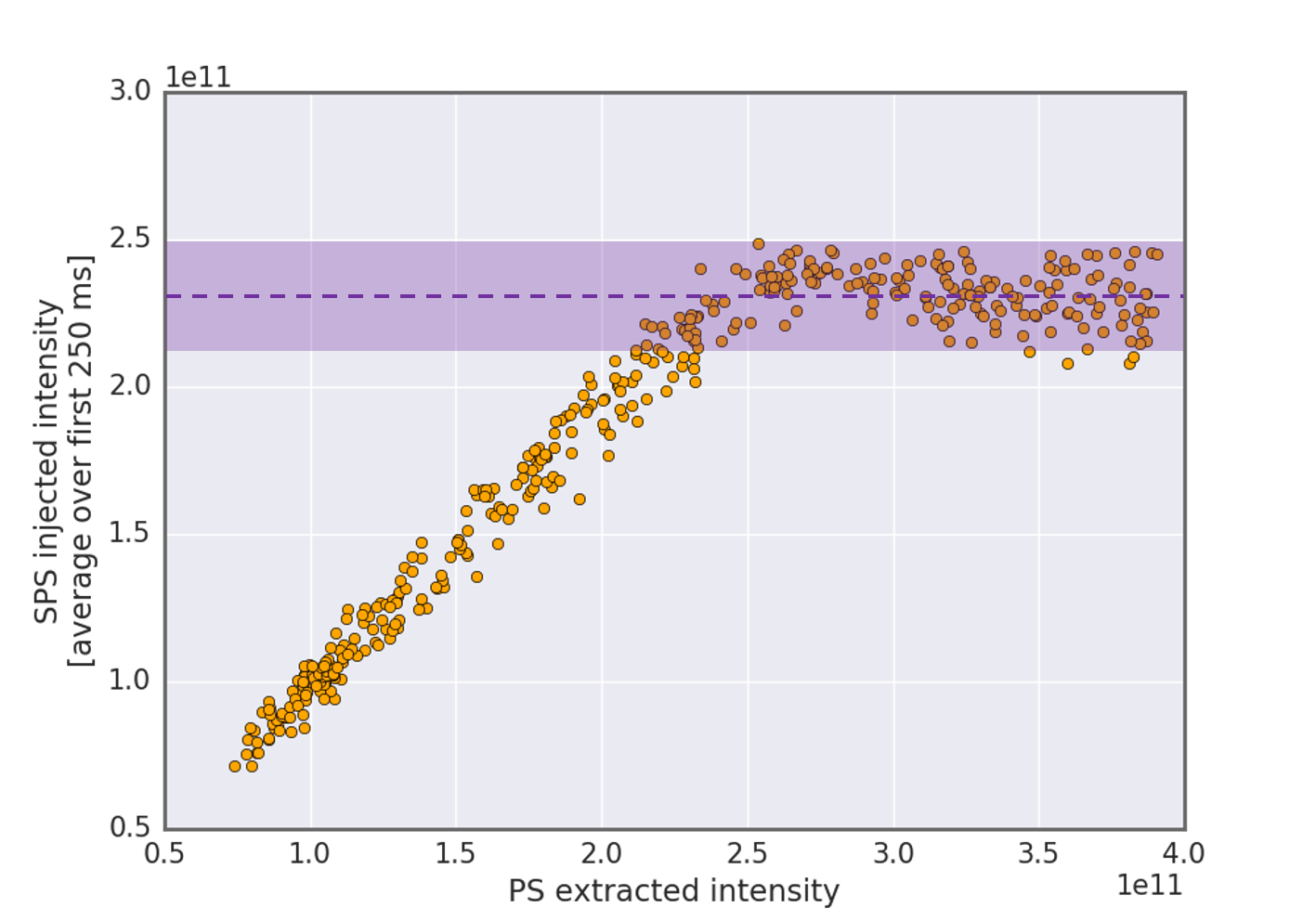}    \label{fig:tmci_sps_injection}}
    \subfigure[BCT curves without and with feedbacks]{\includegraphics[width=0.500\linewidth]{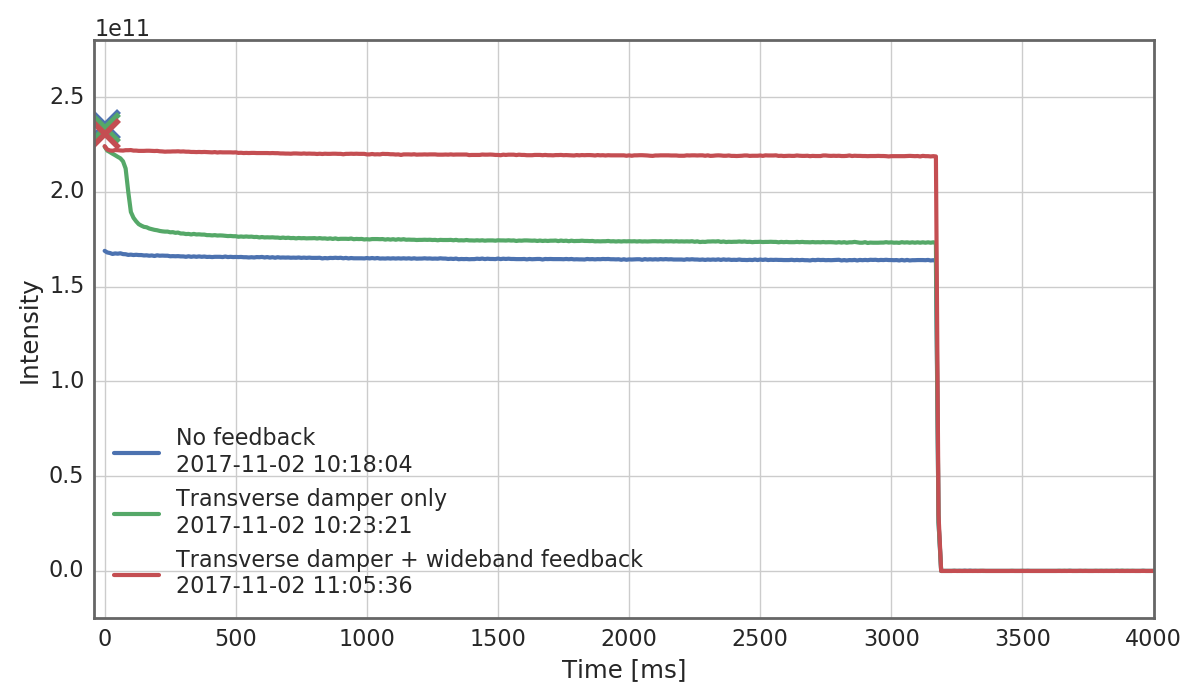}    \label{fig:tmci_wbfb_meas}}
    \caption{The plots above show a measurement of the TMCI threshold in the SPS and the wideband feedback system performance to mitigate these instabilities.}
\end{figure}

Figure~\ref{fig:tmci_sps_injection} shows a systematic measurement to disclose the TMCI threshold in the SPS. Bunches of increasing intensity were extracted from the PS and injected into the SPS. The intensities were measured in the PS just before extraction and in the SPS a few milliseconds after injection. While, initially, the dependency between extracted intensity in the PS and measured intensity in the SPS is linear, it is clearly visible, how the intensity in the SPS saturates at the maximum attainable bunch intensity; this intensity limit is defined by the TMCI threshold which, for the experiment in Fig.~\ref{fig:tmci_sps_injection}, was around 2.4e11~ppb. As the bunch becomes unstable immediately after injection into the SPS, the bunch begins to oscillate wildly which leads to significant particles losses. These losses reduce the overall bunch intensity; the bunch stabilizes, once the total number of lost particles brings back the total bunch intensity below the TMCI threshold. In a follow-up experiment, an attempt was made to breach this intensity limitation and to sustain the originally extracted bunch intensity from the PS after injection in the SPS. It is worth noting, that in this experiment the TMCI threshold was artificially reduced for machine protection reasons to 1.6e11~ppb. Figure~\ref{fig:tmci_wbfb_meas} shows the BCT curve along an SPS cycle exhibiting the evolution of the injected beam intensity for three different machine configurations. The crosses at time 0~ms correspond to the bunch intensity measured at extraction in the PS. With no feedback system active (blue curve) one can clearly observe the very fast and high particle losses taking place right after injection, even before the first BCT measurement 5~ms after injection into the SPS. With the conventional transverse damper active (red curve), the bunch can be stabilized from some time as the bunch centroid motion is damped right after injection. However, as the intra-bunch motion triggered by the TMCI keeps taking off, the particle losses eventually occur and the total bunch intensity settles at a value below the TMCI threshold. It is only with the wideband feedback system active, that the TMCI is completely suppressed and the full injected bunch intensity can be sustained over the full cycle. This has been the first proof-of-concept that TMCI can be suppressed for short bunches using active feedback systems of sufficiently high bandwidth.

To summarize, apart from the (slow) headtail instability, we have seen yet another, particularly violent type of instability, the fast headtail or Transverse Mode Coupling Instability. As opposed to the slow headtail instability, this instability can occur also at zero chromaticity. Furthermore, whereas the slow headtail modes are always excited for non-zero chromaticity at a finite growth rate, the TMCI occurs only when the bunch intensity is large enough to exceed the TMCI threshold. At this intensity, existing separate bunch modes will become degenerate and couple which leads to a very fast and violent instability. TMCI usually poses a hard limit on the maximum attainable intensity of a machine. The TMCI threshold is usually already taken into account during design phase of a machine. Otherwise, impedance reduction campaigns or machine optics optimization can help to increase the TMCI threshold. Recent developments of wideband feedback systems indicate, however, that these limits could probably also be breached by means of active mitigation.

\subsection{Longitudinal instabilities}

In this last subsection of our introductory lecture on collective effects in beam dynamics we will spend a few words on collective effects and instabilities in the longitudinal plane, without going much into details. Of course, as we have seen in the previous section, there are longitudinal wake fields which are equally well capable of exciting longitudinal instabilities. The mechanics are slightly different, though. Any coupling between the transverse and longitudinal planes can be neglected, in general, such that the beam dynamics considerations can be discussed remaining in a single plane only. Moreover, higher order wakefields are not considered; we always stay at the zeroth order wake field term as already pointed out for Eq.~\ref{eq:longitudinal_wake_function}. On the other hand, the longitudinal motion and its instabilities have their own peculiarities and difficulties.

Deriving from the longitudinal equations of motion in presence of the longitudinal wake function and looking at single bunch instabilities one  can identify two distinct regimes of dynamics. At low intensities, the additional wake field term can be regarded as a modification of the longitudinal potential well. Hence, this regime is also called the regime of potential well distortion. This leads to a modification of the equilibrium solution, accompanied by a shift in the stable phase and the synchrotron frequency and a change of the matched distribution (which can lead to a bunch lengthening, especially for lepton machines). The regime of potential well distortion, thus, leads to a new equilibrium solution which is stable and does not feature any growth in amplitude. Beyond a certain intensity, however, this regime switches abruptly to the regime of longitudinal instabilities. Above the threshold intensity, the bunch suddenly becomes unstable, performing dipole oscillations, or forming distorted distributions in longitudinal phase space (solitons) accompanied by bunch lengthening and particle losses. The latter is a signature of the microwave or longitudinal mode coupling instability, which can be seen as the longitudinal pendent to the TMCI.

\begin{figure}[htbp]
    \centering
    \subfigure[Transition between stable and unstable regimes]{\includegraphics[width=0.650\linewidth]{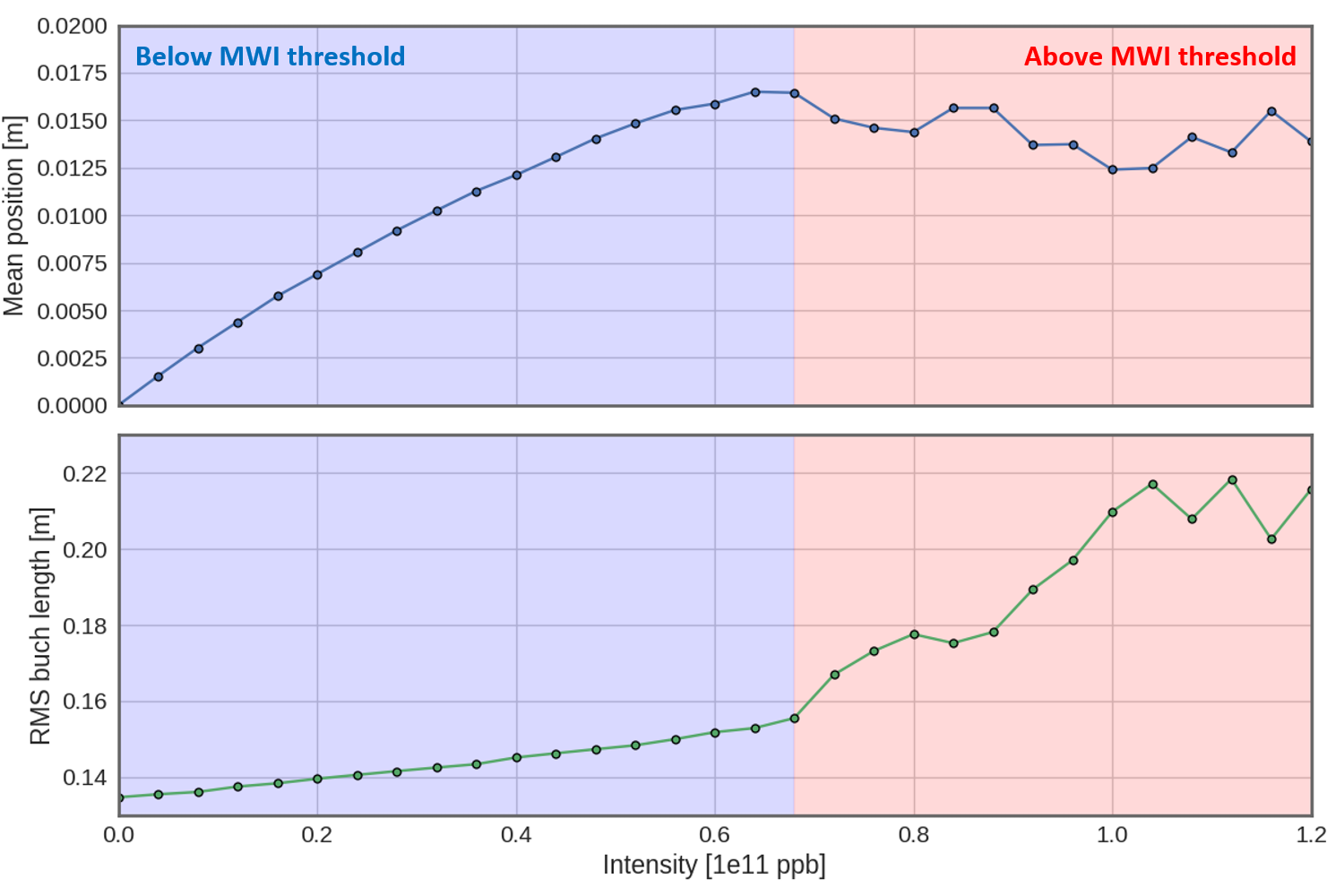}}
    \subfigure[Stable regime phase space]{\includegraphics[width=0.495\linewidth]{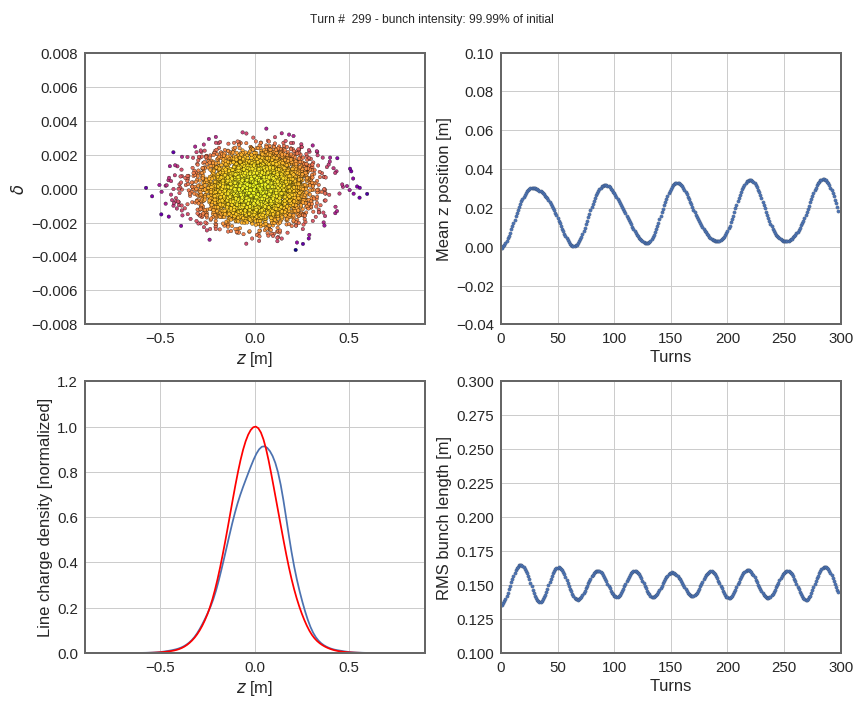}}
    \subfigure[Unstable regime phase space]{\includegraphics[width=0.495\linewidth]{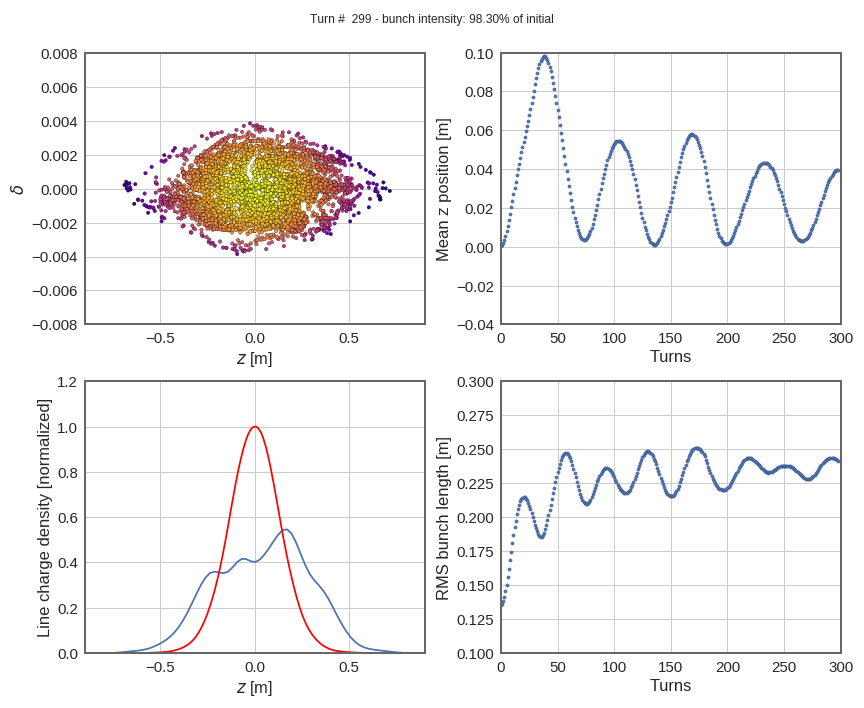}}
    \caption{Illustration of the transition from the stable into the unstable regime for the longitudinal bunch motion; the phase space plots show the chaotic motion and turbulent bunch lengthening above the MWI threshold.}
    \label{fig:microwave}
\end{figure}

A plot obtained from macro-particle simulations using a simple resonator wake field highlights these two regimes and is shown in Fig.~\ref{fig:microwave}. Below the microwave instability threshold, in the bunch lengthening and emittance blow-up regime, the beam dynamics is characterized mainly by a linear increase with intensity of the synchronous phase and of the bunch length. Above this threshold the bunch enters the turbulent bunch lengthening regime. The bunch becomes vastly unstable, high frequency modulations appear on the longitudinal bunch profile and particles are eventually lost out of the RF bucket.  

Of course, the longitudinal plane features many other types of instabilities, which we can not further discuss here. These include longitudinal coupled bunch instabilities, the negative mass instability or the Robinson instability, just to name a few.

\begin{figure}[htbp]
    \centering
    \subfigure[Bunch synchrotron tune spectrum]{\includegraphics[width=0.750\linewidth]{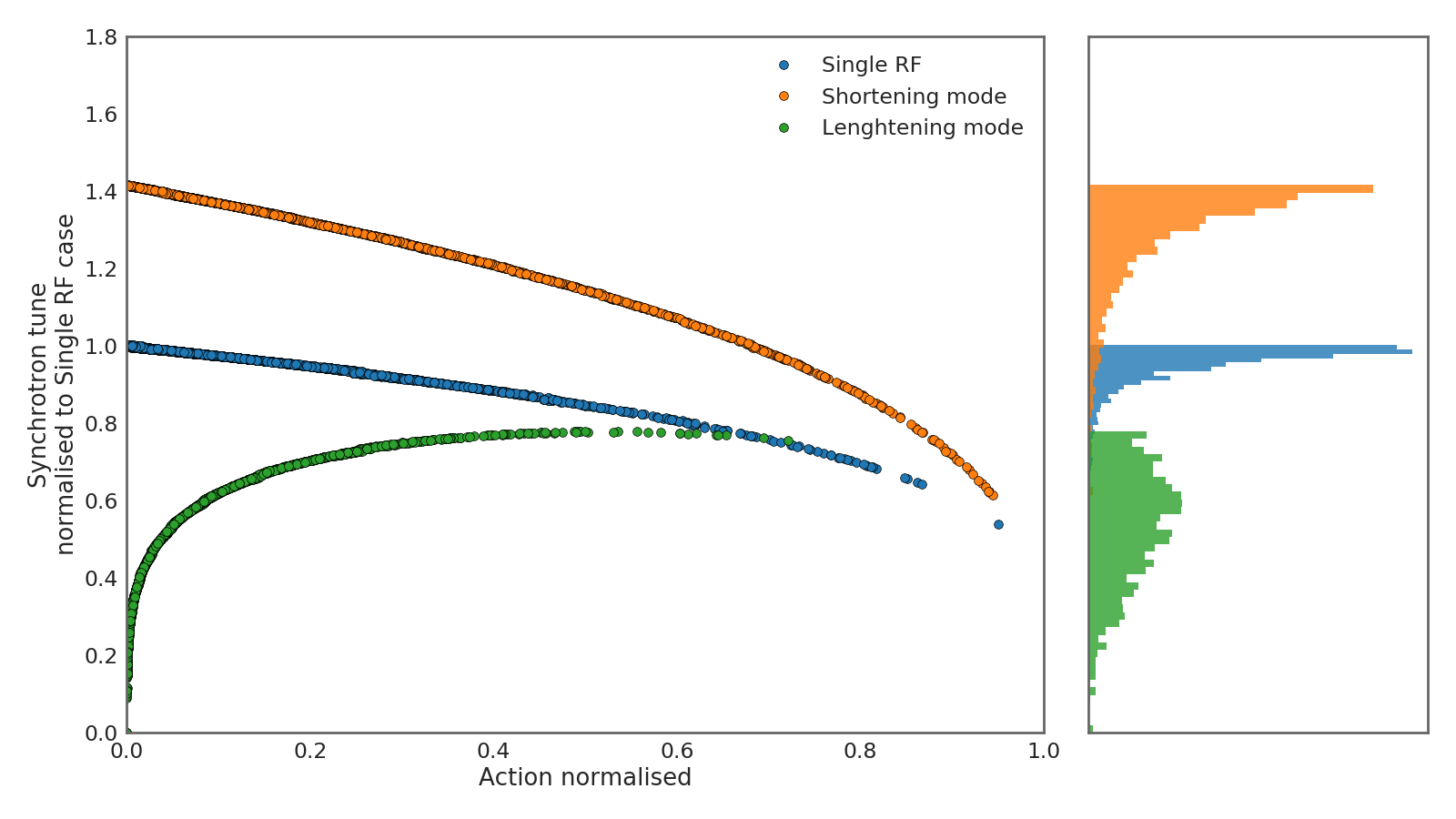}}
    \subfigure[Single harmonic]{\includegraphics[width=0.325\linewidth]{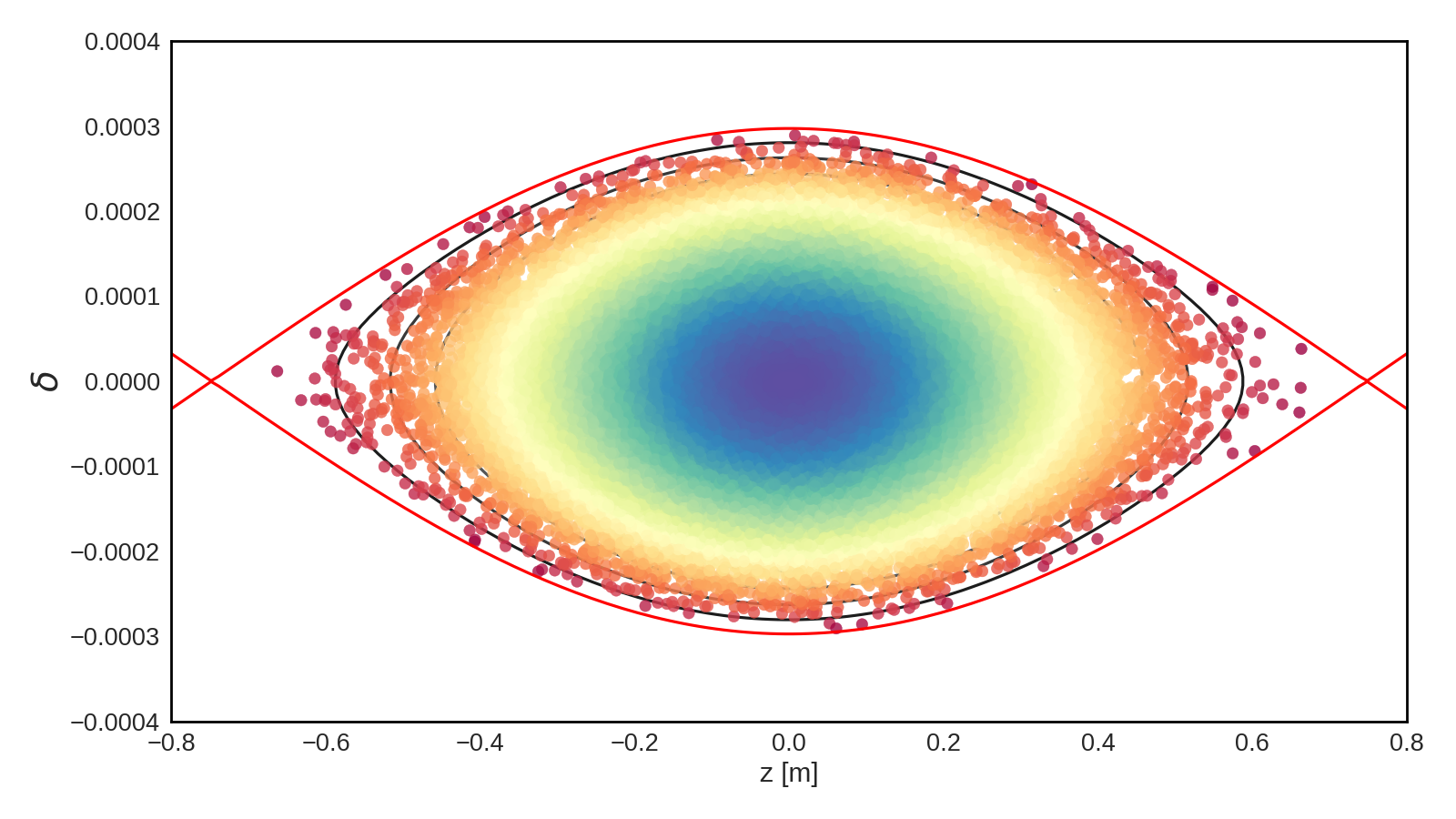}}
    \subfigure[Bunch shortening mode]{\includegraphics[width=0.325\linewidth]{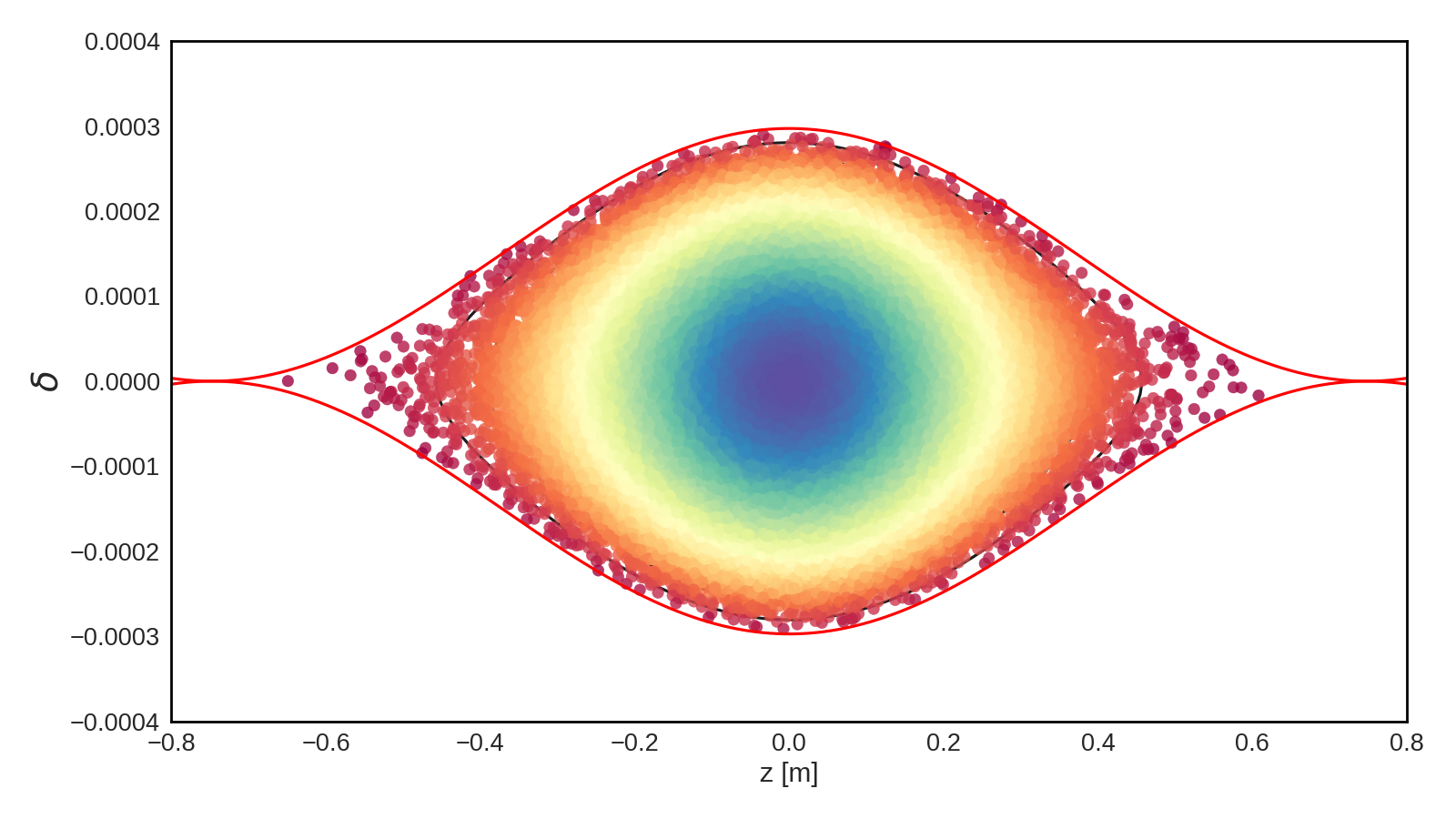}}
    \subfigure[Bunch lengthening mode]{\includegraphics[width=0.325\linewidth]{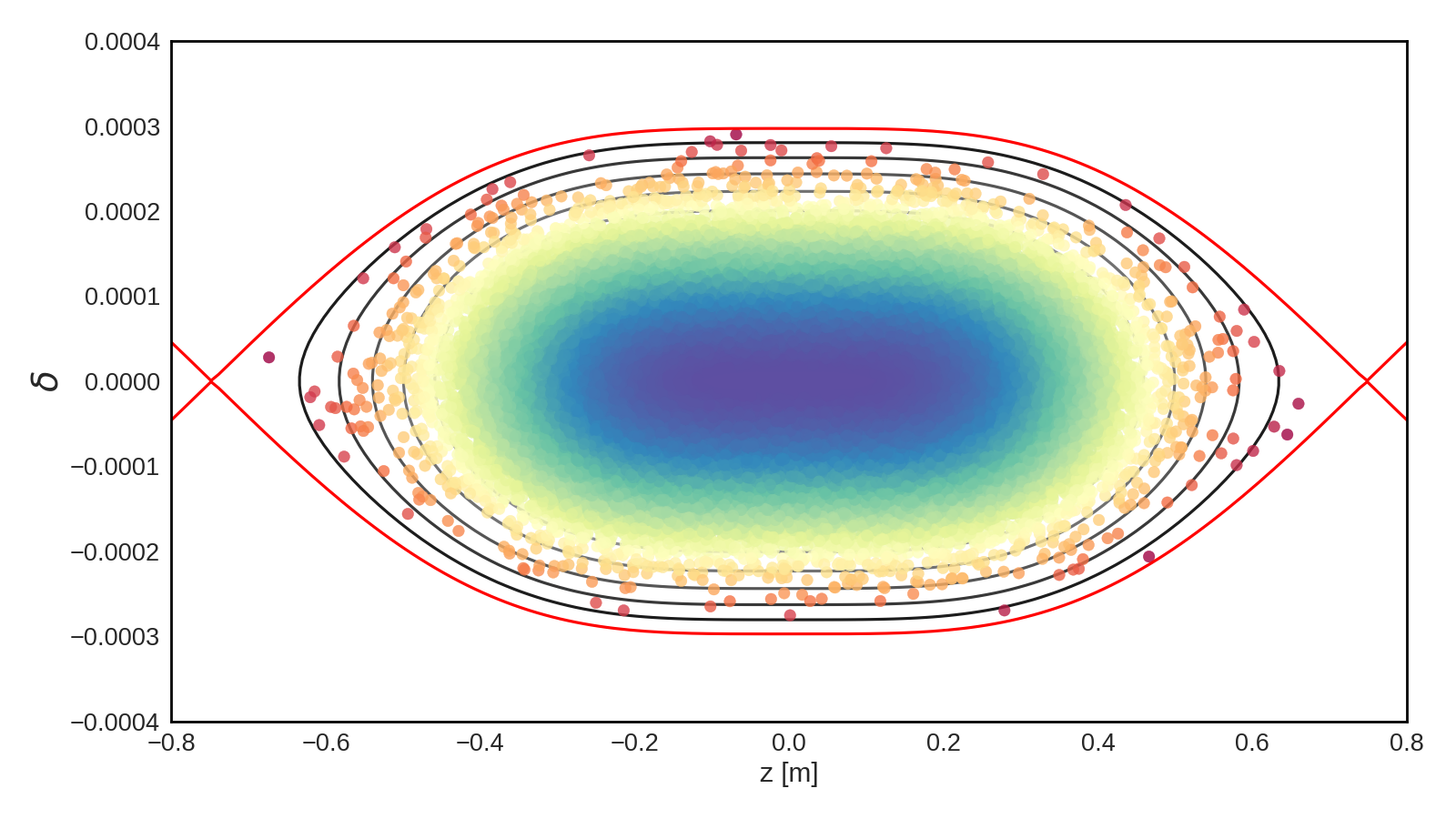}}
    \caption{The effect of a double harmonic RF system on the incoherent synchrotron tune spectrum. Bunch shortening mode provides a good compromise between fast synchrotron tunes and wide and stable tune spreads.}
    \label{fig:synch_tune_spectrum}
\end{figure}

As for the transverse plane, adequate mitigation methods need to be found in order to overcome the limitations imposed also by longitudinal instabilities. Apart from the obvious ones such as respecting an impedance budget or undertaking an impedance reduction campaign, one can also resort to more passive, beam dynamics oriented mitigation techniques. One method often deployed is similar to the use of octupoles in the transverse plane. By deliberately increasing the incoherent tune spectrum of the bunch, one can suppress existing instabilities via Landau damping. This is achieved in the longitudinal plane by the employment of harmonic cavities \cite{Argyropoulos:2285796, Repond:2695204}. Fig.~\ref{fig:synch_tune_spectrum} shows the RF buckets along with the longitudinal phase space and the resulting incoherent synchrotron tune spectrum of the particle distribution for different configurations of fundamental and harmonic RF voltages. There are three modes of operation highlighted: single harmonic, double harmonic in bunch shortening and double harmonic and bunch lengthening mode. Apart from changing the bunch shape and introducing bunch shortening or lengthening, the double harmonic also affects the synchrotron tune spread as can be seen on the top plot. At the SPS, bunch shortening mode is used for high intensity beams as it combines fast synchrotron motion with a decent and stable tune spread. in fact, this higher harmonic RF system is a key component for the stabilization of the future LIU beams in the SPS.

\subsection{Closing remarks}

In the last four lectures we have briefly touched the topic of collective effects in accelerator physics and beam dynamics. We have treated most items phenomenologically to gain an intuitive understanding of the involved mechanisms. By now we should be able to identify the main differences between the dynamics of single particles vs. multi-particle systems. We understand the features of collective effects such as space charge and how instead of being constant forces, instead, they depend on the particle distribution function itself. We know how we can use the concept of wake fields and impedances to model the impact of more complex elements. And we know how to identify an instability, with a rough overview over the different type of instabilities typically observed in synchrotrons along with possible mitigations. More complex analysis involve the Vlasov formalism to analytically model simplified cases or full macroparticle models to simulate the beam dynamics of collective effects.
\clearpage\pagebreak


\medskip
\printbibliography

\end{document}